\documentclass[a4paper,superscriptaddress,floatfix,nofootinbib,onecolumn,longbibliography]{revtex4-1}

\usepackage{bbm}
\usepackage{bbold}
\usepackage{bbm}
\usepackage[pdftex]{graphicx}
\usepackage{latexsym,amsmath,verbatim,amssymb,txfonts}
\usepackage{lipsum}
\usepackage{color}
\usepackage{rotating}
\usepackage{verbatim}
\usepackage{multirow}
\usepackage[english]{babel}
\usepackage{comment}
\usepackage{bm}
\usepackage{amsmath}
\usepackage{amsfonts}
\usepackage{amssymb}
\usepackage{float}
\usepackage{color}
\usepackage{braket}
\usepackage{blkarray, bigstrut}
\usepackage{nicematrix}

\begin{document}
\title{Synchronization induced by directed
higher-order interactions}

\author{Luca Gallo}
\thanks{These two authors contributed equally}
\affiliation{Department of Physics and Astronomy \& INFN, University of Catania, Italy}
\affiliation{naXys, Namur Institute for Complex Systems, University of Namur, Belgium}
\email{luca.gallo@phd.unict.it}

\author{Riccardo Muolo}
\thanks{These two authors contributed equally}
\affiliation{naXys, Namur Institute for Complex Systems, University of Namur, Belgium}
\affiliation{Department of Mathematics, University of Namur, Belgium}

\author{Lucia Valentina Gambuzza}
\affiliation{Department of Electrical, Electronics and Computer Science Engineering, University of Catania, Italy}

\author{Vito Latora}
\affiliation{Department of Physics and Astronomy \& INFN, University of Catania, Italy}
\affiliation{School of Mathematical Sciences, Queen Mary University of London, UK}
\affiliation{Complexity Science Hub Vienna, Austria}

\author{Mattia Frasca}
\affiliation{Department of Electrical, Electronics and Computer Science Engineering, University of Catania, Italy}
\affiliation{Istituto di Analisi dei Sistemi ed Informatica “A. Ruberti”, Consiglio Nazionale delle Ricerche (IASI-CNR), Roma, Italy}

\author{Timoteo Carletti}
\affiliation{naXys, Namur Institute for Complex Systems, University of Namur, Belgium}
\affiliation{Department of Mathematics, University of Namur, Belgium}


\begin{abstract}
Non-reciprocal interactions play a crucial role in many social and biological complex systems. While directionality has been thoroughly accounted for in networks with pairwise interactions, its effects in systems with higher-order interactions have not yet been explored as deserved. Here, we introduce the concept of $M$-directed hypergraphs, a general class of directed higher-order structures, which allow to investigate dynamical systems coupled through directed group interactions. As an application we study the synchronization of nonlinear oscillators on $1$-directed hypergraphs, finding that directed higher-order interactions can destroy synchronization, but also stabilize otherwise unstable synchronized states.

\end{abstract}

\maketitle

\section*{Introduction}\label{sec1}

Network science is a powerful {and effective} tool in modeling natural and artificial {systems with a discrete topology}. The study of dynamical systems on networks has {thus} triggered the interest of scientists and has spread {across} disciplines, from physics and engineering, to social science and ecology~\cite{newmanbook,boccaletti2006complex,latora_nicosia_russo_2017}. Network models rely on the hypothesis that the interactions between the units of a system are pairwise~\cite{battiston2020networks}{. However this is only a first order approximation in many} empirical systems, such as protein interaction networks~\cite{klamt2009hypergraphs,estrada2018centralities}, brain networks~\cite{petri2014homological,giusti2015clique,sizemore2018cliques,Bassett}, social systems~\cite{benson2016higher,patania2017shape} and ecological networks~\cite{billick1994higher,bairey2016high,grilli2017higher}, {where group interactions are widespread and important.}
Recent years have thus witnessed an increasing research interest for more complex mathematical structures, such as simplicial complexes and hypergraphs~\cite{berge1973graphs,battiston2020networks,maxime2020,carletti2020dynamical,de2021phase}, capable of encoding many-body interactions. These systems 
have been used to investigate various dynamical processes, such as epidemic and social contagion \cite{stonge2021universal,iacopini2019simplicial,de2019social}, random walks \cite{carletti2020random,carletti2021random}, synchronization \cite{skardal2019abrupt,skardal2019higher}, consensus \cite{neuhauser2020multibody,neuhauser2021}, to name a few. However, the proposed 
formalism is not general enough to describe systems where the {\em group interactions are 
intrinsically asymmetric}. For instance, group pressure or bullying in social systems have an asymmetric nature, due to the fact that group interactions are addressed {against} one or more individuals but {(often)} not reciprocated \cite{asch1961}. (Bio)chemical reactions are another typical example of higher-order directed processes, as, though some reactions can be reversible, there is often a privileged direction due to thermodynamics~\cite{enzymeskin,klamt2009hypergraphs}. Further examples come from the ecology of microbial communities, where a direct interaction between two species can be mediated by a third {one}~\cite{kelsic2015counteraction,abrudan2015socially}.

Although including some form of directionality in  higher-order  structures is not entirely new \cite{gallo1993directed,klamt2009hypergraphs}, 
the few existing attempts to study the effects of directionality 
on dynamical processes all suffer from a series of limitations. 
For example, in the case of oriented hypergraphs, where the nodes of each hyperedge are partitioned into an input and an output set (not necessarily disjoint), because of the underlying assumptions, one ultimately gets symmetric operators {(e.g., the adjacency or the Laplacian matrix)} despite one would expect directed interactions to yield asymmetric ones 
\cite{mulas_chemical, mulas_oriented, mulas_oriented2}. 
Furthermore, in the case of simplicial complexes~\cite{schaub2018flow,barbarossa2020topological,millan2020explosive,petri_hodge} an orientation has been introduced with the purpose of defining (co-)homology operators, but is not associated to directionality, i.e., the Laplacian matrix is symmetric once again.

Here we introduce the framework of \emph{$M$-directed hypergraphs}, which naturally leads to an asymmetric higher-order Laplacian and allows to study the dynamics of systems (e.g., nonlinear oscillators) with higher-order interactions fully accounting for their directionality. 
We focus, in particular, on synchronization, a phenomenon of utmost importance in many natural and artificial networked systems \cite{boccaletti2018synchronization}. In order to assess the stability of a synchronized state, we determine conditions under which a Master Stability Function (MSF) approach~\cite{pecora1998master,krawiecki2014chaotic,gambuzza2021stability} can be generalized to such directed higher-order structures. As we will show in the following, the complex spectrum of the asymmetric Laplacian operator entering into the MSF has a strong impact on the system behavior. Indeed, we can determine cases where the presence of  directionality in  higher-order interactions can destabilize 
the complete synchronized state of the system, otherwise obtained with reciprocal, i.e., symmetric coupling. 
Analogously, we also find cases where the opposite behavior is observed, i.e., higher-order directionality is the main driver for the onset of synchronization. 

\section*{Results}\label{sec2}

\subsection*{$M$-directed hypergraphs allow to model directionality in higher-order interactions}

\begin{figure*}[t]
\centering
\includegraphics[width=1\linewidth]{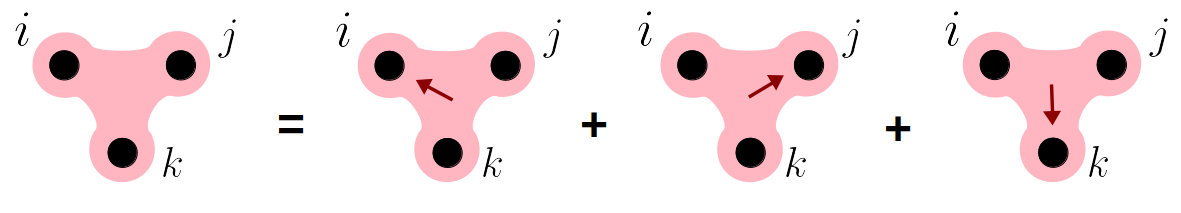}
\includegraphics[width=0.95\linewidth]{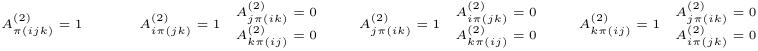}
\caption{An undirected $2$-hyperedge can be seen as the composition of three directed hyperedges. It is important to note that, in each of the directed $2$-hyperedges, the nodes acting as source of the interaction commute, i.e., a permutation of them does not alter the nature of the interaction, as denoted by the adjacency tensors.}
\label{fig:semi_dir}
\end{figure*}

To introduce the framework we start by defining a { \em $1$-directed $d$-hyperedge} as a set of $(d+1)$ nodes, $d$ of which, the {\em source} nodes, ``point'' toward the remaining one; \textcolor{black}{let us observe that we used the notation where a $d$-hyperedge represents the interactions among $d+1$ agents (this is similar to the notation adopted for simplicial complexes, where a $d$-simplex models the interactions of $d+1$ agents, while, often, for hypergraphs a $d$-hyperedge accounts for the interactions among $d$ agents \cite{carletti2020dynamical}).} In this way, an undirected $d$-hyperedge can be seen as the union of $(d+1)$ directed ones (see Fig.~\ref{fig:semi_dir}).
Notice that this is a natural extension of the network framework, in which a pairwise undirected interaction can be decomposed into two directed interactions. {A $1$-directed $d$-hyperedge, where the source nodes $j_1$, $j_2$, $\dots$, $j_d$ point toward node $i$, can be represented by an adjacency tensor $A^{(d)}$ with the following property}
\begin{equation}
A_{ij_1\dots j_d}^{(d)}=1 \Rightarrow A_{i\pi(j_1\dots j_d)}^{(d)}=1\, , 
\label{eq:semidir_def}
\end{equation} 
where $\pi(j_1,\dots ,j_d)$ is any permutation of the indices $j_1,\dots, j_d$ (Fig.~\ref{fig:semi_dir}). Observe that a generic permutation involving also index $i$ does not necessarily imply a nonzero entry in the adjacency tensor,
i.e., $A^{(d)}$ is in general asymmetric. Note however that the $(d-1)$-th rank tensors obtained by fixing the first index of $A^{(d)}$ are symmetric. 
By \emph{$1$-directed $D$-hypergraph} we define a hypergraph formed by $1$-directed $d$-hyperedges of any size $d$ smaller or equal to $D$.  
Note that these definitions provide a formalization in terms of tensors of the concept of \emph{B-arc} and \emph{B-hypergraph} introduced in~\cite{gallo1993directed}. 
Indeed, as it will be clear later on, our results strongly rely on the properties of such tensors.

Following the same reasoning, we can {define a $m$-directed $d$-hyperedge, for some $m\leq d$, as a set of $(d+1)$ nodes, {a subset of which (formed by $s=d+1-m$ units)} point{s} toward the remaining $m$ ones.}
{Resorting again to the adjacency tensor we can write}
\begin{equation}
A_{\pi(i_1,\dots, i_m) \pi'(j_1,\dots, j_{s})}^{(d)}=1 \, ,
\label{eq:n_dir_def}
\end{equation}
where $\pi(i_1,\dots, i_m)$ is any permutation of the indices $i_1,\dots, i_m$ and $\pi'(j_1,\dots, j_s)$ is any permutation of the indices $j_1,\dots, j_s$. In analogy with the former case, a permutation where one or more of the indices $i_1,\dots, i_m$ appear in a position other than the first $m$, may result in a zero entry of the adjacency tensor. By indicating with $M$ the largest value of $m$, and with $D$ the largest value of $d$, we can then define an \emph{$M$-directed $D$-hypergraph} (or $M$-directed hypergraph of order $D$). The framework above can be straightforwardly extended to the case of weighted directed hypergraphs.

\subsection*{$M$-directed hypergraphs are applied to dynamical systems with asymmetric higher-order interactions}
Let us now consider the dynamics of $N$ identical units coupled through a $1$-directed hypergraph of order $D$. The equations governing the system can be written as

\begin{equation}
\begin{array}{lll}
\dot{\vec{x}}_i &=& \vec{f}(\vec{x_i}) + \sum\limits_{d=1}^D \sigma_d\sum\limits_{j_1,\dots,j_d=1}^N A_{ij_1\dots j_d}^{(d)}\vec{g}^{(d)}(\vec{x}_i,\vec{x}_{j_1},\dots,\vec{x}_{j_d}), 

\end{array}
\label{eq:dyn_main}
\end{equation}
where $\vec{x}_i(t)\in\mathbb{R}^{m}$ is the state vector describing the dynamics of unit $i$, $\sigma_1$, $\dots$, $\sigma_D > 0$ are the coupling strengths, $\vec{f}:\mathbb{R}^{m}\rightarrow \mathbb{R}^m$ {is a nonlinear function that} describes the local dynamics, while $\vec{g}^{(d)}:\mathbb{R}^{m\times (d+1)}\rightarrow \mathbb{R}^m$, with $d\in\{1,\dots,D\}$ are {nonlinear} coupling functions encoding the $(d+1)$-body interactions. Let us now assume that the coupling functions at each order $d$ are diffusive-like 
\begin{equation}
\label{eq:difflikeg}
\vec{g}^{(d)}(\vec{x}_i,\vec{x}_{j_1},\dots,\vec{x}_{j_d})=\vec{h}^{(d)}(\vec{x}_{j_1},\dots,\vec{x}_{j_d})-\vec{h}^{(d)}(\vec{x}_{i},\dots,\vec{x}_{i}),
\end{equation} 
with $\vec{h}^{(d)}:\mathbb{R}^{m\times d}\rightarrow \mathbb{R}^m$, to ensure the existence of a synchronized solution $\vec{x}^{s} = \vec{x}_{1} = \dots = \vec{x}_{N}${, i.e., the synchronization manifold}. {\color{black} Diffusive coupling is common in many systems \cite{pikovsky2003synchronization}, being such assumption not particularly restrictive. However, it can be further relaxed to the milder requirement that the coupling functions are non-invasive, i.e., $\vec{g}^{(d)}(\vec{x},\vec{x},\dots,\vec{x}) \equiv 0, \; \forall d$, which still guarantees the existence of the invariant solution \cite{gambuzza2021stability}. In addition, let us also assume} that the coupling functions $\vec{h}^{(d)}$ satisfy the condition of natural coupling \cite{gambuzza2021stability, carletti}, namely
\begin{equation}
\label{eq:natcoupling}
\vec{h}^{(D)}(\vec{x},\dots ,\vec{x}) = \dots = \vec{h}^{(2)}(\vec{x},\vec{x}) = \vec{h}^{(1)}(\vec{x})\, . 
\end{equation}
\textcolor{black}{This second assumption turns out to be crucial to derive a Master Stability Equation to characterize the synchronization (see Methods) and disentangle the effect of the directionality of the higher-order interactions on it.}

For sake of definiteness, in the following we focus on the synchronization of identical oscillators coupled via $1$-directed hypergraphs, \textcolor{black}{whose adjacency tensors $A^{(d)}$ respect the symmetry property \eqref{eq:semidir_def}.} 

{\color{black} Let us thus denote by $\vec{x}^s(t)$ the synchronous state, which is solution of the decoupled systems $\dot{\vec{x}}_i=\vec{f}(\vec{x}_i)$. From Eq.~\eqref{eq:difflikeg}, it immediately follows that the former is also solution of the coupled system. To characterize the synchronization of this system, a linear stability analysis can be performed. To this aim, we linearize Eqs.~\eqref{eq:dyn_main} around $\vec{x}^s(t)$, by considering small perturbations~ $\delta\vec{x}_i = \vec{x}_i-\vec{x}^s$, and, since the time evolution of these variables determines the stability of the synchronous solution, we study their dynamics. In particular, it is convenient to introduce the stack vector $\delta\vec{x} = [\delta\vec{x}_1^{\top},\dots,\delta\vec{x}_N^{\top}]^{\top}$, whose dynamical equation under the hypothesis of natural coupling can be derived with a series of steps detailed in Methods, by obtaining:}

\begin{equation}
\delta\dot{\vec{x}}=\Big[\mathbb{I}_N \otimes JF - \mathcal{M}\otimes JH\Big]\delta\vec{x}
\label{eq:stack_equation}
\end{equation}
where $JF$ (resp. $JH$), is the Jacobian matrix associated to the function $\vec{f}$ (resp. $\vec{h}^{(1)}$), evaluated on the synchronous state $\vec{x}^{s}$, and where $\mathcal{M}$ is the matrix
\begin{equation}
\label{eq:matrixM}
\mathcal{M}=\sigma_1 L^{(1)}+\sigma_2 L^{(2)}+... +\sigma_D L^{(D)}\, .
\end{equation}

Matrix $L^{(d)}$ is the generalized Laplacian matrix for the interactions of order $d$ defined by
\begin{align}
L_{ij}^{(d)}&= 
\begin{cases} d!k_{in}^{(d)}(i) & i=j \\ -(d-1)!k_{in}^{(d)}(i,j) &i\neq j\, ,
\end{cases}
\label{eq:d_laplacian}
\end{align}
where $k_{\mathrm{in}}^{(d)}(i)$ is the generalized $d$-in-degree of node $i$ 
\begin{equation}
k_{in}^{(d)}(i)=\frac{1}{d!}\sum\limits_{j_1,..,j_d=1}^N A_{ij_1\dots j_d}^{(d)}\, ,
\label{eq:node_in_degree}
\end{equation}
namely the number of $d$-hyperedges pointing to node $i$, and $k_{in}^{(d)}(i,j)$ the generalized $d$-in-degree of a couple of nodes $(i,j)$
\begin{equation}
k_{in}^{(d)}(i,j)=\frac{1}{(d-1)!}\sum_{k_1,...,k_{d-1}}^N A_{ijk_1\dots k_{d-1}}^{(d)}\, .
\label{eq:link_in_degree}
\end{equation}
The latter represents the number of $d$-hyperedges pointing to node $i$ and having node $j$ as one of the source nodes. Let us stress that, because the adjacency tensor $A^{(d)}$ is asymmetric, the Laplacian matrix $L^{(d)}$ is asymmetric as well. {\color{black}This matrix represents the generalization to the directed case of the Laplacian matrix introduced in \cite{maxime2020,gambuzza2021stability} for undirected higher-order interactions.}

{\color{black} As an equivalent formulation, we rewrite Eq.~\eqref{eq:stack_equation} as follows
\begin{equation}
\delta\dot{\vec{x}}=\Big[\mathbb{I}_N \otimes JF - \sigma_1\widetilde{\mathcal{M}}\otimes JH\Big]\delta\vec{x}
\label{eq:stack_equation_alt}
\end{equation}
with $\widetilde{\mathcal{M}}$ given by
\begin{equation}
\label{eq:matrixM_alt}
\widetilde{\mathcal{M}}=L^{(1)}+ r_2 L^{(2)}+... + r_D L^{(D)},
\end{equation}
and where $r_i = \sigma_i/\sigma_1$, $i=2,\dots, D$. Eq.~\eqref{eq:stack_equation_alt} highlights the analogy between synchronization in directed hypergraphs with natural coupling functions and networks. In facts, once fixed the parameters $r_i$, the equations governing the dynamics of the perturbations are formally equivalent to those of a system with weighted, directed pairwise interactions among the units, coupling coefficient equal to $\sigma_1$, and a Laplacian matrix given by $\widetilde{\mathcal{M}}$. As both formulations~\eqref{eq:stack_equation} and~\eqref{eq:stack_equation_alt} are equivalent, for convenience hereby we conclude the discussion on the analysis of the linearized system referring back to Eq.~\eqref{eq:stack_equation}, while Eq.~\eqref{eq:stack_equation_alt} will turn out useful in the numerical investigation, where, fixing $r_i$, we can focus the analysis on the behavior as a function of $\sigma_1$.
}

{\color{black}Assuming for simplicity that $\mathcal{M}$ is diagonalizable}, we can project Eq.~\eqref{eq:stack_equation} onto each eigenvector, obtaining in this way $N$ decoupled $m$-dimensional linear equations, parametrized by the corresponding eigenvalue, from which the following generic Master Stability Equation (MSE) can be written
\begin{equation}
    \dot{\vec{\xi}} = [JF(\vec{x}^{s}) - (\alpha+i\beta)JH(\vec{x}^{s})]\vec{\xi}.
    \label{MSE}
\end{equation}
Note that, since the generalized Laplacian matrices are asymmetric, the effective matrix $\mathcal{M}$ will also be asymmetric, therefore it will have in general complex eigenvalues, motivating thus the use of the complex parameter $\alpha+i\beta$. From the MSE, the maximum Lyapunov exponent $\Lambda_{\mathrm{max}}$ can be calculated as a function of the complex parameter $\alpha+i \beta$. \textcolor{black}{Stability requires that $\Lambda_{\mathrm{max}}(\alpha+i\beta)<0$ where $\alpha+i \beta$ is any non-zero eigenvalue of $\mathcal{M}$. The same condition on stability can be also found when $\mathcal{M}$ is not diagonalizable, provided to consider an approach analogous to that introduced in \cite{nishikawa2006synchronization} for networks of directed pairwise interactions and based on Jordan block decomposition in place of diagonalization. Here, the crucial step is to identify the matrix, which in our case is $\mathcal{M}$, that provides the eigenvalues to consider in checking the condition $\Lambda_{\mathrm{max}}(\alpha+i\beta)<0$.}

The linear stability analysis \textcolor{black}{that leads to Eq.~\eqref{MSE} can be carried out following steps similar to those performed in~\cite{gambuzza2021stability} for undirected simplicial complexes. These steps can be straightforwardly generalized to deal with undirected hypergraphs. Instead, for directed hypergraphs the asymmetry of the adjacency tensors must be taken into account. In fact, in this case, the adjacency tensors are not symmetric with respect to all their indices. However, the property \eqref{eq:semidir_def} still allows the derivation of generalized Laplacian matrices, extending the formalism presented in~\cite{gambuzza2021stability}. The interested reader can find the detailed calculations in Methods}.

{\color{black} Despite the formal similarities of the equations for synchronization in hypergraphs and simplicial complexes, we emphasize that in the two scenarios different dynamical behaviors can be obtained. For instance, due to the requirement that, given a simplex of order $d$, all the simplices of lower order included in it are present, the regions of synchronization are not identical in the two higher-order structures. An example of the different dynamics in the case of undirected interactions is provided in Appendix A, showing a larger region of synchronization for the simplicial complex.}

{\color{black} A further important analysis would be to compare the dynamical behaviors of directed hypergraphs and simplicial complexes. However, at variance with hypergraphs, the definition of directed simplicial complexes is disputable. In particular, a crucial aspect to solve is how to deal with the inclusion constraint, establishing whether and how it can be extended to the case of directed interactions. An attempt in this direction has been made for oriented simplicial complexes \cite{millan2020explosive}, where it is highlighted that a simplex and its boundary can have either concordant or opposite orientation. The definition and the study of directed simplicial complexes are beyond the purpose of the present paper, and thus left as future work.}

\subsection*{Directed higher-order interactions can change stability behavior}
Using the above introduced approach, we now illustrate the effect of the higher-order directionality on synchronization by using a paradigmatic example of chaotic oscillator, i.e., the R\"ossler system~\cite{rossler1976equation}. We thus consider a system of $N$ coupled Rössler oscillators, whose parameters have been set to $a=b=0.2$, and $c=9$, so that the dynamics of the isolated system is chaotic. {For sake of clarity we limited our analysis to $1$-directed $2$-hypergraphs, but of course its applicability goes beyond the considered case.} The system equations read
\begin{equation}
    \begin{cases}
    \dot{x}_i = -y_i-z_i +\sigma_1\sum\limits_{j=1}^{N}A_{ij}^{(1)}(x_j^3-x_i^3)
 +\sigma_2\sum\limits_{j,k=1}^{N}A_{ijk}^{(2)}(x_j^2x_k-x_i^3)\\ 
    \dot{y}_i = x_i+ay_i \\ 
    \dot{z}_i = b+z_i(x_i-c),
    \end{cases} 
    \label{eq:rossler_x2x_coupling}
\end{equation} 
with $i\in\{1,\dots,N\}$. We remark that the coupling functions appearing in Eqs. \eqref{eq:rossler_x2x_coupling} are nonlinear and satisfy the natural coupling hypothesis.

\begin{figure}[t]
\centering
\includegraphics[width=\linewidth]{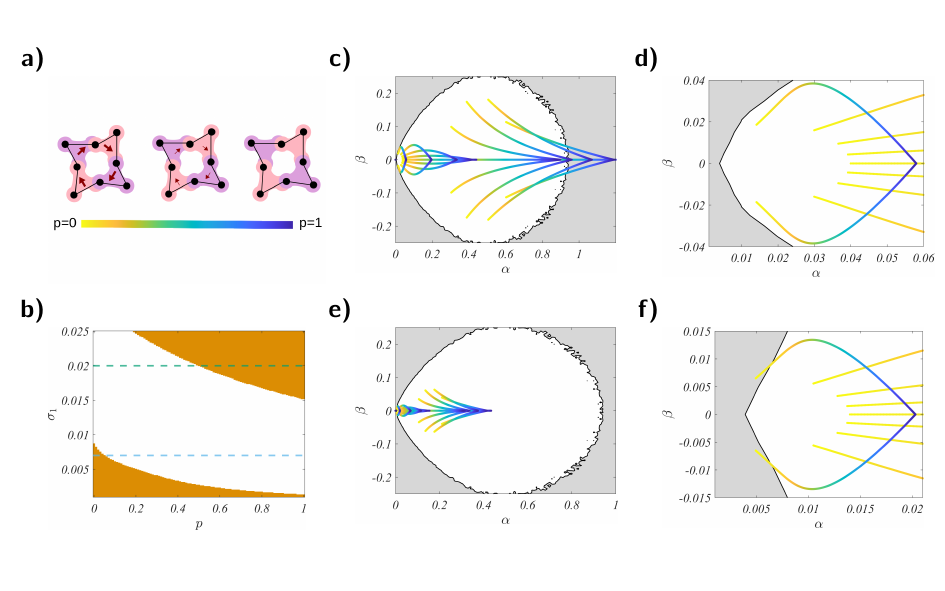}
\caption{\textcolor{black}{\textit{Directionality induced (de)synchronization}. a) Structure of the weighted hypergraph as a function of $p$, controlling the transition of the hyperedges from directed to undirected (the structure is schematically represented for $N=8$ nodes). Each undirected $2$-hyperedge can be seen as the combination of three directed hyperedges, two of which have a weight $p\in[0,1]$. When $p=0$, a triplet of nodes interacts only through a single directed hyperedge, whereas when $p=1$, the hypergraph is symmetric. b) Synchronization diagram in the plane $(p,\sigma_1)$ for a system of Rössler oscillators with $x$-$x$ cubic coupling. The white area indicates the region of stability, while the orange one the region where synchronization is unstable. The horizontal dashed lines represent two values of $\sigma_1$ for which the system transits from a synchronized to an unsynchronized state as a function of $p$ (green line), and the other way around (blue line). Panels c)-f) show the locus of eigenvalues of $\mathcal{M}$ as a function of $p$, for a weighted hypergraph with $N=20$ nodes at two different values of $\sigma_1$ (color coding is such that the directed case $p=0$ is represented in yellow, and the symmetric one $p=1$, in blue). In the background, the white area indicates the region identified by a negative MSF, the black line the boundary of this region, and the gray area the region where the MSF is positive. Panels d) and f) represent a zoom of the area close to the origin of panels c) and e), respectively. Panels c) and d) show a setting where the directed topology drives the system unstable, indeed assuming a symmetric hypergraph the synchronization manifold will result stable. Panels e) and f) display a case for which the directed topology admits a stable synchronization state, while the symmetric hypergraph triggers the instability. The coupling strength for panels c) and d) is set as $\sigma_1=0.2$, while for panels e) and f) as $\sigma_1=0.07$. In both cases $r_2=10$.}}
\label{fig:msf_rossler_x2x}
\end{figure}

We consider the system to be coupled through a directed weighted $2$-hypergraph, whose asymmetry varies with a parameter $p\in[0,1]$, representing the relative weight of the directed hyperedges. {\color{black} The topology of the directed weighted $2$-hypergraph is schematically illustrated in panel a) of Fig.~\ref{fig:msf_rossler_x2x}, where for the purpose of representation we fixed $N=8$.} When $p=0$, a triplet of nodes interacts only through a single $1$-directed hyperedge. As $p$ increases, so does the weight of the other two components, up to $p=1$, where an undirected hypergraph is recovered (see Methods for further details).

To proceed with the analysis, first we calculate the MSF associated to system~\eqref{eq:rossler_x2x_coupling}, by evaluating the maximum Lyapunov exponent, $\Lambda_{\mathrm{max}}(\alpha+i\beta)$ as a function of $\alpha$ and $\beta$ with the Wolf's algorithm \cite{wolf1985determining}. {\color{black} For synchronization to be achieved, it is required that $\Lambda_{\mathrm{max}}(\alpha+i\beta)<0$, where $\alpha +i\beta$ is any non-zero eigenvalue of the matrix $\mathcal{M}$. Conversely, if there is at least a non-zero eigenvalue of $\mathcal{M}$ such that $\Lambda_{\mathrm{max}}>0$, then synchronization is lost. To illustrate the effect of directionality on synchronization, we consider a directed weighted $2$-hypergraph with structure as in Fig.~\ref{fig:msf_rossler_x2x}(a) but $N=20$ nodes, calculate the eigenvalues of $\mathcal{M}$ as a function of the asymmetry parameter $p$ and the coupling strength $\sigma_1$, and check whether the stability condition is satisfied or not, in this way constructing a synchronization diagram in the plane $(p,\sigma_1)$. Fig.~\ref{fig:msf_rossler_x2x}(b) shows this diagram for $r_2=\sigma_2/\sigma_1=10$. 
The white area represents the values $(p,\sigma_1)$ for which the system synchronizes, i.e., $\Lambda_{\mathrm{max}}<0$ for every eigenvalue of $\mathcal{M}$, while the orange area depicts the region where the synchronous state is unstable, i.e., $\Lambda_{\mathrm{max}}>0$ for at least one eigenvalue of $\mathcal{M}$. While there is a region where varying $p$ at fixed values of $\sigma_1$ has no effect on synchronization, there are two other regions where this leads to a transition. In more detail, two different transitions can appear, an example of which is highlighted by the two horizontal dashed lines. For $\sigma_1=0.2$ the system synchronizes for small values of $p$, i.e,  when the hypergraph is strongly directed, and loses synchronization for larger values of $p$, i.e., when the hypergraph becomes symmetric. Conversely, for $\sigma_1=0.07$ we find the opposite scenario, as synchronization is achieved by increasing $p$, while directed hyperedges hamper synchronization. The locus of the eigenvalues of $\mathcal{M}$ as a function of $p$ and for two different values of $\sigma_1$, corresponding to the two types of transitions induced by directionality, is shown in the panels c)-f) of Fig.~\ref{fig:msf_rossler_x2x}. Here, panels c) and d) refer to $\sigma_1=0.2$, while panels e) and f) to $\sigma_1=0.07$. Moreover, panels d) and f) represent a zoom of the area close to the origin in panels c) and e), respectively. In all these panels, the gray area represents the region where the MSF is positive, while the white area portrays the region of stability. Finally, the black line denotes the boundary value $\Lambda_{\mathrm{max}}(\alpha+i\beta)=0$. We remark that the region of the complex plane for which $\Lambda_{\mathrm{max}}$ is negative is bounded, both along the real component, $\alpha$, and the imaginary one, $\beta$. This suggests that either a large value of $\alpha$ or a large value of $\beta$ can lead to instability. In panels c) and d), obtained for $\sigma_1=0.2$, we note that for large enough $p$ the eigenvalues cross the boundary, thus leaving the stability region and inducing the desynchronization of the system. On the other hand, in panels e) and f), which display the case $\sigma_1=0.07$, the eigenvalues of $\mathcal{M}$ leave the stability region for small values of $p$, namely in this case synchronization is observed for symmetric hyperedges, while directed hyperedges move the system in a region where the synchronous state is unstable.
}
{\color{black} To numerically validate this analysis, we monitor the synchronization error defined as follows:
\begin{equation}
    E = \left\langle \sqrt{\frac{1}{N(N-1)}\sum\limits_{i,j=1}^{N}\|\vec{x}-\vec{x}_i\|^2} \right\rangle_T\label{eq:error}
\end{equation} 
where $T$ is a sufficiently large window of time, after discarding the initial transient. In agreement with the analysis of the eigenvalues, for $\sigma_1=0.2$, $E$ vanishes for $p=0$, while for $p=1$ it diverges after a transient. On the other hand, for $\sigma_1=0.07$, the synchronization error goes to zero for $p=1$, while for $p=0$ it again diverges after a transient. Overall, these results suggest that directionality can change the synchronization behavior of a system of coupled chaotic oscillators, either inducing synchronization in the system or desynchronizing it.
}
{\color{black} However, for a different choice of the coupling functions, a diverse synchronization behavior in relation to the structure of interactions may be obtained. For instance, if the coupling functions are $\vec{h}^{(1)}(\vec{x}_j) = [0,y_j^3,0]$ and $\vec{h}^{(2)}(\vec{x}_j,\vec{x}_k) = [0,y_j^2y_k,0]$, then the resulting region of stability is unbounded, making impossible to desynchronize the system by turning the three-body interactions symmetric (Appendix B).}

{\color{black} The results discussed so far refer to a specific example of connectivity between the oscillators. Since, once fixed the oscillator dynamics and the coupling functions (hence the system MSF), the main determinant for synchronization is the position of the eigenvalues of $\mathcal{M}$ with respect to the region of negative values of the MSF, understanding the effect of directionality in other structures requires the study of the spectrum of $\mathcal{M}$. As a systematic characterization of the spectrum as a function of the topological features of the structure is far from trivial, we limited our analysis to two random hypergraph generative models, obtained as higher-order generalization of random network models, namely the well-known Newman-Watts (NW) model and the Erd\H{o}s-Rényi (ER) one. We have found that the impact of directionality on the eigenvalue position (and so ultimately on synchronization) strongly depends on the model adopted for generating the hypergraph, with the NW-like model showing a larger impact of directionality on the spreading of eigenvalues in the complex plane, when compared to the ER-like model (see Appendix C for a detailed analysis of the two models).}

{\color{black}

\section*{Controlling for confounding factors}

In the previous section, we have shown how directionality can induce either the synchronization of a system of coupled chaotic oscillators or its desynchronization. However, there may be confounding factors determining the change of the system behavior. In fact, the way in which $1$-directed hypergraphs are made symmetric, namely by varying the parameter $p$, does not conserve the total strength of the interactions.

To determine whether the observed effects are truly due to the directionality, we proceed with an alternative symmetrization method that keeps constant the total coupling strength. Starting from a $1$-directed $2$-hyperedge, we now add directed hyperedges in the two remaining directions with a weight $q\in[0,1/3]$, while simultaneously decreasing the strength of the initial one, setting the weight to $1-2q$. In this way, for $q=0$ we have a $1$-directed $2$-hyperedge with unitary weights, while for $q=1/3$ we get an undirected $2$-hyperedge with the same total weight, but having all hyperedges with weight equal to $1/3$ (see Methods for further details). We notice that this symmetrization is analogous to that introduced in \cite{Asllani2} for networks, where, starting from a directed link of weight $1$, one obtains a symmetric link with the same total weight, as it is formed by two directed links, each of weight $1/2$.

\begin{figure}[t!]
\centering
\includegraphics[width=\linewidth]{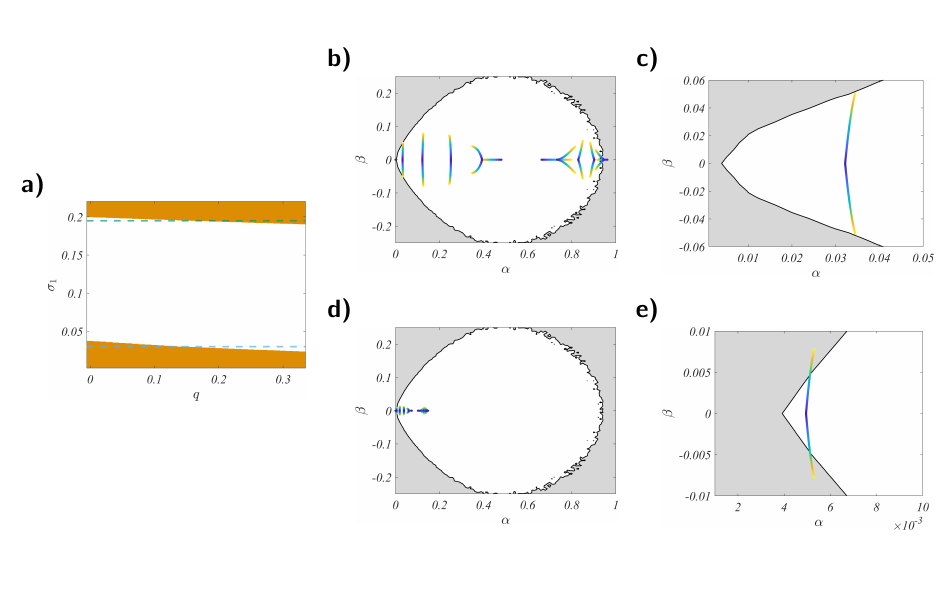}
\caption{\textcolor{black}{\textit{Directionality induced (de)synchronization with an alternative symmetrization method}. a) Synchronization diagram in the plane $(q,\sigma_1)$ for a system of Rössler oscillators with $x$-$x$ cubic coupling. The white area indicates the region of stability, while the orange one the region where synchronization is lost. The horizontal dashed lines represent two values of $\sigma_1$ for which the system transits from a synchronized to an unsynchronized state as a function of $p$ (green line), and the other way around (blue line). Panels b)-e) display the locus of eigenvalues of $\mathcal{M}$ as a function of $q$, for a hypergraph with $N=20$ nodes at two different values of $\sigma_1$ (color coding is such that the directed case $q=0$ is represented in yellow, and the symmetric one $q=1/3$, in blue). In the background, the white area indicates the region where the MSF is negative, the black line the boundary of this region, and the gray area the region with positive MSF. Panels c) and e) represent a zoom of the area close to the origin of panels b) and d), respectively. Panels b) and c) display a setting where the directed topology drives the system unstable, starting from a symmetric hypergraph for which the synchronization manifold is stable. Panels d) and e) show a case for which the directed topology admits a stable synchronization state, while the symmetric hypergraph drives to instability. The coupling strength for panels b) and c) is fixed to $\sigma_1=0.195$, while for panels d) and e) to $\sigma_1=0.03$. In both cases $r_2=0.7$.}}
\label{fig:msf_rossler_x2x_newsymm}
\end{figure}

With this setup, we consider again a system of $N=20$ R\"ossler oscillators coupled through the directed weighted $2$-hypergraph discussed in the previous section.
We then derive the synchronization diagram in the plane $(q,\sigma_1)$. The diagram obtained for $r_2=\sigma_2/\sigma_1=0.7$ is displayed in panel a) of Fig.~\ref{fig:msf_rossler_x2x_newsymm}. Similarly to what observed with the previous symmetrization method, while there is a region where, for fixed $\sigma_1$, varying $q$ does not affect synchronization, there are two areas where changing $q$ leads to a transition in the synchronization behavior. For $\sigma_1=0.195$, highlighted in panel a) of Fig.~\ref{fig:msf_rossler_x2x_newsymm} as a green dashed line, the system synchronizes for small values of $q$, i.e, for a strongly directed hypergraph, whereas it desynchronizes for larger values of $q$, i.e., for a more symmetric structure. Inversely, for $\sigma_1=0.03$, displayed in panel a) as a blue dashed line, we observe the opposite transition, as synchronization is achieved by increasing $q$, while directionality prevents system synchronization.
The locus of the eigenvalues of $\mathcal{M}$ as a function of $q$ and for the two different values of $\sigma_1$ is shown in the panels b)-e) of Fig.~\ref{fig:msf_rossler_x2x_newsymm}. In particular, panels d) and c) refer to $\sigma_1=0.195$, while panels d) and e) to $\sigma_1=0.03$. Panels c) and e) represent a zoom of the area close to the origin in panels b) and d), respectively. 
In panels b) and c), we observe that for large enough $q$ the eigenvalues of $\mathcal{M}$ leave the stability region, thus inducing the desynchronization of the system. Conversely, in panels d) and e), the eigenvalues leave the stability region for small values of $q$, meaning that synchronization is achieved for more symmetric hyperedges, while strongly directed hyperedges make the synchronization manifold unstable.
In conclusion, these results confirm that directionality can change the synchronization behavior of a system of chaotic oscillators coupled through a $1$-directed hypergraph, either inducing system synchronization or its desynchronization. In particular, by using the symmetrization method that preserves the total coupling strength of the interactions, we find that these transitions are due to directionality and not, or at least not only, to confounding factors.

As discussed in the previous section, for a different choice of the coupling functions, namely $\vec{h}^{(1)}(\vec{x}_j) = [0,y_j^3,0]$ and $\vec{h}^{(2)}(\vec{x}_j,\vec{x}_k) = [0,y_j^2y_k,0]$ the resulting region of stability is unbounded. In agreement with the results obtained with the first symmetrization method, turning symmetric the three-body interactions does not desynchronize the system. In this setting, it is only possible to induce desynchronization by making higher-order interactions asymmetric. This further case study is discussed in Appendix B.

}

\section*{Discussion}\label{sec3}
In this paper we have introduced and described the tensor formalism to encode $M$-directed hypergraphs, allowing us to {fully} account for directionality in higher-order structures. {We have then used such directed higher-order structure as coupling substrate for dynamical systems and studied the ensuing synchronization.} We have {shown that the latter can be analyzed by extending the Master Stability Function approach to the present framework} for the particular case of $1$-directed hypergraphs. We have numerically validated our theoretical results for a system of Rössler oscillators and observed that the stability of the synchronized state can be lost or gained as the asymmetry varies. Our results demonstrate that phenomena, previously observed in structures with pairwise interactions \cite{Asllani1,Asllani2,dipatti1,jtb,entropy,carletti}, also appear when directed higher-order interactions are considered. 

 {\color{black}For systems with pairwise interactions, there is a vast literature (see for instance \cite{chavez2005synchronization,hwang2005synchronization,motter2005enhancing}), showing how synchronization is actually enhanced in weighted graphs built using weighting procedures that ultimately result in determining asymmetric interactions in the network links. Few attempts have been already made to extend this study to higher-order topologies, in particular finding that structural symmetric hypergraphs can be optimally synchronizable \cite{tang2022optimizing}. Here, however, we did not aim at using the directionality of the higher-order interactions to optimize the synchronizability of the structure, but focused on introducing the formalism to deal with directionality in higher-order interactions, in order to model systems where there is an evidence of such asymmetric and higher-order coupling, and analyze the effect of directionality on synchronization in these systems.}

Our setting differs from the one recently proposed in~\cite{ABD2022}. In fact, the asymmetry of the higher-order structure is here imposed only on the adjacency tensor, Eq.~\eqref{eq:semidir_def}, and not directly on the higher-order 
coupling function as done in~\cite{ABD2022}. Therefore, our formalism allows for a more general approach, as it leaves more freedom in the choice of the coupling functions.
The new framework and concepts here introduced pave the way to further studies on the effects of directionality in systems where empirical evidence of directed higher-order interactions has been found but not yet systematically investigated, as the proper mathematical setting for their description was lacking. 

\section*{Methods}\label{sec4}

\subsection*{Linear stability analysis of $1$-directed $D$-hypergraphs}
\begin{figure}[h]
\centering
\includegraphics[width=0.6\linewidth]{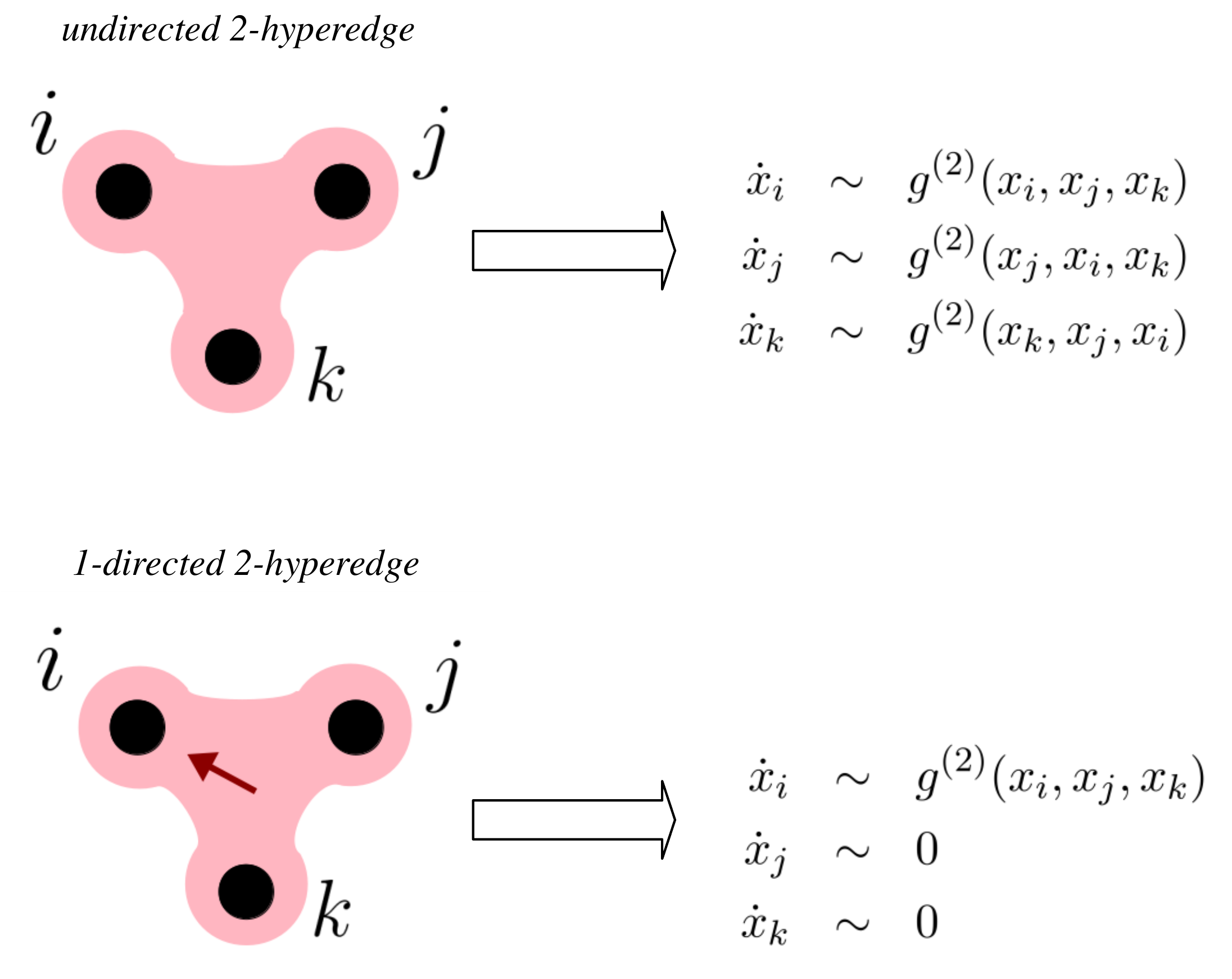}
\caption{From topology to dynamics: difference between the undirected and directed $1$-hyperedge. Top panel: the derivative of the state variables associated to each node $i$, $j$, $k$ receives a contribution from the higher-order interaction. Bottom panel: only the derivative of $\vec{x}_i$ receives a contribution from the source nodes $j$ and $k$, while the derivatives of the state variable of the source nodes, $\vec{x}_j$ and $\vec{x}_k$, do not.}
\label{fig:Rifatta}
\end{figure}

Here we provide the full derivation of the Master Stability Equation, which allows to study the synchronization of a system of $N$ identical oscillators coupled through a $1$-directed $D$-hypergraph. Let us first write the equation describing the dynamics of the system, where, as we previously emphasized, the coupling term associated to the hyperedge provides a contribution only to the growth rate of the state vector of node $i$, i.e., $\vec{x}_i$. This is different from the case of an undirected $d$-hyperedge where the higher-order coupling contributions appear in the derivatives of the state variables of all nodes of the hyperdege (see Fig.~\ref{fig:Rifatta}). 

Taking into account the contributions from all the $1$-directed $d$-hyperedges, $d=1,\dots,D$, we eventually obtain
\begin{equation}
\dot{\vec{x}}_i=\vec{f}(\vec{x_i})+\sum_{d=1}^D\sigma_d\sum\limits_{j_1,\dots,j_d=1}^N A_{ij_1\dots j_d}^{(d)}\vec{g}^{(d)}(\vec{x}_i,\vec{x}_{j_1,\dots,\vec{x}_{j_d}})
\label{dyn1}
\end{equation}
where $\vec{x}_i(t)\in\mathbb{R}^{m}$ is the state vector describing the dynamics of unit $i$, $\sigma_1$, $\dots$, $\sigma_D > 0$ are the coupling strengths, $\vec{f}:\mathbb{R}^{m}\rightarrow \mathbb{R}^m$ describes the local dynamics, while $\vec{g}^{(d)}:\mathbb{R}^{m\times (d+1)}\rightarrow \mathbb{R}^m$, with $d\in\{1,\dots,D\}$ are coupling functions ruling the $(d+1)$-body interactions. Finally, $A^{(d)}_{i j_1\dots j_d}$ are the entries of the adjacency tensors $A^{(d)}$, with $d\in\{1,\dots,D\}$.

Let us now consider diffusive-like coupling functions at each order $d$
\begin{displaymath}
\vec{g}^{(d)}(\vec{x}_i,\vec{x}_{j_1},\vec{x}_{j_2},\dots,\vec{x}_{j_d})=\vec{h}^{(d)}(\vec{x}_{j_1},\dots,\vec{x}_{j_d})-\vec{h}^{(d)}(\vec{x}_{i},\dots,\vec{x}_{i})
\end{displaymath} with \begin{displaymath}
\vec{h}^{(d)}:\mathbb{R}^{m\times d}\rightarrow \mathbb{R}^m
\end{displaymath}
Note that this hypothesis on the form of coupling guarantees the existence of the synchronized solution $\vec{x}_{1} = \dots = \vec{x}_{N} = \vec{x}^{s}$. We remark that, in order to deal with an authentic multibody dynamics, we need to consider nonlinear coupling functions. Indeed, in the case of linear interactions, the three-body dynamical system can be reduced to a two-body dynamical system, by rescaling the adjacency matrix \cite{neuhauser2020multibody}. 

Equation~\eqref{dyn1} becomes then

{\small
\begin{equation}
\dot{\vec{x}}_i =\vec{f}(\vec{x_i})+  \sum_{d=1}^D\sigma_d\sum\limits_{j_1,\dots,j_d=1}^N A_{ij_1\dots j_d}^{(d)}(\vec{h}^{(d)}(\vec{x}_{j_1},\dots,\vec{x}_{j_d})-\vec{h}^{(d)}(\vec{x}_{i},\dots,\vec{x}_{i}))
\label{dyn2}
\end{equation}
}

Let us now perturb the synchronous state $\vec{x}^{s}$ with a spatially inhomogeneous perturbation, meaning that $\forall~i\in\{1,\dots,N\}$ we have $\vec{x}_i=\vec{x}^s+\delta\vec{x}_i$. Substituting into Eq.~\eqref{dyn2} and expanding up to the first order we obtain

{\small
\begin{eqnarray*}
&\delta\dot{\vec{x}}_i&  =   \frac{\partial \vec{f}(\vec{x}_i)}{\partial \vec{x}_i}\mid_{\vec{x}^s}\delta\vec{x}_i+\\
&-&\hspace{-1em}\sum_{d=1}^D\sigma_d\sum\limits_{j_1,\dots,j_d=1}^N T_{ij_1\dots j_d}  \sum_{\ell=1}^{d} \frac{\partial \vec{h}^{(d)}(\vec{x}_{j_1},\dots,\vec{x}_{j_d})}{\partial \vec{x}_{j_\ell}}\mid_{(\vec{x}^s,\dots,\vec{x}^s)}\delta\vec{x}_{j_\ell}\, ,
\end{eqnarray*}
}

where 
\begin{eqnarray*}
T_{ij_1}&=& k_{in}^{(1)}(i)\delta_{ij_1} - A_{ij_1}^{(1)}\, ,\\
T_{ij_1j_2}&=&2k_{in}^{(2)}(i)\delta_{ij_1j_2}-A_{ij_1j_2}^{(2)}\, , \dots\\
T_{ij_1j_2\dots j_D}&=&D!k_{in}^{(D)}(i)\delta_{ij_1j_2\dots j_D}-A_{ij_1j_2\dots j_D}^{(D)}\, ,
\end{eqnarray*} 
being $\delta_{ij_1j_2\dots j_D}$ the generalized multi-indexes Kronecker-$\delta$, and the $d$-\textit{in-degree} $k_{in}^{(d)}(i)$ is here defined as
\begin{displaymath}
k_{in}^{(d)}(i)=\frac{1}{d!}\sum\limits_{j_1,..,j_d=1}^N A_{ij_1\dots j_d}^{(d)},
\end{displaymath} which represents the number of hyperedges of order $d$ pointing to node $i$. \\

Let us now consider the terms relative to the $d$-body interactions
\begin{eqnarray*}
\sum\limits_{j_1=1}^N \frac{\partial \vec{h}^{(d)}(\vec{x}_{j_1},\dots,\vec{x}_{j_d})}{\partial \vec{x}_{j_1}}\mid_{(\vec{x}^s,\dots,\vec{x}^s)}\delta\vec{x}_{j_1} \sum\limits_{j_2=1}^N \dots \sum\limits_{j_d=1}^N T_{ij_1\dots j_d} + \dots \\+
\sum\limits_{j_d=1}^N \frac{\partial \vec{h}^{(d)}(\vec{x}_{j_1},\dots,\vec{x}_{j_d})}{\partial \vec{x}_{j_d}}\mid_{(\vec{x}^s,\dots,\vec{x}^s)}\delta\vec{x}_{j_d} \sum\limits_{j_1=1}^N\dots \sum\limits_{j_{d-1}=1}^N T_{ij_1\dots j_d}.
\end{eqnarray*}
By defining
\begin{displaymath}
k_{in}^{(d)}(i,j)=\frac{1}{(d-1)!}\sum_{k_1,...,k_{d-1}}^N A_{ijk_1\dots k_{d-1}}^{(d)},
\end{displaymath}
which represents the number of hyperedges of order $d$ pointing to node $i$ and having node $j$ as one of the source nodes, and by observing that, given the property of symmetry of $1$-directed hypergraphs, we have
\begin{equation}
T_{ij_1\dots j_d}=T_{i\pi(j_1\dots j_d)}\,,
\label{eq:Tprop}
\end{equation}
for any permutation $\pi$ of the indexes $j_1\dots j_d$, we can write 

\begin{eqnarray*}
\hspace{-2em}&&\sum\limits_{j_1=1}^N L_{ij_1}^{(d)} \frac{\partial \vec{h}^{(d)}}{\partial \vec{x}_{j_1}}\mid_{(\vec{x}^s,\dots,\vec{x}^s)}\delta\vec{x}_{j_1}  + \dots +\sum\limits_{j_d=1}^N L_{ij_d}^{(d)} \frac{\partial \vec{h}^{(d)}}{\partial \vec{x}_{j_d}}\mid_{(\vec{x}^s,\dots,\vec{x}^s)}\delta\vec{x}_{j_d} \\
&~~~~~&=\sum\limits_{j=1}^N L_{ij}^{(d)} \Big( \frac{\partial \vec{h}^{(d)}}{\partial \vec{x}_{j_1}}\mid_{(\vec{x}^s,\dots,\vec{x}^s)}+\dots+\frac{\partial \vec{h}^{(d)}}{\partial \vec{x}_{j_d}}\mid_{(\vec{x}^s,\dots,\vec{x}^s)}\Big)\delta\vec{x}_{j}\, ,
\end{eqnarray*}

\noindent where to lighten the notation we removed the explicit dependence of $\vec{h}^{(d)}$ on $(\vec{x}_{j_1},\dots,\vec{x}_{j_d})$, and we have defined the generalized Laplacian matrix for the interaction of order $d$ as 
\begin{align}
L_{ij}^{(d)}&= 
\begin{cases} d!k_{in}^{(d)}(i) & i=j \\ -(d-1)!k_{in}^{(d)}(i,j) &i\neq j
\end{cases}\, .
\end{align}
It is worth noting that the generalized Laplacian matrices defined above may not be symmetric, hence in general they have complex spectra.

Finally, by denoting 

\begin{equation*}
JH^{(d)}:=
\sum_{\ell=1}^d \frac{\partial \vec{h}^{(d)}(\vec{x}_{j_1},\dots , \vec{x}_{j_d})}{\partial \vec{x}_{j_\ell}}
\mid_{(\vec{x}^*,\dots , \vec{x}^*)}\, , 
\end{equation*} 
and by defining the vector $\vec{x}=(\vec{x}_1^\top, \dots, \vec{x}_N^\top)^\top$, we can rewrite equation \eqref{dyn2} in a more compact form, namely

\begin{equation}
\delta\dot{\vec{x}}=\Big[\mathbb{I}_N \otimes JF -\sum_{d=1}^D \sigma_d L^{(d)}\otimes JH^{(d)}\Big]\delta\vec{x}
\label{dyn3}
\end{equation}

We here assume the hypothesis of natural coupling
\begin{displaymath}
\vec{h}^{(D)}(\vec{x},\dots ,\vec{x}) = \dots = \vec{h}^{(2)}(\vec{x},\vec{x}) = \vec{h}^{(1)}(\vec{x}), \quad ~\forall \vec{x}\in\mathbb{R}^m\end{displaymath}
which leads to \begin{displaymath}
JH^{(D)} = \dots = JH^{(2)} = JH^{(1)}.\end{displaymath}

Under such hypothesis, we can define the matrix \begin{displaymath}
\mathcal{M}=\sigma_1 L^{(1)}+\sigma_2 L^{(2)}+... +\sigma_D L^{(D)},\end{displaymath} allowing us to write the following Master Stability Equation describing the dynamics of the perturbation \begin{equation}
\delta\dot{\vec{x}}=\Big[\mathbb{I}_N \otimes JF - \mathcal{M}\otimes JH\Big]\delta\vec{x}\label{dyn5}
\end{equation}

Assuming that matrix $\mathcal{M}$ is diagonalizable, we can construct a basis made by the eigenvectors $\vec{v}_1,\dots,\vec{v}_N$ of this matrix, and then project Eq.~\eqref{dyn3} onto each eigenvector, obtaining a system of $N$ decoupled linear equations. In more detail, by defining the new variable $\vec{\eta} = (V^{-1}\otimes \mathbb{I}_m)\vec{\delta x}$, where $V=[\vec{v}_1,\dots,\vec{v}_N]$, we can rewrite Eq.~\eqref{dyn3} as
\begin{equation}
    \dot{\vec{\eta}}_i = [JF(\vec{x}^{s}) - \lambda_i JH(\vec{x}^{s})]\vec{\eta}_i\, 
    \label{eq:decoupled_linear_system}
\end{equation}
with $i\in\{1,\dots,N\}$ and where $\lambda_1,\lambda_2,\dots,\lambda_N$ are the eigenvalues of the matrix $\mathcal{M}$. The equation for $i=1$ corresponds to $\lambda_1=0$, representing the linearized motion along the synchronous state $\vec{x}^s(t)$. The other equations describe instead the motion transverse to $\vec{x}^s(t)$. As these equations, except for the eigenvalue $\lambda_i$, have the same form, by considering the generic complex parameter $\alpha+i\beta$, we finally arrive to the Master Stability Equation in~\eqref{MSE}.

\subsection*{Construction of the weighted $1$-directed $2$-hypergraph}
\label{app:spectrum}
We describe here how to construct the $1$-directed hypergraph we have analyzed in Results and give further details about its tensor representation and the resulting generalized Laplacian matrices. 

\begin{figure}[t]
\centering
\includegraphics[width=0.8\linewidth]{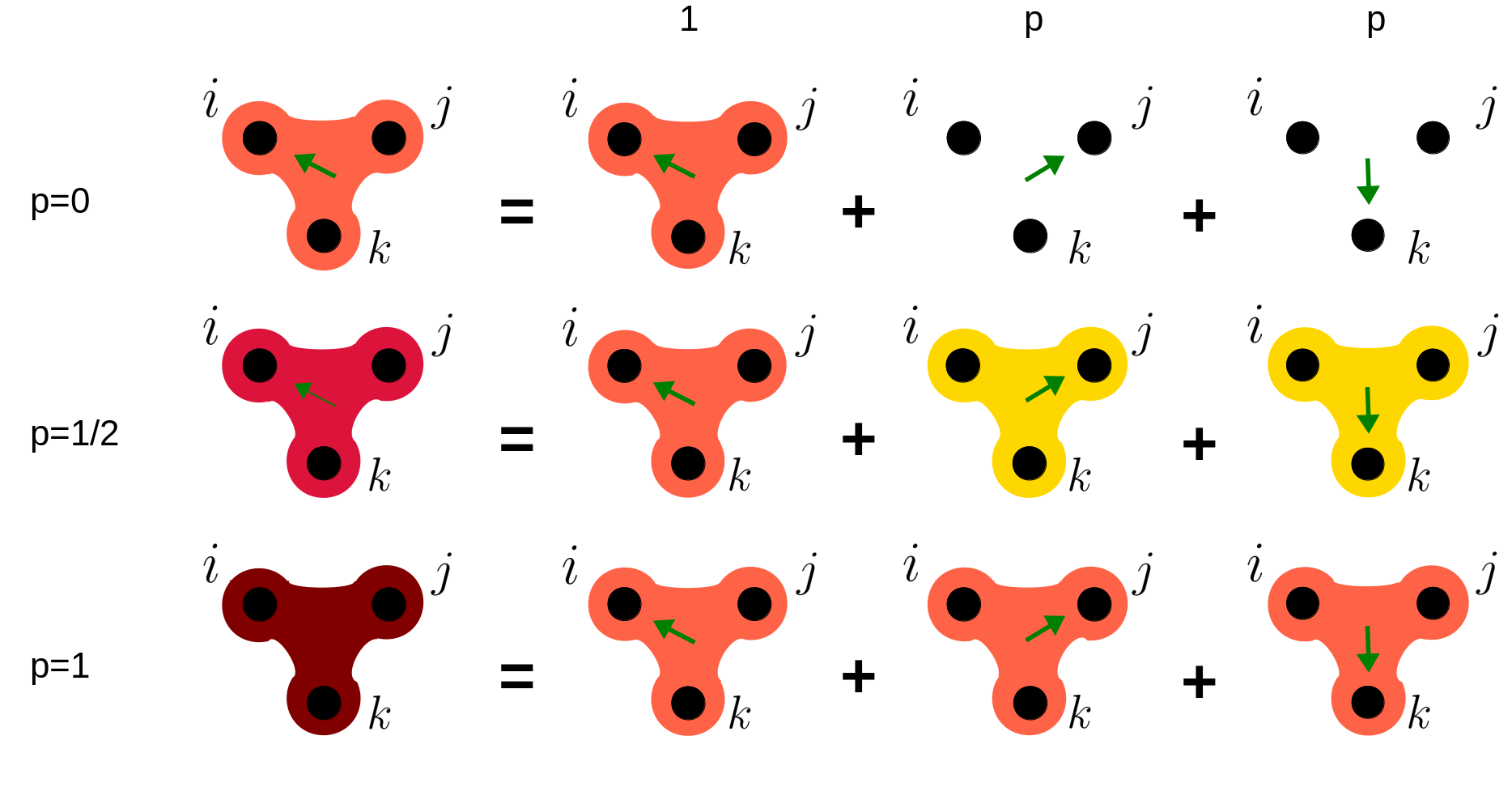}
\caption{{\color{black} Symmetrization of a $1$-directed $2$-hyperedge via the increase of the weight of the hyperdeges associated to the other directions. Starting from a fully directed hyperedge ($p=0$), the strength of the couplings in the other directions grows until all directions of interaction have the same weight ($p=1$).}}
\label{fig:symm1}
\end{figure}

To construct the hypergraph, we start from an undirected ring network of $N$ nodes, where $N$ is even. We consider a consecutive labeling, so that each node $i$ is connected to nodes $i-1$ and $i+1$. We then add $N/2$ {$2$-}hyperedges{, namely containing $3$ nodes}, connecting nodes $(1,2,3)$, $(3,4,5)$, $\dots$, $(N-1,N,1)$. \textcolor{black}{For the first method of symmetrization,} for each triple of nodes $(i,i+1,i+2)$, we set $A^{(2)}_{i+2,i,i+1} = A^{(2)}_{i+2,i+1,i} = 1$, $A^{(2)}_{i,i+1,i+2} = A^{(2)}_{i,i+2,i+1} = p$ and $A^{(2)}_{i+1,i+2,i} = A^{(2)}_{i+1,i,i+2} = p$, where $p\in[0,1]$. {In this way we encode the information that nodes $i$ and $i+1$ point toward node $i+2$ with strength $1$, and we allow a weaker directed  interaction from $(i+1,i+2)$ toward $i$ and $(i,i+2)$ toward $i+1$}. As $p$ increases, so does the weight of the other two directions, until we recover an undirected hypergraph for $p=1$. \textcolor{black}{Observe that this symmetrization does not preserve the total coupling strength of the hyperedges.} \textcolor{black}{A graphical representation of the symmetrization is provided in Fig.~\ref{fig:symm1}}. 

\begin{figure}[t]
\centering
\includegraphics[width=0.8\linewidth]{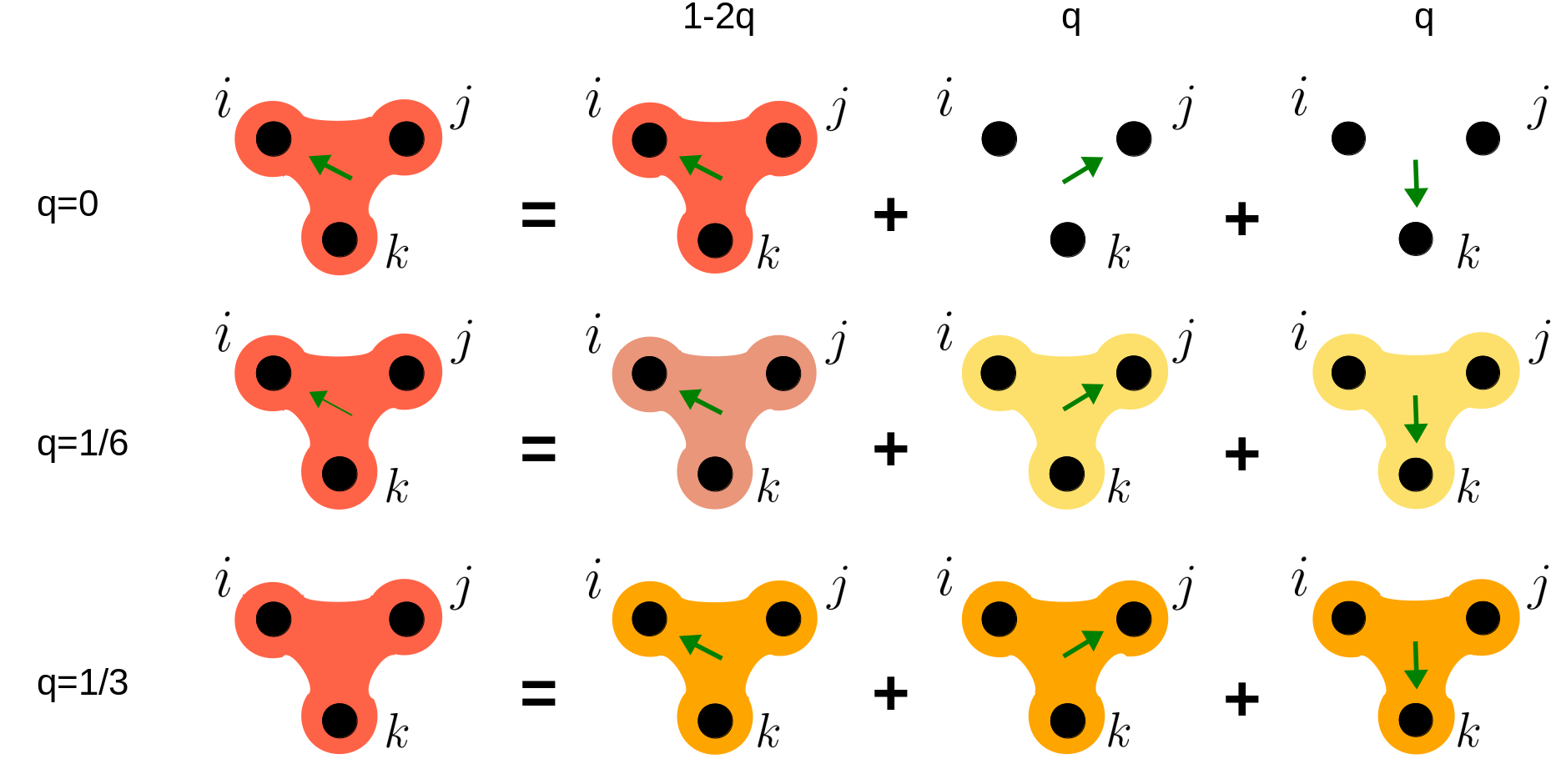}
\caption{{\color{black} Symmetrization of a $1$-directed $2$-hyperedge while preserving the total coupling strengths. Starting from a fully directed hyperedge ($q=0$), as the strength of the couplings in the other directions grows, the weight of the initial directed hyperedge decreases until all directions of interaction have the same weight ($q=1/3$).}}
\label{fig:symm2}
\end{figure}

\textcolor{black}{For what concerns the second method of symmetrization, for each triple of nodes $(i,i+1,i+2)$, we set $A^{(2)}_{i+2,i,i+1} = A^{(2)}_{i+2,i+1,i} = 1-2q$, $A^{(2)}_{i,i+1,i+2} = A^{(2)}_{i,i+2,i+1} = q$ and $A^{(2)}_{i+1,i+2,i} = A^{(2)}_{i+1,i,i+2} = q$, where $q\in[0,1/3]$. As $q$ increases, so does the weight of the hyperedges in the other two directions, until we recover an undirected hypergraph for $q=1/3$. This second method of symmetrization preserves the total coupling strength of the hyperedges, thus allowing to control for confounding factors (see also Results). Fig.~\ref{fig:symm2} displays a graphical representation of the second symmetrization considered.}

\begin{figure}[h]
\centering
\includegraphics[width=0.3\linewidth]{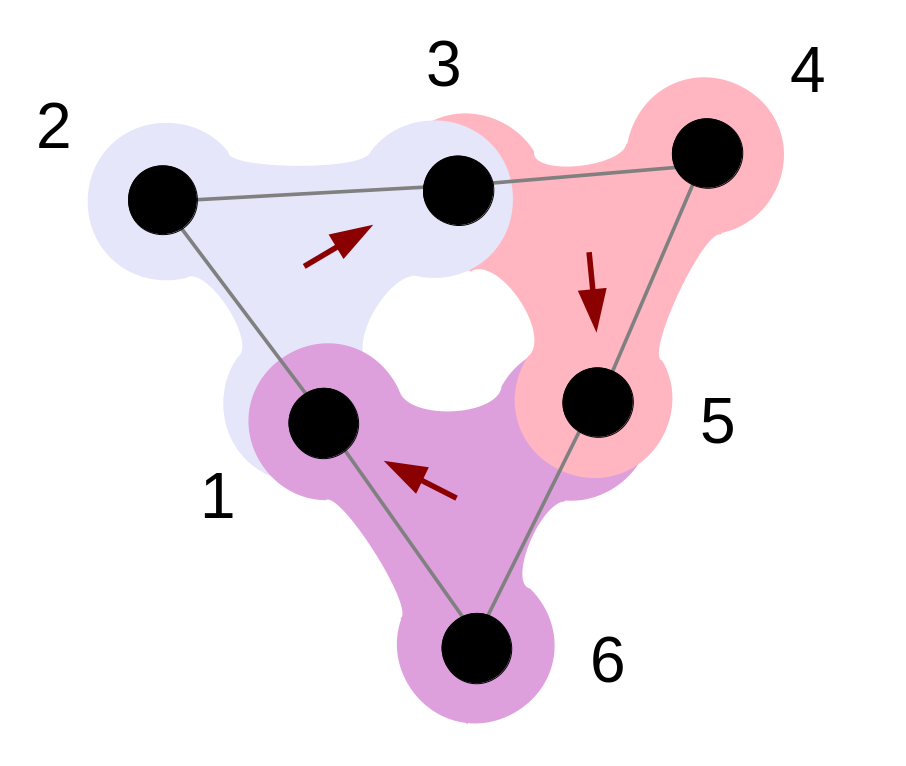}
\caption{Example of a weighted $1$-directed hypegraph with $N=6$ nodes.}
\label{fig:hedgehog3}
\end{figure}

Let us now explicitly characterize the hypergraph of $6$ nodes displayed in Fig.~\ref{fig:hedgehog3} by writing its adjacency tensors and the corresponding Laplacians. First, the adjacency matrix $A^{(1)}$, which {encodes} the standard pairwise interactions, is given by
\begin{equation}
A^{(1)} = \begin{pmatrix}
0 & 1 & 0 & 0 & 0 & 1 \\
1 & 0 & 1 & 0 & 0 & 0 \\
0 & 1 & 0 & 1 & 0 & 0 \\
0 & 0 & 1 & 0 & 1 & 0 \\
0 & 0 & 0 & 1 & 0 & 1 \\
1 & 0 & 0 & 0 & 1 & 0 
\end{pmatrix}.
\end{equation}

From $A^{(1)}$, we can evaluate the Laplacian matrix for the two-body interactions, namely
\begin{equation}
\begin{array}{c}
L^{(1)} = \begin{pNiceMatrix}[columns-width=auto]
 2 & -1 &  0 &  0 &  0 & -1 \\
-1 &  2 & -1 &  0 &  0 &  0 \\
 0 & -1 &  2 & -1 &  0 &  0 \\
 0 &  0 & -1 &  2 & -1 &  0 \\
 0 &  0 &  0 & -1 &  2 & -1 \\
-1 &  0 &  0 &  0 & -1 &  2 
\end{pNiceMatrix},
\end{array}
\end{equation}

\textcolor{black}{For the first method of symmetrization,} the adjacency tensor $A^{(2)}(p)$, which instead describes the three-body interactions, is  
\begin{equation}
\begin{array}{c}
A^{(2)}(p) =
(\{A^{(2)}_{1jk}\},\dots,\{A^{(2)}_{6jk}\})= \\ [8pt]
\left(\begin{pmatrix}
0 & 0 & 0 & 0 & 0 & 0 \\
0 & 0 & p & 0 & 0 & 0 \\
0 & p & 0 & 0 & 0 & 0 \\
0 & 0 & 0 & 0 & 0 & 0 \\
0 & 0 & 0 & 0 & 0 & 1 \\
0 & 0 & 0 & 0 & 1 & 0 
\end{pmatrix}, \begin{pmatrix}
0 & 0 & p & 0 & 0 & 0 \\
0 & 0 & 0 & 0 & 0 & 0 \\
p & 0 & 0 & 0 & 0 & 0 \\
0 & 0 & 0 & 0 & 0 & 0 \\
0 & 0 & 0 & 0 & 0 & 0 \\
0 & 0 & 0 & 0 & 0 & 0 
\end{pmatrix}, \begin{pmatrix}
0 & 1 & 0 & 0 & 0 & 0 \\
1 & 0 & 0 & 0 & 0 & 0 \\
0 & 0 & 0 & 0 & 0 & 0 \\
0 & 0 & 0 & 0 & p & 0 \\
0 & 0 & 0 & p & 0 & 0 \\
0 & 0 & 0 & 0 & 0 & 0 
\end{pmatrix},\right.\\ [45pt]
\,\,\left.\quad\begin{pmatrix}
0 & 0 & 0 & 0 & 0 & 0 \\
0 & 0 & 0 & 0 & 0 & 0 \\
0 & 0 & 0 & 0 & p & 0 \\
0 & 0 & 0 & 0 & 0 & 0 \\
0 & 0 & p & 0 & 0 & 0 \\
0 & 0 & 0 & 0 & 0 & 0 
\end{pmatrix}, \begin{pmatrix}
0 & 0 & 0 & 0 & 0 & p \\
0 & 0 & 0 & 0 & 0 & 0 \\
0 & 0 & 0 & 1 & 0 & 0 \\
0 & 0 & 1 & 0 & 0 & 0 \\
0 & 0 & 0 & 0 & 0 & 0 \\
p & 0 & 0 & 0 & 0 & 0 
\end{pmatrix}, \begin{pmatrix}
0 & 0 & 0 & 0 & p & 0 \\
0 & 0 & 0 & 0 & 0 & 0 \\
0 & 0 & 0 & 0 & 0 & 0 \\
0 & 0 & 0 & 0 & 0 & 0 \\
p & 0 & 0 & 0 & 0 & 0 \\
0 & 0 & 0 & 0 & 0 & 0 
\end{pmatrix}\right).
\end{array}
\end{equation}
We remark that, while the adjacency matrix $A^{(1)}$ is symmetric, the adjacency tensor $A^{(2)}(p)$ is not, as, for example, $A_{123}\neq A_{312}$ for $p\neq 1$. However, one can see that the tensor becomes symmetric ($A^{(2)}_{ijk}=1 \Rightarrow A^{(2)}_{\pi(ijk)}=1$, with $\pi$ a generic permutation of indices) when $p=1$. Furthermore, we note that the matrices resulting from fixing the first index of the tensor, given the property in Eq.~(1), are symmetric for any value of $p$.

Given $A^{(2)}(p)$, it is possible to calculate the generalized in-degrees of the nodes (see Eq.~(\ref{eq:node_in_degree}) for the definition) and the generalized in-degrees of the node couples (Eq.~(\ref{eq:link_in_degree})). Hence, we can evaluate the generalized Laplacian matrix for the three-body interactions (Eq.~(\ref{eq:d_laplacian})). We have
\begin{equation}
\begin{array}{c}
L^{(2)}(p) = \begin{pNiceMatrix}[columns-width=auto]
2(1+p) & -p & -p &  0 & -1 & -1 \\
-p &  2p & -p &  0 &  0 &  0 \\
-1 & -1 & 2(1+p) & -p & -p &  0 \\
 0 &  0 & -p &   2p & -p &  0 \\
-p &  0 & -1 & -1 & 2(1+p) & -p \\
-p &  0 &  0 &  0 & -p &  2p 
\end{pNiceMatrix}.
\end{array}
\end{equation}

Since the adjacency tensor $A^{(2)}(p)$ is asymmetric, consequently $L^{(2)}(p)$ is also asymmetric. Consistently, when $p=1$, which corresponds to the case of an undirected hypergraph, the Laplacian matrix becomes symmetric. \\

{\color{black} For the second method of symmetrization for three-body interactions, the adjacency tensor $A^{(2)}(q)$ is given by
\begin{equation}
\begin{array}{c}
A^{(2)}(q) =
(\{A^{(2)}_{1jk}\},\dots,\{A^{(2)}_{6jk}\})= \\ [8pt]
\left(\begin{pNiceMatrix}[columns-width=auto]
 0 &  0 &  0 &  0 & 0 & 0 \\
 0 &  0 &   q &  0 & 0 & 0 \\
 0 &   q &  0 &  0 & 0 & 0 \\
 0 &  0 &  0 &  0 & 0 & 0 \\
 0 &  0 &  0 &  0 & 0 & 1-2q \\
 0 &  0 &  0 &  0 & 1-2q & 0 
\end{pNiceMatrix}, \begin{pNiceMatrix}[columns-width=auto]
0 & 0 & q & 0 & 0 & 0 \\
0 & 0 & 0 & 0 & 0 & 0 \\
q & 0 & 0 & 0 & 0 & 0 \\
0 & 0 & 0 & 0 & 0 & 0 \\
0 & 0 & 0 & 0 & 0 & 0 \\
0 & 0 & 0 & 0 & 0 & 0 
\end{pNiceMatrix},\right.\\ [45pt]
\,\,\left.\quad \begin{pNiceMatrix}[columns-width=auto]
0 & 1-2q &  0 &  0 &  0 &  0 \\
1-2q & 0 &  0 &  0 &  0 &  0 \\
0 & 0 &  0 &  0 &  0 &  0 \\
0 & 0 &  0 &  0 &  q &  0 \\
0 & 0 &  0 &  q &  0 &  0 \\
0 & 0 &  0 &  0 &  0 &  0 
\end{pNiceMatrix}, \begin{pNiceMatrix}[columns-width=auto]
0 & 0 & 0 & 0 & 0 & 0 \\
0 & 0 & 0 & 0 & 0 & 0 \\
0 & 0 & 0 & 0 & q & 0 \\
0 & 0 & 0 & 0 & 0 & 0 \\
0 & 0 & q & 0 & 0 & 0 \\
0 & 0 & 0 & 0 & 0 & 0 
\end{pNiceMatrix},\right.\\ [45pt]
\,\,\left.\quad \begin{pNiceMatrix}[columns-width=auto]
 0 &  0 & 0 & 0 &  0 &   q \\
 0 &  0 & 0 & 0 &  0 &  0 \\
 0 &  0 & 0 & 1-2q &  0 &  0 \\
 0 &  0 & 1-2q & 0 &  0 &  0 \\
 0 &  0 & 0 & 0 &  0 &  0 \\
  q &  0 & 0 & 0 &  0 &  0 
\end{pNiceMatrix}, \begin{pNiceMatrix}[columns-width=auto]
0 & 0 & 0 & 0 & q & 0 \\
0 & 0 & 0 & 0 & 0 & 0 \\
0 & 0 & 0 & 0 & 0 & 0 \\
0 & 0 & 0 & 0 & 0 & 0 \\
q & 0 & 0 & 0 & 0 & 0 \\
0 & 0 & 0 & 0 & 0 & 0 
\end{pNiceMatrix}\right),
\end{array}
\end{equation}
which, similarly to $A^{(2)}(p)$ is in general asymmetric. From $A^{(2)}(q)$ we can evaluate the generalized Laplacian $L^{(2)}(q)$, which has the following expression \begin{displaymath}
L^{(2)}(q) = \begin{pNiceMatrix}[columns-width=auto]
2(1-q) & -q & -q &  0 & -(1-2q) & -(1-2q) \\
-q &  2q & -q &  0 &  0 &  0 \\
-(1-2q) & -(1-2q) & 2(1-q) & -q & -q &  0 \\
 0 &  0 & -q &   2q & -q &  0 \\
-q &  0 & -(1-2q) & -(1-2q) & 2(1-q) & -q \\
-q &  0 &  0 &  0 & -q &  2q 
\end{pNiceMatrix}.
\end{displaymath}}
\textcolor{black}{As the adjacency tensor $A^{(2)}(q)$ is asymmetric, so the generalized Laplacian matrix $L^{(2)}(q)$ is asymmetric. Nonetheless, when $q=1/3$, corresponding to the case of an undirected hypergraph, $L^{(2)}(q)$ becomes symmetric}. 

\subsubsection*{Acknowledgements} R.M. is supported by a FRIA-FNRS PhD fellowship, Grant FC 33443, funded by the Walloon region. R.M. acknowledges the Erasmus+ program for funding his visit in the group of Professor M.F.

\bibliography{sn-bibliography}

\begin{thebibliography}{61}%
\makeatletter
\providecommand \@ifxundefined [1]{%
 \@ifx{#1\undefined}
}%
\providecommand \@ifnum [1]{%
 \ifnum #1\expandafter \@firstoftwo
 \else \expandafter \@secondoftwo
 \fi
}%
\providecommand \@ifx [1]{%
 \ifx #1\expandafter \@firstoftwo
 \else \expandafter \@secondoftwo
 \fi
}%
\providecommand \natexlab [1]{#1}%
\providecommand \enquote  [1]{``#1''}%
\providecommand \bibnamefont  [1]{#1}%
\providecommand \bibfnamefont [1]{#1}%
\providecommand \citenamefont [1]{#1}%
\providecommand \href@noop [0]{\@secondoftwo}%
\providecommand \href [0]{\begingroup \@sanitize@url \@href}%
\providecommand \@href[1]{\@@startlink{#1}\@@href}%
\providecommand \@@href[1]{\endgroup#1\@@endlink}%
\providecommand \@sanitize@url [0]{\catcode `\\12\catcode `\$12\catcode
  `\&12\catcode `\#12\catcode `\^12\catcode `\_12\catcode `\%12\relax}%
\providecommand \@@startlink[1]{}%
\providecommand \@@endlink[0]{}%
\providecommand \url  [0]{\begingroup\@sanitize@url \@url }%
\providecommand \@url [1]{\endgroup\@href {#1}{\urlprefix }}%
\providecommand \urlprefix  [0]{URL }%
\providecommand \Eprint [0]{\href }%
\providecommand \doibase [0]{http://dx.doi.org/}%
\providecommand \selectlanguage [0]{\@gobble}%
\providecommand \bibinfo  [0]{\@secondoftwo}%
\providecommand \bibfield  [0]{\@secondoftwo}%
\providecommand \translation [1]{[#1]}%
\providecommand \BibitemOpen [0]{}%
\providecommand \bibitemStop [0]{}%
\providecommand \bibitemNoStop [0]{.\EOS\space}%
\providecommand \EOS [0]{\spacefactor3000\relax}%
\providecommand \BibitemShut  [1]{\csname bibitem#1\endcsname}%
\let\auto@bib@innerbib\@empty
\bibitem [{\citenamefont {Newman}(2010)}]{newmanbook}%
  \BibitemOpen
  \bibfield  {author} {\bibinfo {author} {\bibfnamefont {Mark~EJ}\ \bibnamefont
  {Newman}},\ }\href@noop {} {\emph {\bibinfo {title} {Networks: An
  Introduction}}}\ (\bibinfo  {publisher} {Oxford University Press},\ \bibinfo
  {address} {Oxford},\ \bibinfo {year} {2010})\BibitemShut {NoStop}%
\bibitem [{\citenamefont {Boccaletti}\ \emph {et~al.}(2006)\citenamefont
  {Boccaletti}, \citenamefont {Latora}, \citenamefont {Moreno}, \citenamefont
  {Chavez},\ and\ \citenamefont {Hwang}}]{boccaletti2006complex}%
  \BibitemOpen
  \bibfield  {author} {\bibinfo {author} {\bibfnamefont {Stefano}\ \bibnamefont
  {Boccaletti}}, \bibinfo {author} {\bibfnamefont {Vito}\ \bibnamefont
  {Latora}}, \bibinfo {author} {\bibfnamefont {Yamir}\ \bibnamefont {Moreno}},
  \bibinfo {author} {\bibfnamefont {Martin}\ \bibnamefont {Chavez}}, \ and\
  \bibinfo {author} {\bibfnamefont {D-U}\ \bibnamefont {Hwang}},\ }\bibfield
  {title} {\enquote {\bibinfo {title} {Complex networks: Structure and
  dynamics},}\ }\href@noop {} {\bibfield  {journal} {\bibinfo  {journal}
  {Physics Reports}\ }\textbf {\bibinfo {volume} {424}},\ \bibinfo {pages}
  {175--308} (\bibinfo {year} {2006})}\BibitemShut {NoStop}%
\bibitem [{\citenamefont {Latora}\ \emph {et~al.}(2017)\citenamefont {Latora},
  \citenamefont {Nicosia},\ and\ \citenamefont
  {Russo}}]{latora_nicosia_russo_2017}%
  \BibitemOpen
  \bibfield  {author} {\bibinfo {author} {\bibfnamefont {Vito}\ \bibnamefont
  {Latora}}, \bibinfo {author} {\bibfnamefont {Vincenzo}\ \bibnamefont
  {Nicosia}}, \ and\ \bibinfo {author} {\bibfnamefont {Giovanni}\ \bibnamefont
  {Russo}},\ }\href@noop {} {\emph {\bibinfo {title} {Complex Networks:
  Principles, Methods and Applications}}}\ (\bibinfo  {publisher} {Cambridge
  University Press},\ \bibinfo {address} {Cambridge},\ \bibinfo {year}
  {2017})\BibitemShut {NoStop}%
\bibitem [{\citenamefont {Battiston}\ \emph {et~al.}(2020)\citenamefont
  {Battiston}, \citenamefont {Cencetti}, \citenamefont {Iacopini},
  \citenamefont {Latora}, \citenamefont {Lucas}, \citenamefont {Patania},
  \citenamefont {Young},\ and\ \citenamefont {Petri}}]{battiston2020networks}%
  \BibitemOpen
  \bibfield  {author} {\bibinfo {author} {\bibfnamefont {Federico}\
  \bibnamefont {Battiston}}, \bibinfo {author} {\bibfnamefont {Giulia}\
  \bibnamefont {Cencetti}}, \bibinfo {author} {\bibfnamefont {Iacopo}\
  \bibnamefont {Iacopini}}, \bibinfo {author} {\bibfnamefont {Vito}\
  \bibnamefont {Latora}}, \bibinfo {author} {\bibfnamefont {Maxime}\
  \bibnamefont {Lucas}}, \bibinfo {author} {\bibfnamefont {Alice}\ \bibnamefont
  {Patania}}, \bibinfo {author} {\bibfnamefont {Jean-Gabriel}\ \bibnamefont
  {Young}}, \ and\ \bibinfo {author} {\bibfnamefont {Giovanni}\ \bibnamefont
  {Petri}},\ }\bibfield  {title} {\enquote {\bibinfo {title} {Networks beyond
  pairwise interactions: structure and dynamics},}\ }\href@noop {} {\bibfield
  {journal} {\bibinfo  {journal} {Physics Reports}\ } (\bibinfo {year}
  {2020})}\BibitemShut {NoStop}%
\bibitem [{\citenamefont {Klamt}\ \emph {et~al.}(2009)\citenamefont {Klamt},
  \citenamefont {Haus},\ and\ \citenamefont {Theis}}]{klamt2009hypergraphs}%
  \BibitemOpen
  \bibfield  {author} {\bibinfo {author} {\bibfnamefont {Steffen}\ \bibnamefont
  {Klamt}}, \bibinfo {author} {\bibfnamefont {Utz-Uwe}\ \bibnamefont {Haus}}, \
  and\ \bibinfo {author} {\bibfnamefont {Fabian}\ \bibnamefont {Theis}},\
  }\bibfield  {title} {\enquote {\bibinfo {title} {Hypergraphs and cellular
  networks},}\ }\href@noop {} {\bibfield  {journal} {\bibinfo  {journal} {PLoS
  computational biology}\ }\textbf {\bibinfo {volume} {5}},\ \bibinfo {pages}
  {e1000385} (\bibinfo {year} {2009})}\BibitemShut {NoStop}%
\bibitem [{\citenamefont {Estrada}\ and\ \citenamefont
  {Ross}(2018)}]{estrada2018centralities}%
  \BibitemOpen
  \bibfield  {author} {\bibinfo {author} {\bibfnamefont {Ernesto}\ \bibnamefont
  {Estrada}}\ and\ \bibinfo {author} {\bibfnamefont {Grant~J}\ \bibnamefont
  {Ross}},\ }\bibfield  {title} {\enquote {\bibinfo {title} {Centralities in
  simplicial complexes. applications to protein interaction networks},}\
  }\href@noop {} {\bibfield  {journal} {\bibinfo  {journal} {J. Theor. Biol.}\
  }\textbf {\bibinfo {volume} {438}},\ \bibinfo {pages} {46--60} (\bibinfo
  {year} {2018})}\BibitemShut {NoStop}%
\bibitem [{\citenamefont {Petri}\ \emph {et~al.}(2014)\citenamefont {Petri},
  \citenamefont {Expert}, \citenamefont {Turkheimer}, \citenamefont
  {Carhart-Harris}, \citenamefont {Nutt}, \citenamefont {Hellyer},\ and\
  \citenamefont {Vaccarino}}]{petri2014homological}%
  \BibitemOpen
  \bibfield  {author} {\bibinfo {author} {\bibfnamefont {Giovanni}\
  \bibnamefont {Petri}}, \bibinfo {author} {\bibfnamefont {Paul}\ \bibnamefont
  {Expert}}, \bibinfo {author} {\bibfnamefont {Federico}\ \bibnamefont
  {Turkheimer}}, \bibinfo {author} {\bibfnamefont {Robin}\ \bibnamefont
  {Carhart-Harris}}, \bibinfo {author} {\bibfnamefont {David}\ \bibnamefont
  {Nutt}}, \bibinfo {author} {\bibfnamefont {Peter~J}\ \bibnamefont {Hellyer}},
  \ and\ \bibinfo {author} {\bibfnamefont {Francesco}\ \bibnamefont
  {Vaccarino}},\ }\bibfield  {title} {\enquote {\bibinfo {title} {Homological
  scaffolds of brain functional networks},}\ }\href@noop {} {\bibfield
  {journal} {\bibinfo  {journal} {Journal of The Royal Society Interface}\
  }\textbf {\bibinfo {volume} {11}},\ \bibinfo {pages} {20140873} (\bibinfo
  {year} {2014})}\BibitemShut {NoStop}%
\bibitem [{\citenamefont {Giusti}\ \emph {et~al.}(2015)\citenamefont {Giusti},
  \citenamefont {Pastalkova}, \citenamefont {Curto},\ and\ \citenamefont
  {Itskov}}]{giusti2015clique}%
  \BibitemOpen
  \bibfield  {author} {\bibinfo {author} {\bibfnamefont {Chad}\ \bibnamefont
  {Giusti}}, \bibinfo {author} {\bibfnamefont {Eva}\ \bibnamefont
  {Pastalkova}}, \bibinfo {author} {\bibfnamefont {Carina}\ \bibnamefont
  {Curto}}, \ and\ \bibinfo {author} {\bibfnamefont {Vladimir}\ \bibnamefont
  {Itskov}},\ }\bibfield  {title} {\enquote {\bibinfo {title} {Clique topology
  reveals intrinsic geometric structure in neural correlations},}\ }\href@noop
  {} {\bibfield  {journal} {\bibinfo  {journal} {Pro. Natl. Acad Sci. U.S.A.}\
  }\textbf {\bibinfo {volume} {112}},\ \bibinfo {pages} {13455--13460}
  (\bibinfo {year} {2015})}\BibitemShut {NoStop}%
\bibitem [{\citenamefont {Sizemore}\ \emph {et~al.}(2018)\citenamefont
  {Sizemore}, \citenamefont {Giusti}, \citenamefont {Kahn}, \citenamefont
  {Vettel}, \citenamefont {Betzel},\ and\ \citenamefont
  {Bassett}}]{sizemore2018cliques}%
  \BibitemOpen
  \bibfield  {author} {\bibinfo {author} {\bibfnamefont {Ann~E}\ \bibnamefont
  {Sizemore}}, \bibinfo {author} {\bibfnamefont {Chad}\ \bibnamefont {Giusti}},
  \bibinfo {author} {\bibfnamefont {Ari}\ \bibnamefont {Kahn}}, \bibinfo
  {author} {\bibfnamefont {Jean~M}\ \bibnamefont {Vettel}}, \bibinfo {author}
  {\bibfnamefont {Richard~F}\ \bibnamefont {Betzel}}, \ and\ \bibinfo {author}
  {\bibfnamefont {Danielle~S}\ \bibnamefont {Bassett}},\ }\bibfield  {title}
  {\enquote {\bibinfo {title} {Cliques and cavities in the human connectome},}\
  }\href@noop {} {\bibfield  {journal} {\bibinfo  {journal} {J. Comp.
  Neurosci.}\ }\textbf {\bibinfo {volume} {44}},\ \bibinfo {pages} {115--145}
  (\bibinfo {year} {2018})}\BibitemShut {NoStop}%
\bibitem [{\citenamefont {Giusti}\ \emph {et~al.}(2016)\citenamefont {Giusti},
  \citenamefont {Ghrist},\ and\ \citenamefont {Bassett}}]{Bassett}%
  \BibitemOpen
  \bibfield  {author} {\bibinfo {author} {\bibfnamefont {C}~\bibnamefont
  {Giusti}}, \bibinfo {author} {\bibfnamefont {R}~\bibnamefont {Ghrist}}, \
  and\ \bibinfo {author} {\bibfnamefont {D~S}\ \bibnamefont {Bassett}},\
  }\bibfield  {title} {\enquote {\bibinfo {title} {Two’s company, three (or
  more) is a simplex. algebraic-topological tools for understanding
  higher-order structure in neural data},}\ }\href@noop {} {\bibfield
  {journal} {\bibinfo  {journal} {J Comput Neurosci}\ }\textbf {\bibinfo
  {volume} {41}},\ \bibinfo {pages} {1--14} (\bibinfo {year}
  {2016})}\BibitemShut {NoStop}%
\bibitem [{\citenamefont {Benson}\ \emph {et~al.}(2016)\citenamefont {Benson},
  \citenamefont {Gleich},\ and\ \citenamefont {Leskovec}}]{benson2016higher}%
  \BibitemOpen
  \bibfield  {author} {\bibinfo {author} {\bibfnamefont {Austin~R}\
  \bibnamefont {Benson}}, \bibinfo {author} {\bibfnamefont {David~F}\
  \bibnamefont {Gleich}}, \ and\ \bibinfo {author} {\bibfnamefont {Jure}\
  \bibnamefont {Leskovec}},\ }\bibfield  {title} {\enquote {\bibinfo {title}
  {Higher-order organization of complex networks},}\ }\href@noop {} {\bibfield
  {journal} {\bibinfo  {journal} {Science}\ }\textbf {\bibinfo {volume}
  {353}},\ \bibinfo {pages} {163--166} (\bibinfo {year} {2016})}\BibitemShut
  {NoStop}%
\bibitem [{\citenamefont {Patania}\ \emph {et~al.}(2017)\citenamefont
  {Patania}, \citenamefont {Petri},\ and\ \citenamefont
  {Vaccarino}}]{patania2017shape}%
  \BibitemOpen
  \bibfield  {author} {\bibinfo {author} {\bibfnamefont {Alice}\ \bibnamefont
  {Patania}}, \bibinfo {author} {\bibfnamefont {Giovanni}\ \bibnamefont
  {Petri}}, \ and\ \bibinfo {author} {\bibfnamefont {Francesco}\ \bibnamefont
  {Vaccarino}},\ }\bibfield  {title} {\enquote {\bibinfo {title} {The shape of
  collaborations},}\ }\href@noop {} {\bibfield  {journal} {\bibinfo  {journal}
  {EPJ Data Sci.}\ }\textbf {\bibinfo {volume} {6}},\ \bibinfo {pages} {18}
  (\bibinfo {year} {2017})}\BibitemShut {NoStop}%
\bibitem [{\citenamefont {Billick}\ and\ \citenamefont
  {Case}(1994)}]{billick1994higher}%
  \BibitemOpen
  \bibfield  {author} {\bibinfo {author} {\bibfnamefont {Ian}\ \bibnamefont
  {Billick}}\ and\ \bibinfo {author} {\bibfnamefont {Ted~J}\ \bibnamefont
  {Case}},\ }\bibfield  {title} {\enquote {\bibinfo {title} {Higher order
  interactions in ecological communities: what are they and how can they be
  detected?}}\ }\href@noop {} {\bibfield  {journal} {\bibinfo  {journal}
  {Ecology}\ }\textbf {\bibinfo {volume} {75}},\ \bibinfo {pages} {1529--1543}
  (\bibinfo {year} {1994})}\BibitemShut {NoStop}%
\bibitem [{\citenamefont {Bairey}\ \emph {et~al.}(2016)\citenamefont {Bairey},
  \citenamefont {Kelsic},\ and\ \citenamefont {Kishony}}]{bairey2016high}%
  \BibitemOpen
  \bibfield  {author} {\bibinfo {author} {\bibfnamefont {Eyal}\ \bibnamefont
  {Bairey}}, \bibinfo {author} {\bibfnamefont {Eric~D}\ \bibnamefont {Kelsic}},
  \ and\ \bibinfo {author} {\bibfnamefont {Roy}\ \bibnamefont {Kishony}},\
  }\bibfield  {title} {\enquote {\bibinfo {title} {High-order species
  interactions shape ecosystem diversity},}\ }\href@noop {} {\bibfield
  {journal} {\bibinfo  {journal} {Nature communications}\ }\textbf {\bibinfo
  {volume} {7}},\ \bibinfo {pages} {1--7} (\bibinfo {year} {2016})}\BibitemShut
  {NoStop}%
\bibitem [{\citenamefont {Grilli}\ \emph {et~al.}(2017)\citenamefont {Grilli},
  \citenamefont {Barab{\'a}s}, \citenamefont {Michalska-Smith},\ and\
  \citenamefont {Allesina}}]{grilli2017higher}%
  \BibitemOpen
  \bibfield  {author} {\bibinfo {author} {\bibfnamefont {Jacopo}\ \bibnamefont
  {Grilli}}, \bibinfo {author} {\bibfnamefont {Gy{\"o}rgy}\ \bibnamefont
  {Barab{\'a}s}}, \bibinfo {author} {\bibfnamefont {Matthew~J}\ \bibnamefont
  {Michalska-Smith}}, \ and\ \bibinfo {author} {\bibfnamefont {Stefano}\
  \bibnamefont {Allesina}},\ }\bibfield  {title} {\enquote {\bibinfo {title}
  {Higher-order interactions stabilize dynamics in competitive network
  models},}\ }\href@noop {} {\bibfield  {journal} {\bibinfo  {journal}
  {Nature}\ }\textbf {\bibinfo {volume} {548}},\ \bibinfo {pages} {210--213}
  (\bibinfo {year} {2017})}\BibitemShut {NoStop}%
\bibitem [{\citenamefont {Berge}(1973)}]{berge1973graphs}%
  \BibitemOpen
  \bibfield  {author} {\bibinfo {author} {\bibfnamefont {Claude}\ \bibnamefont
  {Berge}},\ }\href@noop {} {\emph {\bibinfo {title} {{Graphs and
  hypergraphs}}}},\ North-Holl Math. Libr.\ (\bibinfo  {publisher}
  {North-Holland},\ \bibinfo {address} {Amsterdam},\ \bibinfo {year}
  {1973})\BibitemShut {NoStop}%
\bibitem [{\citenamefont {Lucas}\ \emph {et~al.}(2020)\citenamefont {Lucas},
  \citenamefont {Cencetti},\ and\ \citenamefont {Battiston}}]{maxime2020}%
  \BibitemOpen
  \bibfield  {author} {\bibinfo {author} {\bibfnamefont {Maxime}\ \bibnamefont
  {Lucas}}, \bibinfo {author} {\bibfnamefont {Giulia}\ \bibnamefont
  {Cencetti}}, \ and\ \bibinfo {author} {\bibfnamefont {Federico}\ \bibnamefont
  {Battiston}},\ }\bibfield  {title} {\enquote {\bibinfo {title} {A multi-order
  laplacian framework for the stability of higher-order synchronization},}\
  }\href@noop {} {\bibfield  {journal} {\bibinfo  {journal} {Physical Review
  Research}\ }\textbf {\bibinfo {volume} {2}},\ \bibinfo {pages} {033410}
  (\bibinfo {year} {2020})}\BibitemShut {NoStop}%
\bibitem [{\citenamefont {Carletti}\ \emph
  {et~al.}(2020{\natexlab{a}})\citenamefont {Carletti}, \citenamefont
  {Fanelli},\ and\ \citenamefont {Nicoletti}}]{carletti2020dynamical}%
  \BibitemOpen
  \bibfield  {author} {\bibinfo {author} {\bibfnamefont {Timoteo}\ \bibnamefont
  {Carletti}}, \bibinfo {author} {\bibfnamefont {Duccio}\ \bibnamefont
  {Fanelli}}, \ and\ \bibinfo {author} {\bibfnamefont {Sara}\ \bibnamefont
  {Nicoletti}},\ }\bibfield  {title} {\enquote {\bibinfo {title} {Dynamical
  systems on hypergraphs},}\ }\href@noop {} {\bibfield  {journal} {\bibinfo
  {journal} {Journal of Physics: Complexity}\ }\textbf {\bibinfo {volume}
  {1}},\ \bibinfo {pages} {035006} (\bibinfo {year}
  {2020}{\natexlab{a}})}\BibitemShut {NoStop}%
\bibitem [{\citenamefont {de~Arruda}\ \emph {et~al.}(2021)\citenamefont
  {de~Arruda}, \citenamefont {Tizzani},\ and\ \citenamefont
  {Moreno}}]{de2021phase}%
  \BibitemOpen
  \bibfield  {author} {\bibinfo {author} {\bibfnamefont {Guilherme~Ferraz}\
  \bibnamefont {de~Arruda}}, \bibinfo {author} {\bibfnamefont {Michele}\
  \bibnamefont {Tizzani}}, \ and\ \bibinfo {author} {\bibfnamefont {Yamir}\
  \bibnamefont {Moreno}},\ }\bibfield  {title} {\enquote {\bibinfo {title}
  {Phase transitions and stability of dynamical processes on hypergraphs},}\
  }\href@noop {} {\bibfield  {journal} {\bibinfo  {journal} {Communications
  Physics}\ }\textbf {\bibinfo {volume} {4}},\ \bibinfo {pages} {1--9}
  (\bibinfo {year} {2021})}\BibitemShut {NoStop}%
\bibitem [{\citenamefont {St-Onge}\ \emph {et~al.}(2021)\citenamefont
  {St-Onge}, \citenamefont {Sun}, \citenamefont {Allard}, \citenamefont
  {H\'ebert-Dufresne},\ and\ \citenamefont {Bianconi}}]{stonge2021universal}%
  \BibitemOpen
  \bibfield  {author} {\bibinfo {author} {\bibfnamefont {Guillaume}\
  \bibnamefont {St-Onge}}, \bibinfo {author} {\bibfnamefont {Hanlin}\
  \bibnamefont {Sun}}, \bibinfo {author} {\bibfnamefont {Antoine}\ \bibnamefont
  {Allard}}, \bibinfo {author} {\bibfnamefont {Laurent}\ \bibnamefont
  {H\'ebert-Dufresne}}, \ and\ \bibinfo {author} {\bibfnamefont {Ginestra}\
  \bibnamefont {Bianconi}},\ }\bibfield  {title} {\enquote {\bibinfo {title}
  {Universal nonlinear infection kernel from heterogeneous exposure on
  higher-order networks},}\ }\href@noop {} {\bibfield  {journal} {\bibinfo
  {journal} {Phys. Rev. Lett.}\ }\textbf {\bibinfo {volume} {127}},\ \bibinfo
  {pages} {158301} (\bibinfo {year} {2021})}\BibitemShut {NoStop}%
\bibitem [{\citenamefont {Iacopini}\ \emph {et~al.}(2019)\citenamefont
  {Iacopini}, \citenamefont {Petri}, \citenamefont {Barrat},\ and\
  \citenamefont {Latora}}]{iacopini2019simplicial}%
  \BibitemOpen
  \bibfield  {author} {\bibinfo {author} {\bibfnamefont {Iacopo}\ \bibnamefont
  {Iacopini}}, \bibinfo {author} {\bibfnamefont {Giovanni}\ \bibnamefont
  {Petri}}, \bibinfo {author} {\bibfnamefont {Alain}\ \bibnamefont {Barrat}}, \
  and\ \bibinfo {author} {\bibfnamefont {Vito}\ \bibnamefont {Latora}},\
  }\bibfield  {title} {\enquote {\bibinfo {title} {Simplicial models of social
  contagion},}\ }\href@noop {} {\bibfield  {journal} {\bibinfo  {journal}
  {Nature Communications}\ }\textbf {\bibinfo {volume} {10}},\ \bibinfo {pages}
  {2485} (\bibinfo {year} {2019})}\BibitemShut {NoStop}%
\bibitem [{\citenamefont {de~Arruda}\ \emph {et~al.}(2020)\citenamefont
  {de~Arruda}, \citenamefont {Petri},\ and\ \citenamefont
  {Moreno}}]{de2019social}%
  \BibitemOpen
  \bibfield  {author} {\bibinfo {author} {\bibfnamefont {Guilherme~Ferraz}\
  \bibnamefont {de~Arruda}}, \bibinfo {author} {\bibfnamefont {Giovanni}\
  \bibnamefont {Petri}}, \ and\ \bibinfo {author} {\bibfnamefont {Yamir}\
  \bibnamefont {Moreno}},\ }\bibfield  {title} {\enquote {\bibinfo {title}
  {Social contagion models on hypergraphs},}\ }\href@noop {} {\bibfield
  {journal} {\bibinfo  {journal} {Phys. Rev. Research}\ }\textbf {\bibinfo
  {volume} {2}},\ \bibinfo {pages} {023032} (\bibinfo {year}
  {2020})}\BibitemShut {NoStop}%
\bibitem [{\citenamefont {Carletti}\ \emph
  {et~al.}(2020{\natexlab{b}})\citenamefont {Carletti}, \citenamefont
  {Battiston}, \citenamefont {Cencetti},\ and\ \citenamefont
  {Fanelli}}]{carletti2020random}%
  \BibitemOpen
  \bibfield  {author} {\bibinfo {author} {\bibfnamefont {Timoteo}\ \bibnamefont
  {Carletti}}, \bibinfo {author} {\bibfnamefont {Federico}\ \bibnamefont
  {Battiston}}, \bibinfo {author} {\bibfnamefont {Giulia}\ \bibnamefont
  {Cencetti}}, \ and\ \bibinfo {author} {\bibfnamefont {Duccio}\ \bibnamefont
  {Fanelli}},\ }\bibfield  {title} {\enquote {\bibinfo {title} {Random walks on
  hypergraphs},}\ }\href@noop {} {\bibfield  {journal} {\bibinfo  {journal}
  {Physical Review E}\ }\textbf {\bibinfo {volume} {101}},\ \bibinfo {pages}
  {022308} (\bibinfo {year} {2020}{\natexlab{b}})}\BibitemShut {NoStop}%
\bibitem [{\citenamefont {Carletti}\ \emph {et~al.}(2021)\citenamefont
  {Carletti}, \citenamefont {Fanelli},\ and\ \citenamefont
  {Lambiotte}}]{carletti2021random}%
  \BibitemOpen
  \bibfield  {author} {\bibinfo {author} {\bibfnamefont {Timoteo}\ \bibnamefont
  {Carletti}}, \bibinfo {author} {\bibfnamefont {Duccio}\ \bibnamefont
  {Fanelli}}, \ and\ \bibinfo {author} {\bibfnamefont {Renaud}\ \bibnamefont
  {Lambiotte}},\ }\bibfield  {title} {\enquote {\bibinfo {title} {Random walks
  and community detection in hypergraphs},}\ }\href@noop {} {\bibfield
  {journal} {\bibinfo  {journal} {Journal of Physics: Complexity}\ }\textbf
  {\bibinfo {volume} {2}},\ \bibinfo {pages} {015011} (\bibinfo {year}
  {2021})}\BibitemShut {NoStop}%
\bibitem [{\citenamefont {Skardal}\ and\ \citenamefont
  {Arenas}(2019)}]{skardal2019abrupt}%
  \BibitemOpen
  \bibfield  {author} {\bibinfo {author} {\bibfnamefont {Per~Sebastian}\
  \bibnamefont {Skardal}}\ and\ \bibinfo {author} {\bibfnamefont {Alex}\
  \bibnamefont {Arenas}},\ }\bibfield  {title} {\enquote {\bibinfo {title}
  {Abrupt desynchronization and extensive multistability in globally coupled
  oscillator simplexes},}\ }\href@noop {} {\bibfield  {journal} {\bibinfo
  {journal} {Physical Review Letters}\ }\textbf {\bibinfo {volume} {122}},\
  \bibinfo {pages} {248301} (\bibinfo {year} {2019})}\BibitemShut {NoStop}%
\bibitem [{\citenamefont {Skardal}\ and\ \citenamefont
  {Arenas}(2020)}]{skardal2019higher}%
  \BibitemOpen
  \bibfield  {author} {\bibinfo {author} {\bibfnamefont {Per~Sebastian}\
  \bibnamefont {Skardal}}\ and\ \bibinfo {author} {\bibfnamefont {Alex}\
  \bibnamefont {Arenas}},\ }\bibfield  {title} {\enquote {\bibinfo {title}
  {Higher-order interactions in complex networks of phase oscillators promote
  abrupt synchronization switching},}\ }\href@noop {} {\bibfield  {journal}
  {\bibinfo  {journal} {Communications Physics}\ }\textbf {\bibinfo {volume}
  {3}} (\bibinfo {year} {2020})}\BibitemShut {NoStop}%
\bibitem [{\citenamefont {Neuh{\"a}user}\ \emph {et~al.}(2020)\citenamefont
  {Neuh{\"a}user}, \citenamefont {Mellor},\ and\ \citenamefont
  {Lambiotte}}]{neuhauser2020multibody}%
  \BibitemOpen
  \bibfield  {author} {\bibinfo {author} {\bibfnamefont {Leonie}\ \bibnamefont
  {Neuh{\"a}user}}, \bibinfo {author} {\bibfnamefont {Andrew}\ \bibnamefont
  {Mellor}}, \ and\ \bibinfo {author} {\bibfnamefont {Renaud}\ \bibnamefont
  {Lambiotte}},\ }\bibfield  {title} {\enquote {\bibinfo {title} {Multibody
  interactions and nonlinear consensus dynamics on networked systems},}\
  }\href@noop {} {\bibfield  {journal} {\bibinfo  {journal} {Physical Review
  E}\ }\textbf {\bibinfo {volume} {101}},\ \bibinfo {pages} {032310} (\bibinfo
  {year} {2020})}\BibitemShut {NoStop}%
\bibitem [{\citenamefont {Neuh{\"a}user}\ \emph {et~al.}(2021)\citenamefont
  {Neuh{\"a}user}, \citenamefont {Lambiotte},\ and\ \citenamefont
  {Schaub}}]{neuhauser2021}%
  \BibitemOpen
  \bibfield  {author} {\bibinfo {author} {\bibfnamefont {Leonie}\ \bibnamefont
  {Neuh{\"a}user}}, \bibinfo {author} {\bibfnamefont {Renaud}\ \bibnamefont
  {Lambiotte}}, \ and\ \bibinfo {author} {\bibfnamefont {Michael}\ \bibnamefont
  {Schaub}},\ }\bibfield  {title} {\enquote {\bibinfo {title} {Consensus
  dynamics on temporal hypergraphs},}\ }\href@noop {} {\bibfield  {journal}
  {\bibinfo  {journal} {Physical Review E}\ }\textbf {\bibinfo {volume}
  {104}},\ \bibinfo {pages} {064305} (\bibinfo {year} {2021})}\BibitemShut
  {NoStop}%
\bibitem [{\citenamefont {Asch}(1951)}]{asch1961}%
  \BibitemOpen
  \bibfield  {author} {\bibinfo {author} {\bibfnamefont {S~E}\ \bibnamefont
  {Asch}},\ }\bibfield  {title} {\enquote {\bibinfo {title} {Effects of group
  pressure on the modification and distortion of judgments},}\ }\href@noop {}
  {\bibfield  {journal} {\bibinfo  {journal} {Groups, Leadership and Men}\ ,\
  \bibinfo {pages} {177--190}} (\bibinfo {year} {1951})}\BibitemShut {NoStop}%
\bibitem [{\citenamefont {Cornish-Bowden}(2012)}]{enzymeskin}%
  \BibitemOpen
  \bibfield  {author} {\bibinfo {author} {\bibfnamefont {Athel}\ \bibnamefont
  {Cornish-Bowden}},\ }\href@noop {} {\emph {\bibinfo {title} {Fundamentals of
  Enzyme Kinetics}}}\ (\bibinfo  {publisher} {Wiley-Blackwell},\ \bibinfo
  {address} {Hoboken, New Jersey},\ \bibinfo {year} {2012})\BibitemShut
  {NoStop}%
\bibitem [{\citenamefont {Kelsic}\ \emph {et~al.}(2015)\citenamefont {Kelsic},
  \citenamefont {Zhao}, \citenamefont {Vetsigian},\ and\ \citenamefont
  {Kishony}}]{kelsic2015counteraction}%
  \BibitemOpen
  \bibfield  {author} {\bibinfo {author} {\bibfnamefont {Eric~D}\ \bibnamefont
  {Kelsic}}, \bibinfo {author} {\bibfnamefont {Jeffrey}\ \bibnamefont {Zhao}},
  \bibinfo {author} {\bibfnamefont {Kalin}\ \bibnamefont {Vetsigian}}, \ and\
  \bibinfo {author} {\bibfnamefont {Roy}\ \bibnamefont {Kishony}},\ }\bibfield
  {title} {\enquote {\bibinfo {title} {Counteraction of antibiotic production
  and degradation stabilizes microbial communities},}\ }\href@noop {}
  {\bibfield  {journal} {\bibinfo  {journal} {Nature}\ }\textbf {\bibinfo
  {volume} {521}},\ \bibinfo {pages} {516--519} (\bibinfo {year}
  {2015})}\BibitemShut {NoStop}%
\bibitem [{\citenamefont {Abrudan}\ \emph {et~al.}(2015)\citenamefont
  {Abrudan}, \citenamefont {Smakman}, \citenamefont {Grimbergen}, \citenamefont
  {Westhoff}, \citenamefont {Miller}, \citenamefont {Van~Wezel},\ and\
  \citenamefont {Rozen}}]{abrudan2015socially}%
  \BibitemOpen
  \bibfield  {author} {\bibinfo {author} {\bibfnamefont {Monica~I}\
  \bibnamefont {Abrudan}}, \bibinfo {author} {\bibfnamefont {Fokko}\
  \bibnamefont {Smakman}}, \bibinfo {author} {\bibfnamefont {Ard~Jan}\
  \bibnamefont {Grimbergen}}, \bibinfo {author} {\bibfnamefont {Sanne}\
  \bibnamefont {Westhoff}}, \bibinfo {author} {\bibfnamefont {Eric~L}\
  \bibnamefont {Miller}}, \bibinfo {author} {\bibfnamefont {Gilles~P}\
  \bibnamefont {Van~Wezel}}, \ and\ \bibinfo {author} {\bibfnamefont
  {Daniel~E}\ \bibnamefont {Rozen}},\ }\bibfield  {title} {\enquote {\bibinfo
  {title} {Socially mediated induction and suppression of antibiosis during
  bacterial coexistence},}\ }\href@noop {} {\bibfield  {journal} {\bibinfo
  {journal} {Proceedings of the National Academy of Sciences}\ }\textbf
  {\bibinfo {volume} {112}},\ \bibinfo {pages} {11054--11059} (\bibinfo {year}
  {2015})}\BibitemShut {NoStop}%
\bibitem [{\citenamefont {Gallo}\ \emph {et~al.}(1993)\citenamefont {Gallo},
  \citenamefont {Longo}, \citenamefont {Pallottino},\ and\ \citenamefont
  {Nguyen}}]{gallo1993directed}%
  \BibitemOpen
  \bibfield  {author} {\bibinfo {author} {\bibfnamefont {Giorgio}\ \bibnamefont
  {Gallo}}, \bibinfo {author} {\bibfnamefont {Giustino}\ \bibnamefont {Longo}},
  \bibinfo {author} {\bibfnamefont {Stefano}\ \bibnamefont {Pallottino}}, \
  and\ \bibinfo {author} {\bibfnamefont {Sang}\ \bibnamefont {Nguyen}},\
  }\bibfield  {title} {\enquote {\bibinfo {title} {Directed hypergraphs and
  applications},}\ }\href@noop {} {\bibfield  {journal} {\bibinfo  {journal}
  {Discrete applied mathematics}\ }\textbf {\bibinfo {volume} {42}},\ \bibinfo
  {pages} {177--201} (\bibinfo {year} {1993})}\BibitemShut {NoStop}%
\bibitem [{\citenamefont {Jost}\ and\ \citenamefont
  {Mulas}(2019)}]{mulas_chemical}%
  \BibitemOpen
  \bibfield  {author} {\bibinfo {author} {\bibfnamefont {J}~\bibnamefont
  {Jost}}\ and\ \bibinfo {author} {\bibfnamefont {R}~\bibnamefont {Mulas}},\
  }\bibfield  {title} {\enquote {\bibinfo {title} {Hypergraphs laplace
  operators for chemical reaction networks},}\ }\href@noop {} {\bibfield
  {journal} {\bibinfo  {journal} {Advances in Mathematics}\ }\textbf {\bibinfo
  {volume} {351}},\ \bibinfo {pages} {870–896} (\bibinfo {year}
  {2019})}\BibitemShut {NoStop}%
\bibitem [{\citenamefont {Andreotti}\ and\ \citenamefont
  {Mulas}(2020)}]{mulas_oriented}%
  \BibitemOpen
  \bibfield  {author} {\bibinfo {author} {\bibfnamefont {E}~\bibnamefont
  {Andreotti}}\ and\ \bibinfo {author} {\bibfnamefont {R}~\bibnamefont
  {Mulas}},\ }\bibfield  {title} {\enquote {\bibinfo {title} {Spectra of
  signless normalized laplace operators for hypergraphs},}\ }\href@noop {}
  {\bibfield  {journal} {\bibinfo  {journal} {arXiv preprint
  arXiv:2005.144840}\ } (\bibinfo {year} {2020})}\BibitemShut {NoStop}%
\bibitem [{\citenamefont {Abiad}\ \emph {et~al.}(2021)\citenamefont {Abiad},
  \citenamefont {Mulas},\ and\ \citenamefont {Zhang}}]{mulas_oriented2}%
  \BibitemOpen
  \bibfield  {author} {\bibinfo {author} {\bibfnamefont {A}~\bibnamefont
  {Abiad}}, \bibinfo {author} {\bibfnamefont {R}~\bibnamefont {Mulas}}, \ and\
  \bibinfo {author} {\bibfnamefont {D}~\bibnamefont {Zhang}},\ }\bibfield
  {title} {\enquote {\bibinfo {title} {Coloring the normalized laplacian for
  oriented hypergraphs},}\ }\href@noop {} {\bibfield  {journal} {\bibinfo
  {journal} {Linear Algebra and its Applications}\ }\textbf {\bibinfo {volume}
  {629}},\ \bibinfo {pages} {192--207} (\bibinfo {year} {2021})}\BibitemShut
  {NoStop}%
\bibitem [{\citenamefont {Schaub}\ and\ \citenamefont
  {Segarra}(2018)}]{schaub2018flow}%
  \BibitemOpen
  \bibfield  {author} {\bibinfo {author} {\bibfnamefont {Michael~T}\
  \bibnamefont {Schaub}}\ and\ \bibinfo {author} {\bibfnamefont {Santiago}\
  \bibnamefont {Segarra}},\ }\bibfield  {title} {\enquote {\bibinfo {title}
  {Flow smoothing and denoising: Graph signal processing in the edge-space},}\
  }in\ \href@noop {} {\emph {\bibinfo {booktitle} {2018 IEEE Global Conference
  on Signal and Information Processing (GlobalSIP)}}}\ (\bibinfo {organization}
  {IEEE},\ \bibinfo {year} {2018})\ pp.\ \bibinfo {pages}
  {735--739}\BibitemShut {NoStop}%
\bibitem [{\citenamefont {Barbarossa}\ and\ \citenamefont
  {Sardellitti}(2020)}]{barbarossa2020topological}%
  \BibitemOpen
  \bibfield  {author} {\bibinfo {author} {\bibfnamefont {Sergio}\ \bibnamefont
  {Barbarossa}}\ and\ \bibinfo {author} {\bibfnamefont {Stefania}\ \bibnamefont
  {Sardellitti}},\ }\bibfield  {title} {\enquote {\bibinfo {title} {Topological
  signal processing over simplicial complexes},}\ }\href@noop {} {\bibfield
  {journal} {\bibinfo  {journal} {IEEE Transactions on Signal Processing}\
  }\textbf {\bibinfo {volume} {68}},\ \bibinfo {pages} {2992--3007} (\bibinfo
  {year} {2020})}\BibitemShut {NoStop}%
\bibitem [{\citenamefont {Mill{\'a}n}\ \emph {et~al.}(2020)\citenamefont
  {Mill{\'a}n}, \citenamefont {Torres},\ and\ \citenamefont
  {Bianconi}}]{millan2020explosive}%
  \BibitemOpen
  \bibfield  {author} {\bibinfo {author} {\bibfnamefont {Ana~P}\ \bibnamefont
  {Mill{\'a}n}}, \bibinfo {author} {\bibfnamefont {Joaqu{\'\i}n~J}\
  \bibnamefont {Torres}}, \ and\ \bibinfo {author} {\bibfnamefont {Ginestra}\
  \bibnamefont {Bianconi}},\ }\bibfield  {title} {\enquote {\bibinfo {title}
  {Explosive higher-order kuramoto dynamics on simplicial complexes},}\
  }\href@noop {} {\bibfield  {journal} {\bibinfo  {journal} {Physical Review
  Letters}\ }\textbf {\bibinfo {volume} {124}},\ \bibinfo {pages} {218301}
  (\bibinfo {year} {2020})}\BibitemShut {NoStop}%
\bibitem [{\citenamefont {Arnaudon}\ \emph {et~al.}(2021)\citenamefont
  {Arnaudon}, \citenamefont {Peach}, \citenamefont {Petri},\ and\ \citenamefont
  {Expert}}]{petri_hodge}%
  \BibitemOpen
  \bibfield  {author} {\bibinfo {author} {\bibfnamefont {A}~\bibnamefont
  {Arnaudon}}, \bibinfo {author} {\bibfnamefont {R~L}\ \bibnamefont {Peach}},
  \bibinfo {author} {\bibfnamefont {G}~\bibnamefont {Petri}}, \ and\ \bibinfo
  {author} {\bibfnamefont {P}~\bibnamefont {Expert}},\ }\bibfield  {title}
  {\enquote {\bibinfo {title} {Connecting hodge and sakaguchi-kuramoto: a
  mathematical framework for coupled oscillators on simplicial complexes},}\
  }\href@noop {} {\bibfield  {journal} {\bibinfo  {journal} {arXiv preprint
  arXiv:2111.11073}\ } (\bibinfo {year} {2021})}\BibitemShut {NoStop}%
\bibitem [{\citenamefont {Boccaletti}\ \emph {et~al.}(2018)\citenamefont
  {Boccaletti}, \citenamefont {Pisarchik}, \citenamefont {Del~Genio},\ and\
  \citenamefont {Amann}}]{boccaletti2018synchronization}%
  \BibitemOpen
  \bibfield  {author} {\bibinfo {author} {\bibfnamefont {Stefano}\ \bibnamefont
  {Boccaletti}}, \bibinfo {author} {\bibfnamefont {Alexander~N}\ \bibnamefont
  {Pisarchik}}, \bibinfo {author} {\bibfnamefont {Charo~I}\ \bibnamefont
  {Del~Genio}}, \ and\ \bibinfo {author} {\bibfnamefont {Andreas}\ \bibnamefont
  {Amann}},\ }\href@noop {} {\emph {\bibinfo {title} {Synchronization: from
  coupled systems to complex networks}}}\ (\bibinfo  {publisher} {Cambridge
  University Press},\ \bibinfo {address} {Cambridge},\ \bibinfo {year}
  {2018})\BibitemShut {NoStop}%
\bibitem [{\citenamefont {Pecora}\ and\ \citenamefont
  {Carroll}(1998)}]{pecora1998master}%
  \BibitemOpen
  \bibfield  {author} {\bibinfo {author} {\bibfnamefont {Louis~M}\ \bibnamefont
  {Pecora}}\ and\ \bibinfo {author} {\bibfnamefont {Thomas~L}\ \bibnamefont
  {Carroll}},\ }\bibfield  {title} {\enquote {\bibinfo {title} {Master
  stability functions for synchronized coupled systems},}\ }\href@noop {}
  {\bibfield  {journal} {\bibinfo  {journal} {Physical Review Letters}\
  }\textbf {\bibinfo {volume} {80}},\ \bibinfo {pages} {2109} (\bibinfo {year}
  {1998})}\BibitemShut {NoStop}%
\bibitem [{\citenamefont {Krawiecki}(2014)}]{krawiecki2014chaotic}%
  \BibitemOpen
  \bibfield  {author} {\bibinfo {author} {\bibfnamefont {A}~\bibnamefont
  {Krawiecki}},\ }\bibfield  {title} {\enquote {\bibinfo {title} {Chaotic
  synchronization on complex hypergraphs},}\ }\href@noop {} {\bibfield
  {journal} {\bibinfo  {journal} {Chaos, Solitons \& Fractals}\ }\textbf
  {\bibinfo {volume} {65}},\ \bibinfo {pages} {44--50} (\bibinfo {year}
  {2014})}\BibitemShut {NoStop}%
\bibitem [{\citenamefont {Gambuzza}\ \emph {et~al.}(2021)\citenamefont
  {Gambuzza}, \citenamefont {Di~Patti}, \citenamefont {Gallo}, \citenamefont
  {Lepri}, \citenamefont {Romance}, \citenamefont {Criado}, \citenamefont
  {Frasca}, \citenamefont {Latora},\ and\ \citenamefont
  {Boccaletti}}]{gambuzza2021stability}%
  \BibitemOpen
  \bibfield  {author} {\bibinfo {author} {\bibfnamefont {Lucia~Valentina}\
  \bibnamefont {Gambuzza}}, \bibinfo {author} {\bibfnamefont {Francesca}\
  \bibnamefont {Di~Patti}}, \bibinfo {author} {\bibfnamefont {Luca}\
  \bibnamefont {Gallo}}, \bibinfo {author} {\bibfnamefont {Stefano}\
  \bibnamefont {Lepri}}, \bibinfo {author} {\bibfnamefont {Miguel}\
  \bibnamefont {Romance}}, \bibinfo {author} {\bibfnamefont {Regino}\
  \bibnamefont {Criado}}, \bibinfo {author} {\bibfnamefont {Mattia}\
  \bibnamefont {Frasca}}, \bibinfo {author} {\bibfnamefont {Vito}\ \bibnamefont
  {Latora}}, \ and\ \bibinfo {author} {\bibfnamefont {Stefano}\ \bibnamefont
  {Boccaletti}},\ }\bibfield  {title} {\enquote {\bibinfo {title} {Stability of
  synchronization in simplicial complexes},}\ }\href@noop {} {\bibfield
  {journal} {\bibinfo  {journal} {Nature communications}\ }\textbf {\bibinfo
  {volume} {12}},\ \bibinfo {pages} {1--13} (\bibinfo {year}
  {2021})}\BibitemShut {NoStop}%
\bibitem [{\citenamefont {Pikovsky}\ \emph {et~al.}(2003)\citenamefont
  {Pikovsky}, \citenamefont {Kurths}, \citenamefont {Rosenblum},\ and\
  \citenamefont {Kurths}}]{pikovsky2003synchronization}%
  \BibitemOpen
  \bibfield  {author} {\bibinfo {author} {\bibfnamefont {Arkady}\ \bibnamefont
  {Pikovsky}}, \bibinfo {author} {\bibfnamefont {Jurgen}\ \bibnamefont
  {Kurths}}, \bibinfo {author} {\bibfnamefont {Michael}\ \bibnamefont
  {Rosenblum}}, \ and\ \bibinfo {author} {\bibfnamefont {J{\"u}rgen}\
  \bibnamefont {Kurths}},\ }\href@noop {} {\emph {\bibinfo {title}
  {Synchronization: a universal concept in nonlinear sciences}}},\
  Vol.~\bibinfo {volume} {12}\ (\bibinfo  {publisher} {Cambridge university
  press},\ \bibinfo {year} {2003})\BibitemShut {NoStop}%
\bibitem [{\citenamefont {Carletti}\ and\ \citenamefont
  {Muolo}(2021)}]{carletti}%
  \BibitemOpen
  \bibfield  {author} {\bibinfo {author} {\bibfnamefont {Timoteo}\ \bibnamefont
  {Carletti}}\ and\ \bibinfo {author} {\bibfnamefont {Riccardo}\ \bibnamefont
  {Muolo}},\ }\bibfield  {title} {\enquote {\bibinfo {title} {Non-reciprocal
  interactions enhance heterogeneity},}\ }\href@noop {} {\bibfield  {journal}
  {\bibinfo  {journal} {arXiv preprint arXiv:2112.02549}\ } (\bibinfo {year}
  {2021})}\BibitemShut {NoStop}%
\bibitem [{\citenamefont {Nishikawa}\ and\ \citenamefont
  {Motter}(2006)}]{nishikawa2006synchronization}%
  \BibitemOpen
  \bibfield  {author} {\bibinfo {author} {\bibfnamefont {Takashi}\ \bibnamefont
  {Nishikawa}}\ and\ \bibinfo {author} {\bibfnamefont {Adilson~E}\ \bibnamefont
  {Motter}},\ }\bibfield  {title} {\enquote {\bibinfo {title} {Synchronization
  is optimal in nondiagonalizable networks},}\ }\href@noop {} {\bibfield
  {journal} {\bibinfo  {journal} {Physical Review E}\ }\textbf {\bibinfo
  {volume} {73}},\ \bibinfo {pages} {065106} (\bibinfo {year}
  {2006})}\BibitemShut {NoStop}%
\bibitem [{\citenamefont {R{\"o}ssler}(1976)}]{rossler1976equation}%
  \BibitemOpen
  \bibfield  {author} {\bibinfo {author} {\bibfnamefont {Otto~E}\ \bibnamefont
  {R{\"o}ssler}},\ }\bibfield  {title} {\enquote {\bibinfo {title} {An equation
  for continuous chaos},}\ }\href@noop {} {\bibfield  {journal} {\bibinfo
  {journal} {Physics Letters A}\ }\textbf {\bibinfo {volume} {57}},\ \bibinfo
  {pages} {397--398} (\bibinfo {year} {1976})}\BibitemShut {NoStop}%
\bibitem [{\citenamefont {Wolf}\ \emph {et~al.}(1985)\citenamefont {Wolf},
  \citenamefont {Swift}, \citenamefont {Swinney},\ and\ \citenamefont
  {Vastano}}]{wolf1985determining}%
  \BibitemOpen
  \bibfield  {author} {\bibinfo {author} {\bibfnamefont {Alan}\ \bibnamefont
  {Wolf}}, \bibinfo {author} {\bibfnamefont {Jack~B}\ \bibnamefont {Swift}},
  \bibinfo {author} {\bibfnamefont {Harry~L}\ \bibnamefont {Swinney}}, \ and\
  \bibinfo {author} {\bibfnamefont {John~A}\ \bibnamefont {Vastano}},\
  }\bibfield  {title} {\enquote {\bibinfo {title} {Determining lyapunov
  exponents from a time series},}\ }\href@noop {} {\bibfield  {journal}
  {\bibinfo  {journal} {Physica D: Nonlinear Phenomena}\ }\textbf {\bibinfo
  {volume} {16}},\ \bibinfo {pages} {285--317} (\bibinfo {year}
  {1985})}\BibitemShut {NoStop}%
\bibitem [{\citenamefont {Asllani}\ \emph {et~al.}(2020)\citenamefont
  {Asllani}, \citenamefont {Carletti}, \citenamefont {Fanelli},\ and\
  \citenamefont {Maini}}]{Asllani2}%
  \BibitemOpen
  \bibfield  {author} {\bibinfo {author} {\bibfnamefont {M}~\bibnamefont
  {Asllani}}, \bibinfo {author} {\bibfnamefont {T}~\bibnamefont {Carletti}},
  \bibinfo {author} {\bibfnamefont {D}~\bibnamefont {Fanelli}}, \ and\ \bibinfo
  {author} {\bibfnamefont {P~K}\ \bibnamefont {Maini}},\ }\bibfield  {title}
  {\enquote {\bibinfo {title} {A universal route to pattern formation in
  multicellular systems},}\ }\href@noop {} {\bibfield  {journal} {\bibinfo
  {journal} {The European Physics Journal B}\ }\textbf {\bibinfo {volume} {93}}
  (\bibinfo {year} {2020})}\BibitemShut {NoStop}%
\bibitem [{\citenamefont {Asllani}\ \emph {et~al.}(2014)\citenamefont
  {Asllani}, \citenamefont {Challenger}, \citenamefont {Pavone}, \citenamefont
  {Sacconi},\ and\ \citenamefont {Fanelli}}]{Asllani1}%
  \BibitemOpen
  \bibfield  {author} {\bibinfo {author} {\bibfnamefont {M}~\bibnamefont
  {Asllani}}, \bibinfo {author} {\bibfnamefont {Joseph~D}\ \bibnamefont
  {Challenger}}, \bibinfo {author} {\bibfnamefont {F~S}\ \bibnamefont
  {Pavone}}, \bibinfo {author} {\bibfnamefont {L}~\bibnamefont {Sacconi}}, \
  and\ \bibinfo {author} {\bibfnamefont {D}~\bibnamefont {Fanelli}},\
  }\bibfield  {title} {\enquote {\bibinfo {title} {The theory of pattern
  formation on directed networks},}\ }\href@noop {} {\bibfield  {journal}
  {\bibinfo  {journal} {Nature Communication}\ }\textbf {\bibinfo {volume} {5}}
  (\bibinfo {year} {2014})}\BibitemShut {NoStop}%
\bibitem [{\citenamefont {Di~Patti}\ \emph {et~al.}(2017)\citenamefont
  {Di~Patti}, \citenamefont {Fanelli}, \citenamefont {Miele},\ and\
  \citenamefont {Carletti}}]{dipatti1}%
  \BibitemOpen
  \bibfield  {author} {\bibinfo {author} {\bibfnamefont {Francesca}\
  \bibnamefont {Di~Patti}}, \bibinfo {author} {\bibfnamefont {Duccio}\
  \bibnamefont {Fanelli}}, \bibinfo {author} {\bibfnamefont {Filippo}\
  \bibnamefont {Miele}}, \ and\ \bibinfo {author} {\bibfnamefont {Timoteo}\
  \bibnamefont {Carletti}},\ }\bibfield  {title} {\enquote {\bibinfo {title}
  {Benjamin–feir instabilities on directed networks},}\ }\href@noop {}
  {\bibfield  {journal} {\bibinfo  {journal} {Chaos, Solitons \& Fractals}\
  }\textbf {\bibinfo {volume} {96}},\ \bibinfo {pages} {8 -- 16} (\bibinfo
  {year} {2017})}\BibitemShut {NoStop}%
\bibitem [{\citenamefont {Muolo}\ \emph {et~al.}(2019)\citenamefont {Muolo},
  \citenamefont {Asllani}, \citenamefont {Fanelli}, \citenamefont {Maini},\
  and\ \citenamefont {Carletti}}]{jtb}%
  \BibitemOpen
  \bibfield  {author} {\bibinfo {author} {\bibfnamefont {Riccardo}\
  \bibnamefont {Muolo}}, \bibinfo {author} {\bibfnamefont {Malbor}\
  \bibnamefont {Asllani}}, \bibinfo {author} {\bibfnamefont {Duccio}\
  \bibnamefont {Fanelli}}, \bibinfo {author} {\bibfnamefont {Ph~K}\
  \bibnamefont {Maini}}, \ and\ \bibinfo {author} {\bibfnamefont {Timoteo}\
  \bibnamefont {Carletti}},\ }\bibfield  {title} {\enquote {\bibinfo {title}
  {Patterns of non-normality in networked systems},}\ }\href@noop {} {\bibfield
   {journal} {\bibinfo  {journal} {Journal of Theoretical Biology}\ }\textbf
  {\bibinfo {volume} {480}},\ \bibinfo {pages} {81} (\bibinfo {year}
  {2019})}\BibitemShut {NoStop}%
\bibitem [{\citenamefont {Muolo}\ \emph {et~al.}(2021)\citenamefont {Muolo},
  \citenamefont {Carletti}, \citenamefont {Gleeson},\ and\ \citenamefont
  {Asllani}}]{entropy}%
  \BibitemOpen
  \bibfield  {author} {\bibinfo {author} {\bibfnamefont {Riccardo}\
  \bibnamefont {Muolo}}, \bibinfo {author} {\bibfnamefont {Timoteo}\
  \bibnamefont {Carletti}}, \bibinfo {author} {\bibfnamefont {James~P}\
  \bibnamefont {Gleeson}}, \ and\ \bibinfo {author} {\bibfnamefont {Malbor}\
  \bibnamefont {Asllani}},\ }\bibfield  {title} {\enquote {\bibinfo {title}
  {Synchronization dynamics in non-normal networks: the trade-off for
  optimality},}\ }\href@noop {} {\bibfield  {journal} {\bibinfo  {journal}
  {Entropy}\ }\textbf {\bibinfo {volume} {23}},\ \bibinfo {pages} {36}
  (\bibinfo {year} {2021})}\BibitemShut {NoStop}%
\bibitem [{\citenamefont {Chavez}\ \emph {et~al.}(2005)\citenamefont {Chavez},
  \citenamefont {Hwang}, \citenamefont {Amann}, \citenamefont {Hentschel},\
  and\ \citenamefont {Boccaletti}}]{chavez2005synchronization}%
  \BibitemOpen
  \bibfield  {author} {\bibinfo {author} {\bibfnamefont {M}~\bibnamefont
  {Chavez}}, \bibinfo {author} {\bibfnamefont {D-U}\ \bibnamefont {Hwang}},
  \bibinfo {author} {\bibfnamefont {Arno}\ \bibnamefont {Amann}}, \bibinfo
  {author} {\bibfnamefont {HGE}\ \bibnamefont {Hentschel}}, \ and\ \bibinfo
  {author} {\bibfnamefont {Stefano}\ \bibnamefont {Boccaletti}},\ }\bibfield
  {title} {\enquote {\bibinfo {title} {Synchronization is enhanced in weighted
  complex networks},}\ }\href@noop {} {\bibfield  {journal} {\bibinfo
  {journal} {Physical Review Letters}\ }\textbf {\bibinfo {volume} {94}},\
  \bibinfo {pages} {218701} (\bibinfo {year} {2005})}\BibitemShut {NoStop}%
\bibitem [{\citenamefont {Hwang}\ \emph {et~al.}(2005)\citenamefont {Hwang},
  \citenamefont {Chavez}, \citenamefont {Amann},\ and\ \citenamefont
  {Boccaletti}}]{hwang2005synchronization}%
  \BibitemOpen
  \bibfield  {author} {\bibinfo {author} {\bibfnamefont {D-U}\ \bibnamefont
  {Hwang}}, \bibinfo {author} {\bibfnamefont {M}~\bibnamefont {Chavez}},
  \bibinfo {author} {\bibfnamefont {A}~\bibnamefont {Amann}}, \ and\ \bibinfo
  {author} {\bibfnamefont {S}~\bibnamefont {Boccaletti}},\ }\bibfield  {title}
  {\enquote {\bibinfo {title} {Synchronization in complex networks with age
  ordering},}\ }\href@noop {} {\bibfield  {journal} {\bibinfo  {journal}
  {Physical review letters}\ }\textbf {\bibinfo {volume} {94}},\ \bibinfo
  {pages} {138701} (\bibinfo {year} {2005})}\BibitemShut {NoStop}%
\bibitem [{\citenamefont {Motter}\ \emph {et~al.}(2005)\citenamefont {Motter},
  \citenamefont {Zhou},\ and\ \citenamefont {Kurths}}]{motter2005enhancing}%
  \BibitemOpen
  \bibfield  {author} {\bibinfo {author} {\bibfnamefont {Adilson~E}\
  \bibnamefont {Motter}}, \bibinfo {author} {\bibfnamefont {CS}~\bibnamefont
  {Zhou}}, \ and\ \bibinfo {author} {\bibfnamefont {J{\"u}rgen}\ \bibnamefont
  {Kurths}},\ }\bibfield  {title} {\enquote {\bibinfo {title} {Enhancing
  complex-network synchronization},}\ }\href@noop {} {\bibfield  {journal}
  {\bibinfo  {journal} {EPL (Europhysics Letters)}\ }\textbf {\bibinfo {volume}
  {69}},\ \bibinfo {pages} {334} (\bibinfo {year} {2005})}\BibitemShut
  {NoStop}%
\bibitem [{\citenamefont {Tang}\ \emph {et~al.}(2022)\citenamefont {Tang},
  \citenamefont {Shi},\ and\ \citenamefont {L{\"u}}}]{tang2022optimizing}%
  \BibitemOpen
  \bibfield  {author} {\bibinfo {author} {\bibfnamefont {Ying}\ \bibnamefont
  {Tang}}, \bibinfo {author} {\bibfnamefont {Dinghua}\ \bibnamefont {Shi}}, \
  and\ \bibinfo {author} {\bibfnamefont {Linyuan}\ \bibnamefont {L{\"u}}},\
  }\bibfield  {title} {\enquote {\bibinfo {title} {Optimizing higher-order
  network topology for synchronization of coupled phase oscillators},}\
  }\href@noop {} {\bibfield  {journal} {\bibinfo  {journal} {Communications
  Physics}\ }\textbf {\bibinfo {volume} {5}},\ \bibinfo {pages} {1--12}
  (\bibinfo {year} {2022})}\BibitemShut {NoStop}%
\bibitem [{\citenamefont {Aguiar}\ \emph {et~al.}(2022)\citenamefont {Aguiar},
  \citenamefont {Bick},\ and\ \citenamefont {Dias}}]{ABD2022}%
  \BibitemOpen
  \bibfield  {author} {\bibinfo {author} {\bibfnamefont {Manuela}\ \bibnamefont
  {Aguiar}}, \bibinfo {author} {\bibfnamefont {Christian}\ \bibnamefont
  {Bick}}, \ and\ \bibinfo {author} {\bibfnamefont {Ana}\ \bibnamefont
  {Dias}},\ }\bibfield  {title} {\enquote {\bibinfo {title} {Network dynamics
  with higher-order interactions: Coupled cell hypernetworks for identical
  cells and synchrony},}\ }\href@noop {} {\bibfield  {journal} {\bibinfo
  {journal} {arXiv preprint arXiv:2201.0937}\ } (\bibinfo {year}
  {2022})}\BibitemShut {NoStop}%
\bibitem [{\citenamefont {Newman}\ and\ \citenamefont
  {Watts}(1999)}]{newman1999scaling}%
  \BibitemOpen
  \bibfield  {author} {\bibinfo {author} {\bibfnamefont {Mark~EJ}\ \bibnamefont
  {Newman}}\ and\ \bibinfo {author} {\bibfnamefont {Duncan~J}\ \bibnamefont
  {Watts}},\ }\bibfield  {title} {\enquote {\bibinfo {title} {Scaling and
  percolation in the small-world network model},}\ }\href@noop {} {\bibfield
  {journal} {\bibinfo  {journal} {Physical review E}\ }\textbf {\bibinfo
  {volume} {60}},\ \bibinfo {pages} {7332} (\bibinfo {year}
  {1999})}\BibitemShut {NoStop}%
\bibitem [{\citenamefont {Erd\H{o}s}\ and\ \citenamefont
  {R{\'e}nyi}(1960)}]{erdos1960evolution}%
  \BibitemOpen
  \bibfield  {author} {\bibinfo {author} {\bibfnamefont {Paul}\ \bibnamefont
  {Erd\H{o}s}}\ and\ \bibinfo {author} {\bibfnamefont {Alfr{\'e}d}\
  \bibnamefont {R{\'e}nyi}},\ }\bibfield  {title} {\enquote {\bibinfo {title}
  {On the evolution of random graphs},}\ }\href@noop {} {\bibfield  {journal}
  {\bibinfo  {journal} {Publ. Math. Inst. Hung. Acad. Sci}\ }\textbf {\bibinfo
  {volume} {5}},\ \bibinfo {pages} {17--60} (\bibinfo {year}
  {1960})}\BibitemShut {NoStop}%
\end{thebibliography}%

\newpage

\appendix

\section{Synchronization in symmetric hypergraphs}
\label{app:hyp_vs_simp}

\setcounter{equation}{0}
\renewcommand{\theequation}{A\arabic{equation}}
\setcounter{figure}{0}
\renewcommand{\thefigure}{A\arabic{figure}}

The stability analysis presented in Results for directed hypergraphs also applies to undirected hypergraphs (this latter case can also be seen as an extension of the method presented in~\emph{(44)} developed for simplicial complexes), so that we here briefly discuss an example of synchronization in the presence of undirected higher-order interactions. Notice that, at variance with the derivation outlined in Methods, in the undirected case the adjacency tensor is symmetric, as the generalized Laplacian matrix of order $d$ does. This latter is in fact given by
\begin{align}
L_{ij}^{(d)}&= 
\begin{cases} d!k^{(d)}(i) & i=j \\ -(d-1)!k^{(d)}(i,j) &i\neq j
\end{cases}.
\end{align} 

To illustrate our results, we consider again a system of $N$ coupled Rössler oscillators, whose parameters have been set to $a=b=0.2$, and $c=9$, so that the dynamics of the isolated system is chaotic. The system is coupled via the $x$ component, through the coupling functions $\vec{h}^{(1)}(\vec{x}_j) = [x_j^3,0,0]$ and $\vec{h}^{(2)}(\vec{x}_j,\vec{x}_k) = [x_j^2x_k,0,0]$. The equations governing the system read
\begin{equation}
    \begin{cases}
    \dot{x}_i = -y_i-z_i + \sigma_1\sum\limits_{j=1}^{N}A_{ij}^{(1)}(x_j^3-x_i^3)+
    \sigma_2\sum\limits_{j,k=1}^{N}A_{ijk}^{(2)}(x_j^2x_k-x_i^3)\\ 
    \dot{y}_i = x_i+ay_i \\ 
    \dot{z}_i = b+z_i(x_i-c),
    \end{cases} 
    \label{eq:rossler_x2x_coupling_yet_again}
\end{equation} 
with $i\in\{1,\dots,N\}$.

\begin{figure}[t!]
\centering
\includegraphics[width=0.9\linewidth]{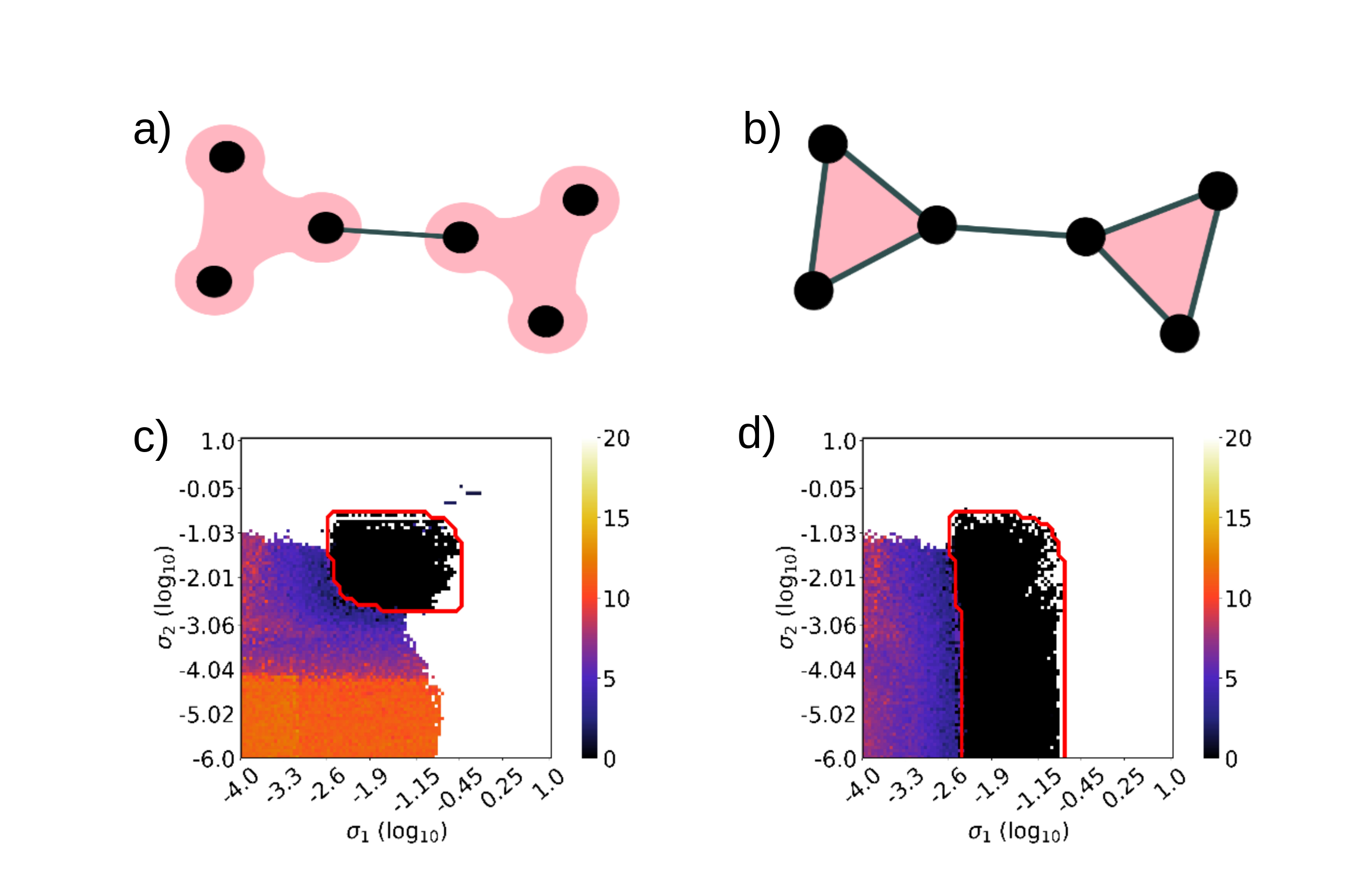}
\caption{\textit{Synchronization in hypergraphs and in simplicial complexes}. In panel a) we depict an undirected hypergraph, in which two $2$-hyperedges are connected through a pairwise link (black line), while in panel b) an undirected simplicial complex formed by two $2$-simplices connected through a pairwise link (black line) is reported; let us observe that each $2$-simplex contains also the three pairwise interactions (black lines on the boundary of the triangles). Let us stress that the latter fact determines the key difference between hypegraphs and simplicial complexes. In panels c) and d), we report the averaged synchronization error $E(\sigma_1, \sigma_2)$ as defined in Eq.~\eqref{eq:error} by using the shown color code, with the solid red line depicting the theoretical prediction of the boundary of the stability region provided by the MSF.}
\label{fig:pallozzo}
\end{figure}

In particular, we analyzed the undirected hypergraph with $N = 6$ nodes shown in panel a) of Fig.~\ref{fig:pallozzo}. Note that no link exists between the nodes in the $2$-hyperedges, meaning that the higher-order structure is not a simplicial complex. We simulate Eqs.~\eqref{eq:rossler_x2x_coupling_yet_again} on top of this structure, for different value of the coupling strengths $\sigma_1$ and $\sigma_2$. The state of the system is monitored by the average synchronization error defined as in Eq.~\eqref{eq:error}. Panel c) of Fig.~\ref{fig:pallozzo} displays the synchronization error $E(\sigma_1, \sigma_2)$ (colormap), along with the theoretical prediction of the boundary of the stability region provided by the MSF (solid red line). As one can see, the numerical simulations are in very good agreement with the theoretical predictions for the synchronization thresholds. 

To fully appreciate the difference between (undirected) hypergraphs and simplicial complexes, let us consider a simplicial complex having the same $2$-hyperedges as the structure in panel a) of Fig.~\ref{fig:pallozzo}, but different links, as shown in panel b). Comparing panel c) with panel d), illustrating the synchronization error, one can conclude that the presence of pairwise interactions in the simplicial complex preserves the stability of the synchronized state even when the higher-order coupling $\sigma_2$ is small, while in the case of the hypergraph, under such conditions, synchronization is lost.

\section{Directed hypergraph of R\"ossler systems with $y-y$ coupling}
\setcounter{equation}{0}
\renewcommand{\theequation}{B\arabic{equation}}
\setcounter{figure}{0}
\renewcommand{\thefigure}{B\arabic{figure}}
\label{app:y_y}

In Results, we have considered a system of Rössler oscillators coupled through a $1$-directed hypergraph, with the coupling functions being $\vec{h}^{(1)}(\vec{x}_j) = [x_j^3,0,0]$ and $\vec{h}^{(2)}(\vec{x}_j,\vec{x}_k) = [x_j^2x_k,0,0]$. As a further example, we here account for a different choice of the coupling functions, namely $\vec{h}^{(1)}(\vec{x}_j) = [0,y_j^3,0]$ and $\vec{h}^{(2)}(\vec{x}_j,\vec{x}_k) = [0,y_j^2y_k,0]$, which also satisfy the natural coupling hypothesis. The equations for the coupled system read
\begin{equation}
    \begin{cases} 
    \dot{x}_i = -y_i-z_i\\ 
    \dot{y}_i = x_i+ay_i + \sigma_1\sum\limits_{j=1}^{N}A_{ij}^{(1)}(y_j^3-y_i^3)+\sigma_2\sum\limits_{j,k=1}^{N}A_{ijk}^{(2)}(y_j^2y_k-y_i^3)\\ 
    \dot{z}_i = b+z_i(x_i-c).
    \end{cases} 
    \label{eq:rossler_y2y_coupling}
\end{equation}
for $i=\{1,\dots, N\}$.

\begin{figure}[t!]
\centering
\includegraphics[width=0.4\linewidth]{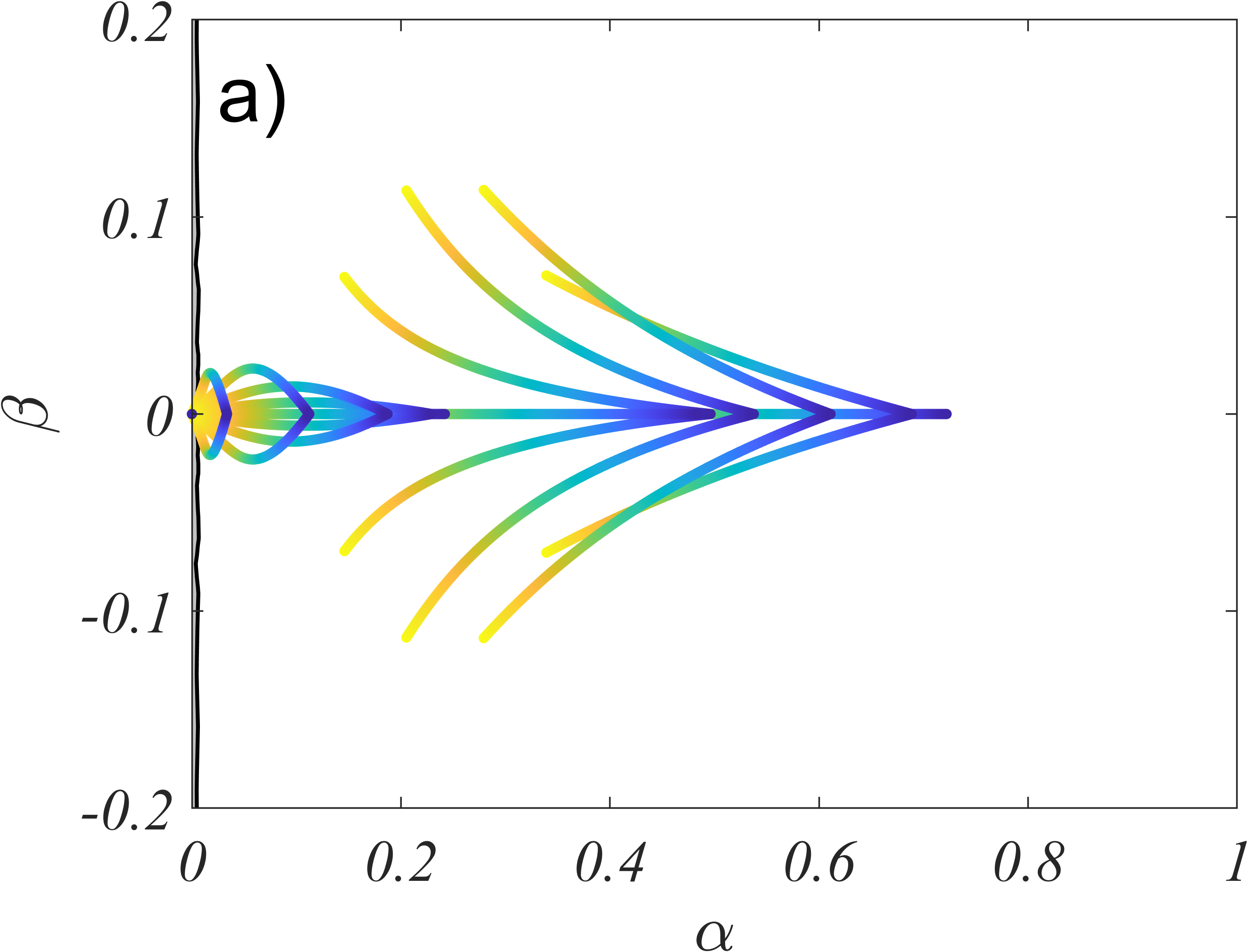}
\includegraphics[width=0.4\linewidth]{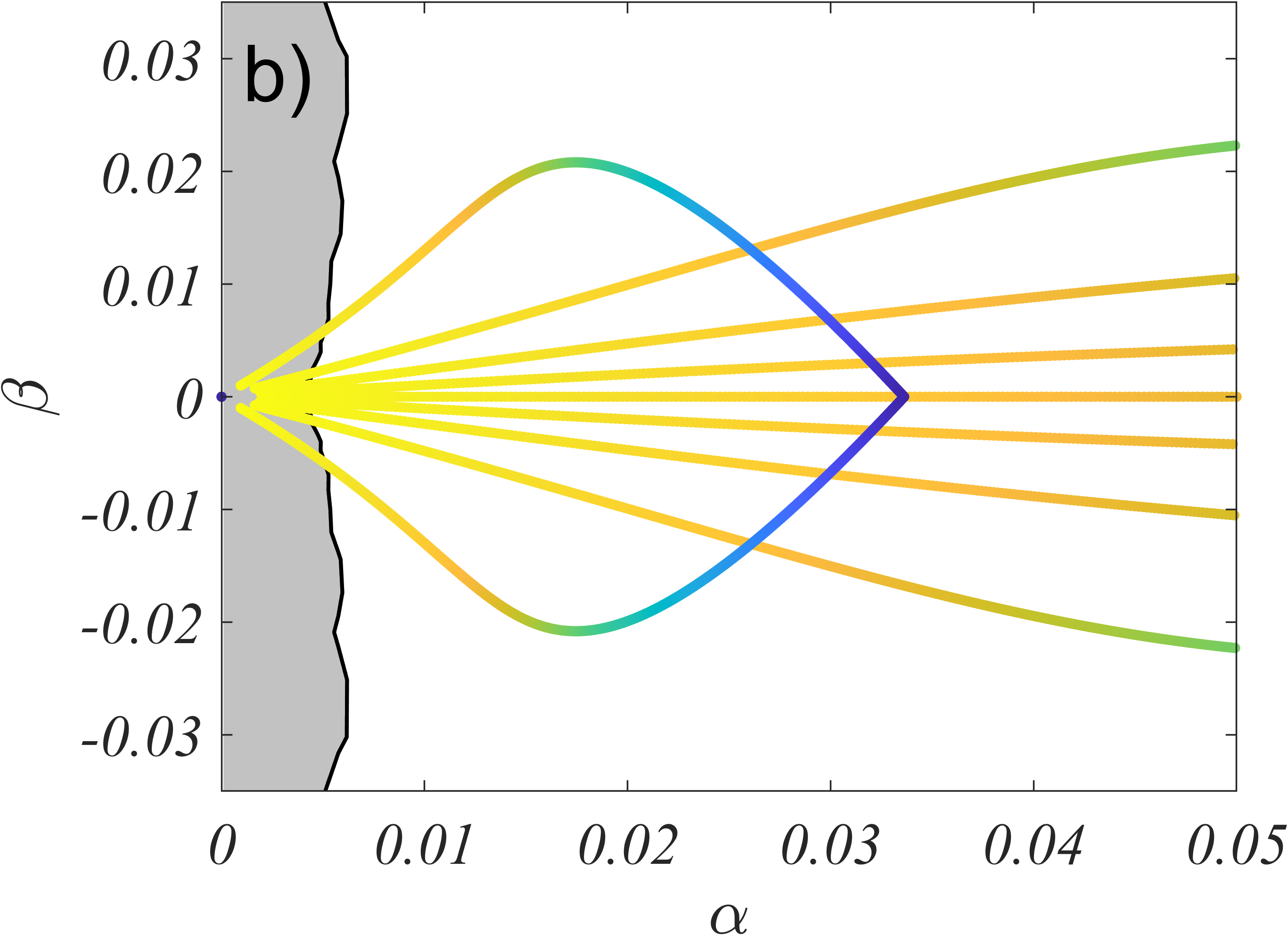}
\includegraphics[width=0.4\linewidth]{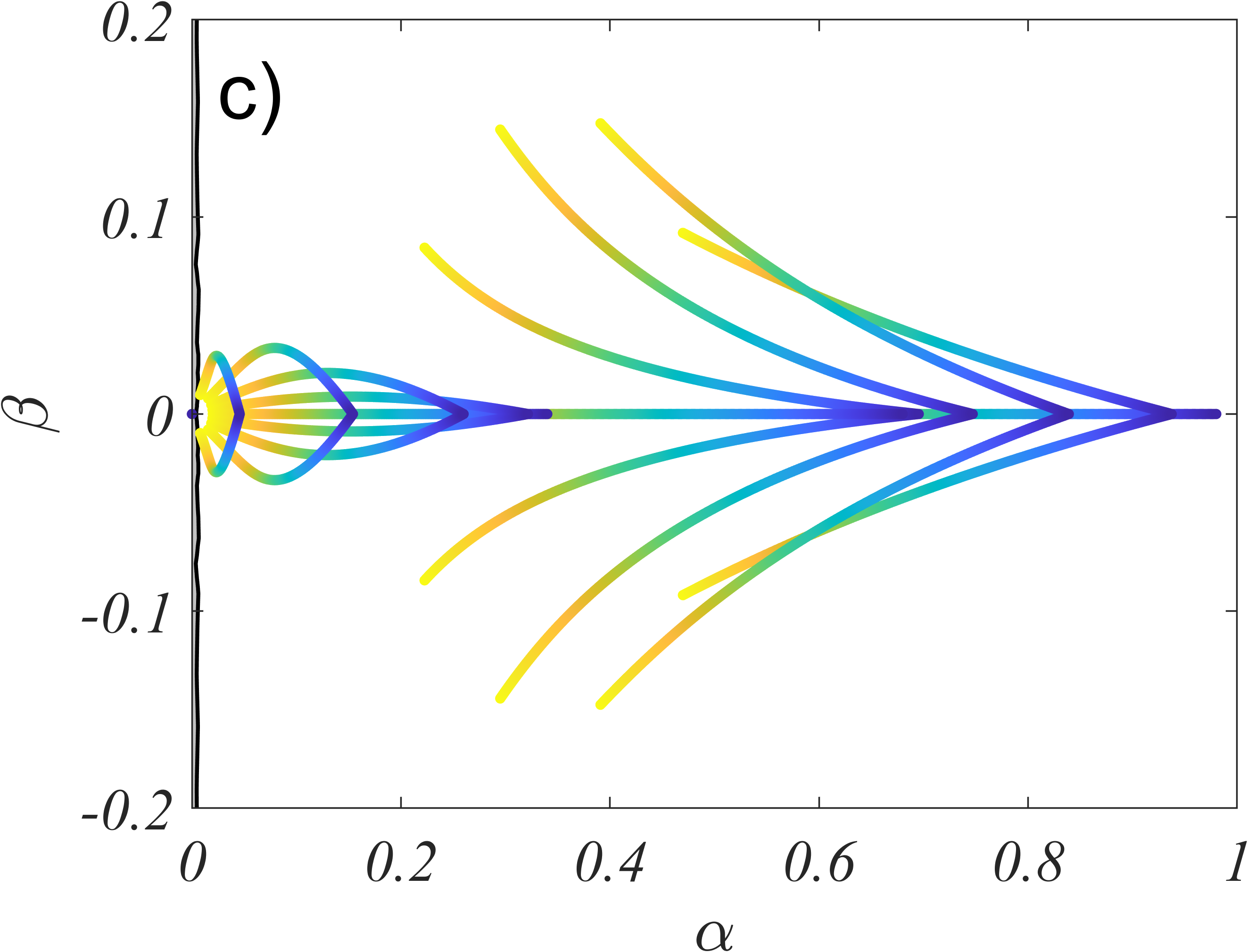}
\includegraphics[width=0.4\linewidth]{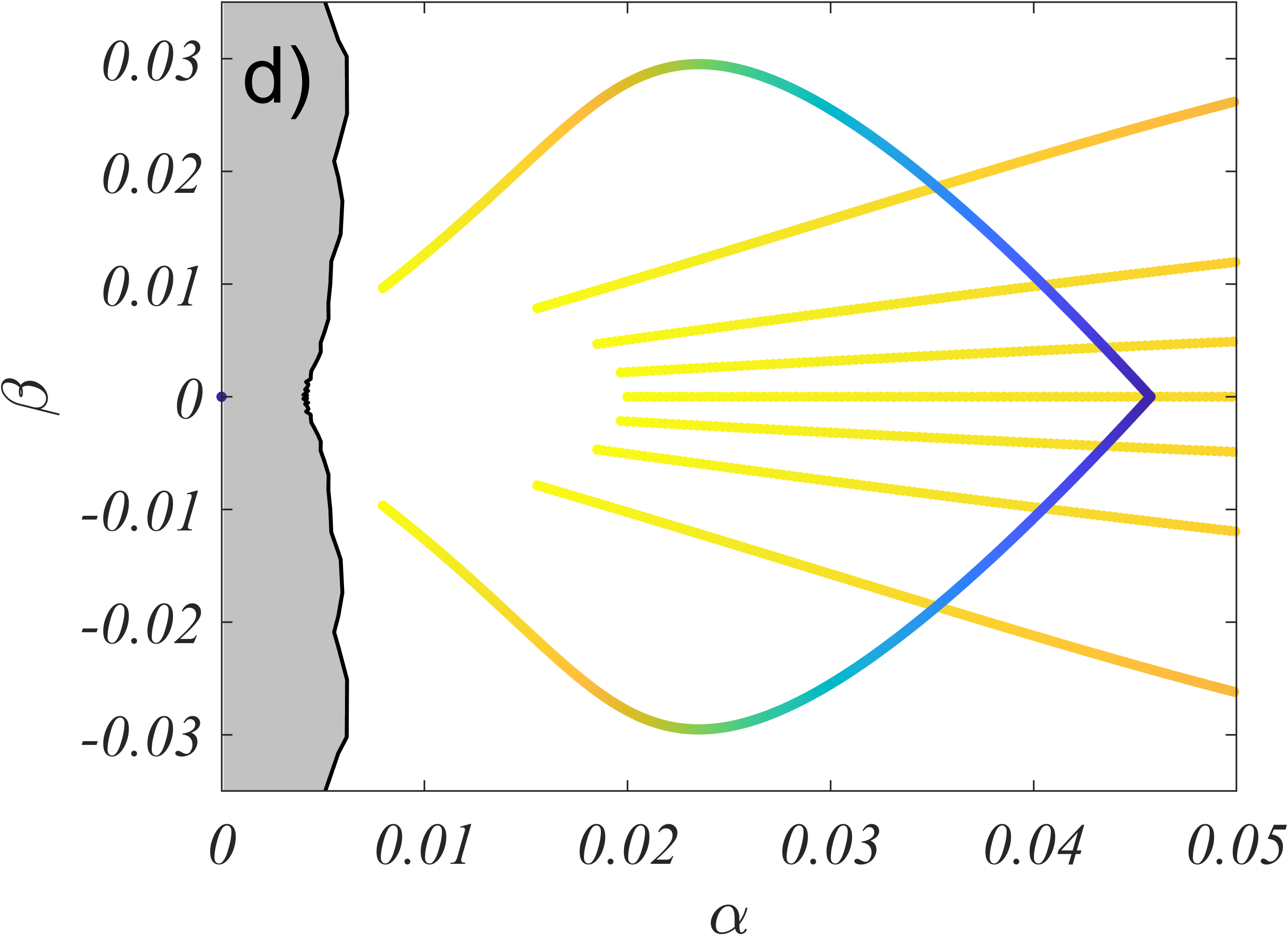}
\caption{\textcolor{black}{\textit{Directionality induced (de)synchronization with $y$-$y$ cubic coupling}. Panels a)-d) show the locus of the eigenvalues of $\mathcal{M}$ as a function of $p\in[0,1]$, for a weighted hypergraph with $N=20$ nodes (color coding is such that the directed case $p=0$ is represented in yellow, and the symmetric one $p=1$, in blue). In the background, the white area indicates the region identified by a negative MSF, the black line the boundary of this region, and the gray area the region where the MSF is positive. Panels b) and d) represent a zoom of the area close to the origin of panels a) and c), respectively. Panels a) and b) show how the directed topology drives the system unstable, while for the symmetric hypergraph the synchronization manifold is stable. Panels c) and d) display how, admitting the directed topology a stable synchronization state, giving the shape of the MSF, it is not possible to desynchronize the system of oscillators by making the hypergraph symmetric. The coupling strengths for panels a) and b) are $\sigma_1=0.001$ and $\sigma_2=0.12$, while for panel c) and d) they are fixed to $\sigma_1=0.01$ and $\sigma_2=0.16$.}}
\label{fig:msf_rossler_y2y}
\end{figure}

\begin{figure}[t!]
\centering
\includegraphics[width=0.4\linewidth]{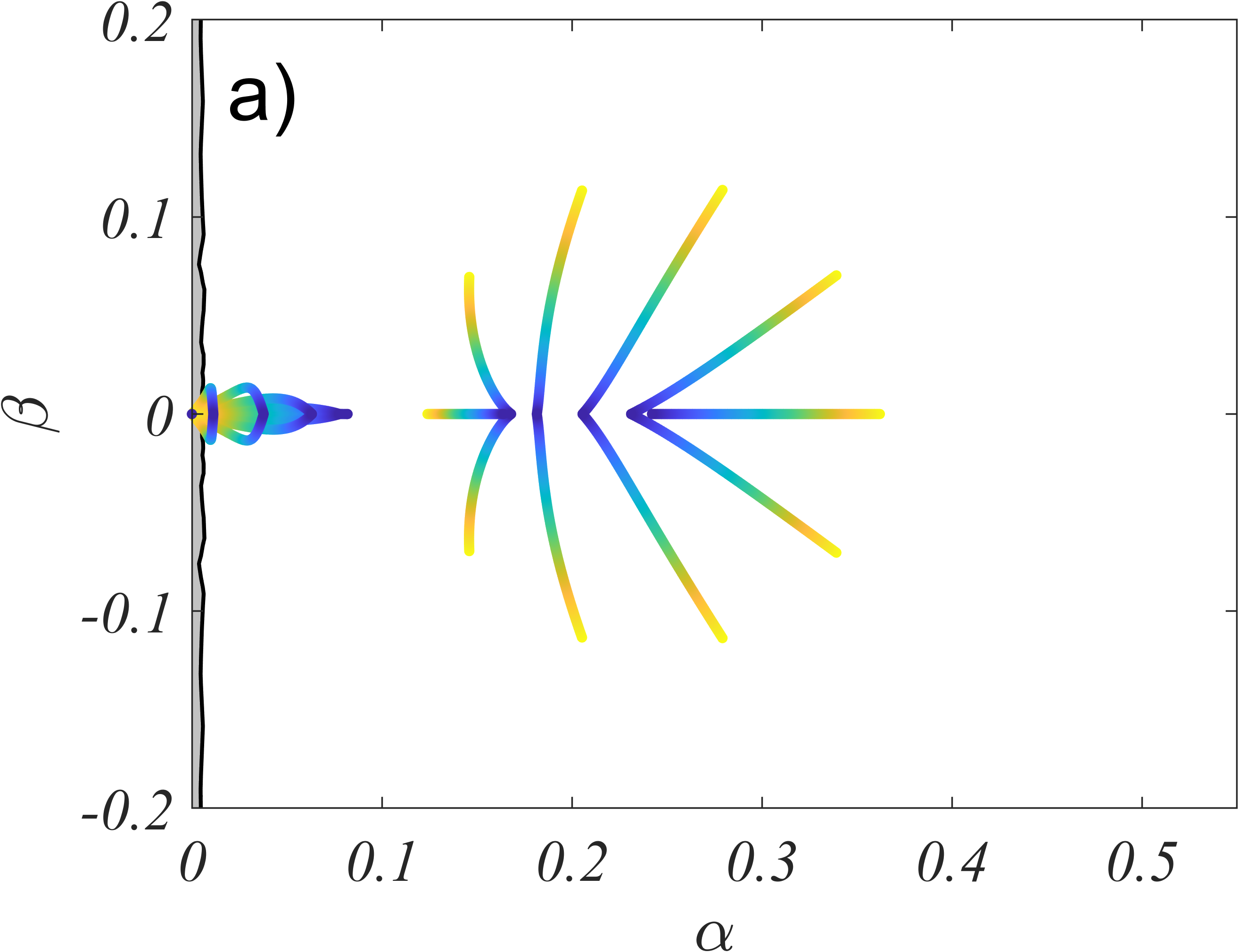}
\includegraphics[width=0.4\linewidth]{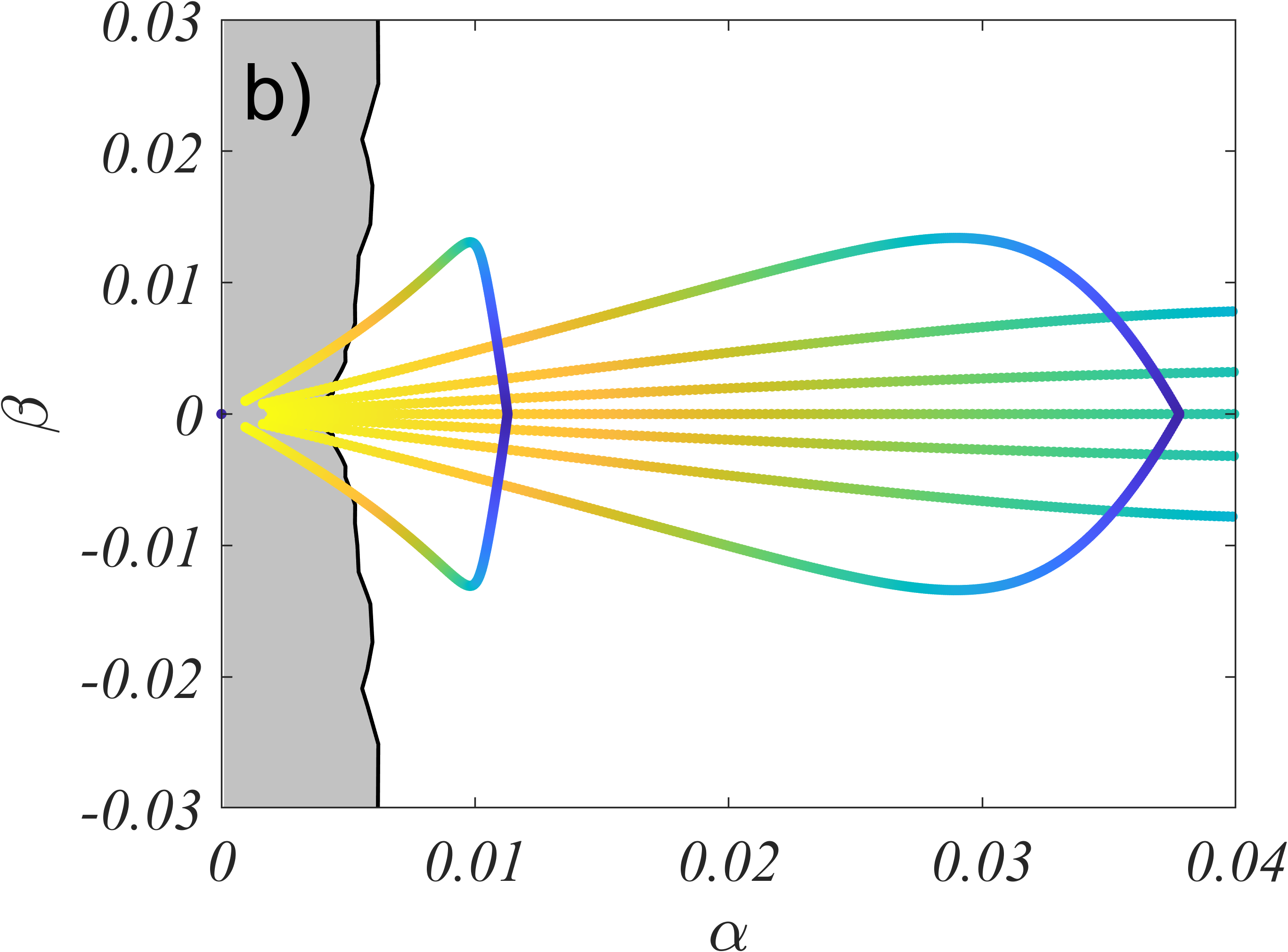}
\includegraphics[width=0.4\linewidth]{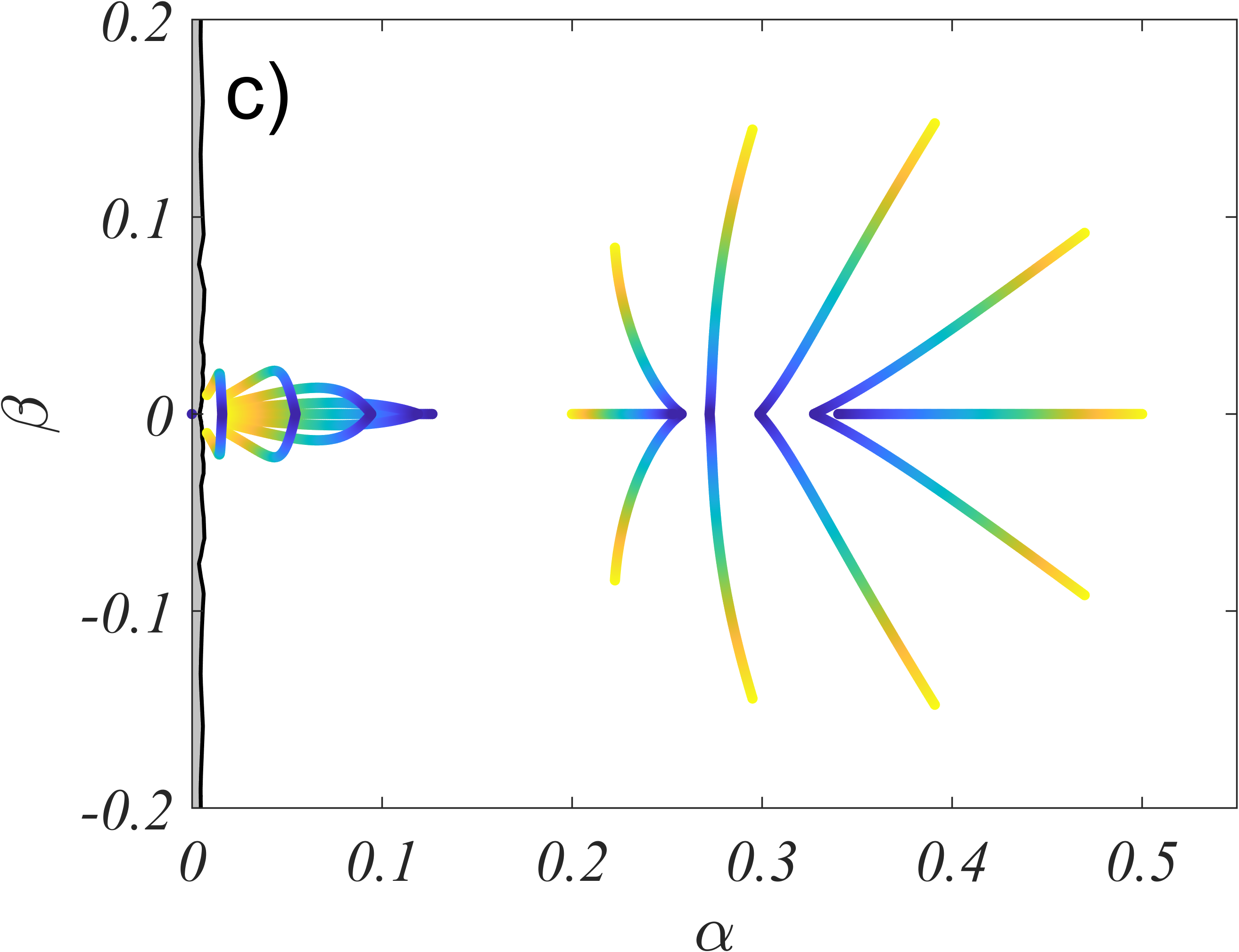}
\includegraphics[width=0.4\linewidth]{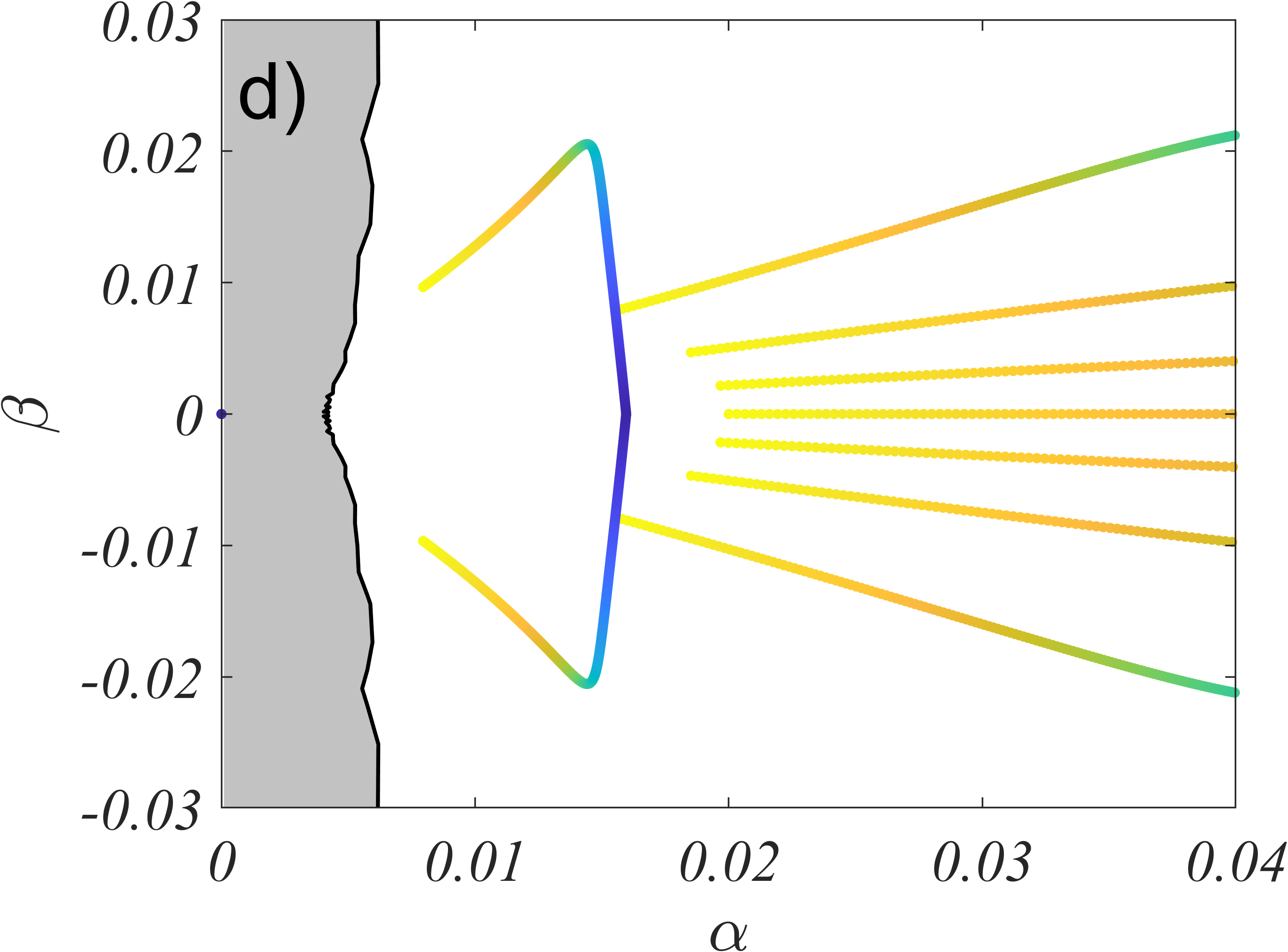}
\caption{\textcolor{black}{\textit{Directionality induced (de)synchronization with $y$-$y$ cubic coupling, considering an alternative symmetrization method}. Panels a)-d) display the locus of the eigenvalues of $\mathcal{M}$ as a function of $q\in[0,1/3]$, for a weighted hypergraph with $N=20$ nodes (color coding is such that the directed case $q=0$ is represented in yellow, and the symmetric one $q=1/3$, in blue). In the background, the white area indicates the region identified by a negative MSF, the black line the boundary of this region, and the gray area the region where the MSF is positive. Panels b) and d) represent a zoom of the area close to the origin of panels a) and c), respectively. Panels a) and b) display that directionality can drive the system unstable, while for the symmetric hypergraph the synchronous state is achieved. Panels c) and d) show that, admitting the directed topology a stable synchronization state, giving the shape of the MSF, the system of oscillators does not lose synchronization as the hypergraph is made symmetric. The coupling strengths for panels a) and b) are $\sigma_1=0.001$ and $\sigma_2=0.12$, while for panel c) and d) they are fixed to $\sigma_1=0.01$ and $\sigma_2=0.16$.}}
\label{fig:msf_rossler_y2y_alternative}
\end{figure}

As we have done for the coupling on the $x$ component, we study the effects of directed topology on synchronization by varying the directionality of the $2$-hyperedges. 
%
\textcolor{black}{Fig.~\ref{eq:rossler_y2y_coupling} shows the variation of the eigenvalues of $\mathcal{M}$ as a function of $p$ for two different sets $(\sigma_1,\sigma_2)$, namely $\sigma_1=0.001$ and $\sigma_2=0.12$, in panels a) and b), and $\sigma_1=0.01$ and $\sigma_2=0.16$, in panels c) and d). Observe that panels b) and d) represent a zoom of the area close to the origin in panels a) and c), respectively. In the background in each panel, the MSF for system \eqref{eq:rossler_y2y_coupling} is represented. In particular, the gray area represents the region where the MSF is positive, the white area portrays the region of stability, while the black line denotes the boundary value $\Lambda_{\mathrm{max}}(\alpha+i\beta)=0$.}
%
{\color{black} From the Figure, it can be noted that the shape of the MSF in this setting allows the system to go unstable only for low values of the parameter $p$, in contrast to the case shown in the main text.} Panels a) and b) show the case where the directed hypergraph ($p=0$) leads to the desynchronization of the system, while the symmetric structure ($p=1$) admits a stable solution. On the other hand, panels c) and d) display the case where, starting from a synchronous state that is stable for $p=0$, by varying the value of $p$ the eigenvalues remain in the area of the complex plane for which the MSF is negative. This means that, given the shape of the MSF, it is not possible to desynchronize the system by making the interactions among triplets of nodes more symmetric. {\color{black} A qualitatively similar behavior is obtained for a system where the method of symmetrization preserving the total coupling strength is applied. In Fig.~\ref{fig:msf_rossler_y2y_alternative}, we show that the synchronous state can be unstable when the higher-order topology is directed ($q=0$) and stabilize as symmetry increases ($q\rightarrow 1/3$), while if the former is already stable, due to the shape of the MSF, the stability is preserved during the symmetrization.}

{\color{black}
\section{Synchronization in random higher-order structures}
\label{app:random_hyper}
\setcounter{equation}{0}
\renewcommand{\theequation}{C\arabic{equation}}
\setcounter{figure}{0}
\renewcommand{\thefigure}{C\arabic{figure}}

Once fixed the MSF, it is the structure of the interactions that, determining the matrix $\mathcal{M}$, and so its eigenvalues, ultimately controls how directionality will impact synchronization. To investigate how the emerging dynamics is connected to the higher-order structure, here we analyze and compare two models for generating random hypergraphs. First, we consider a higher-order structure inspired by the Newman-Watts (NW) model \cite{newman1999scaling}. In particular, we start from an undirected nonlocal ring of $N$ nodes, where each unit is connected to its $m$ nearest neighbors. Then, for each couple of nodes in the network we add a $1$-directed $2$-hyperedge pointing to a third randomly chosen node with probability $\phi$. Second, we take into account a hypergraph version of the Erd\H{o}s-Rényi (ER) model \cite{erdos1960evolution} ruled by two parameters. The first, as in the classical ER model for networks is the probability $\rho_1$ of connecting two nodes with an undirected link, while the second is the probability $\rho_2$ of adding a $1$-directed $2$-hyperedge among three nodes. 

\begin{figure}[t!]
\centering
\includegraphics[width=0.4\linewidth]{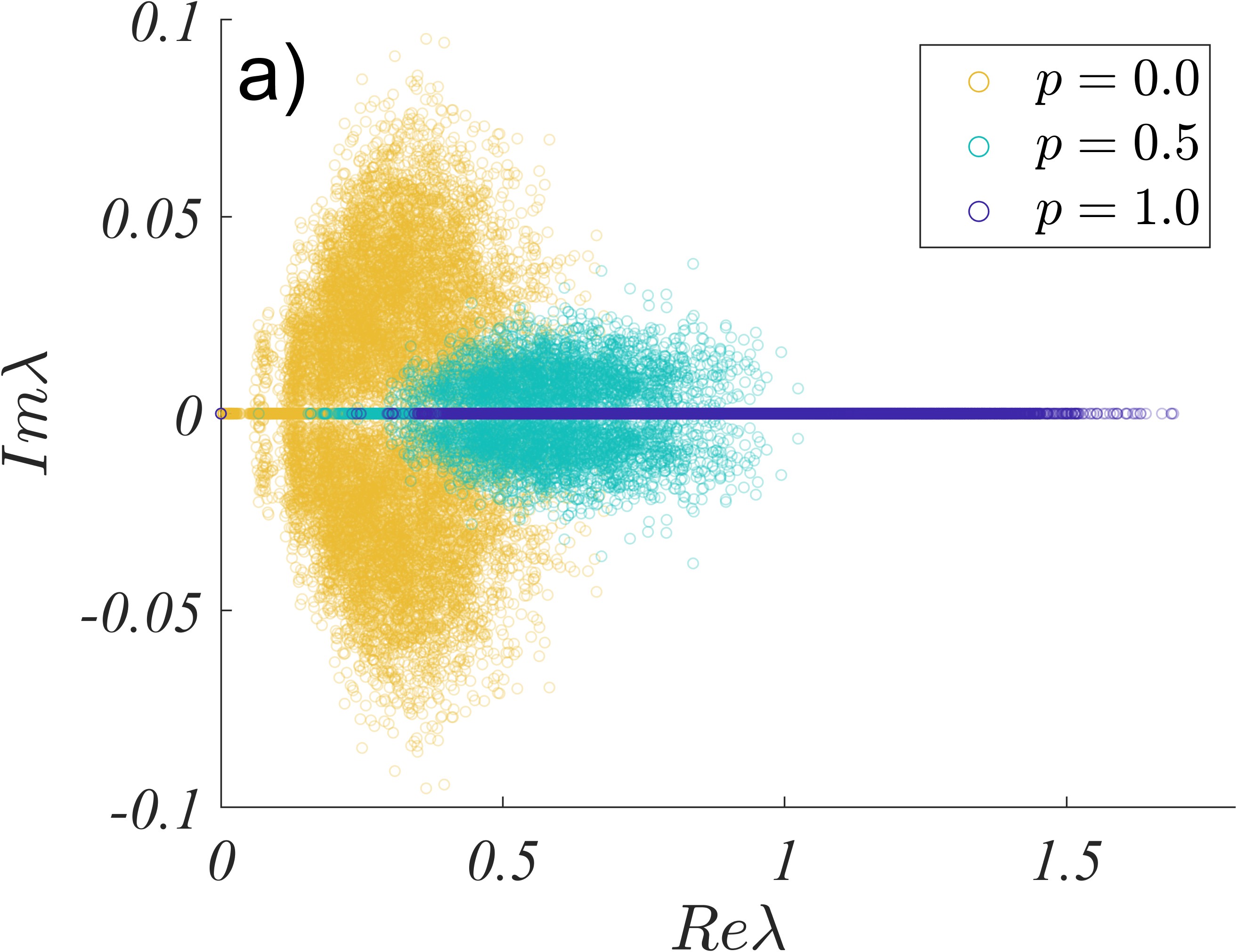}
\includegraphics[width=0.4\linewidth]{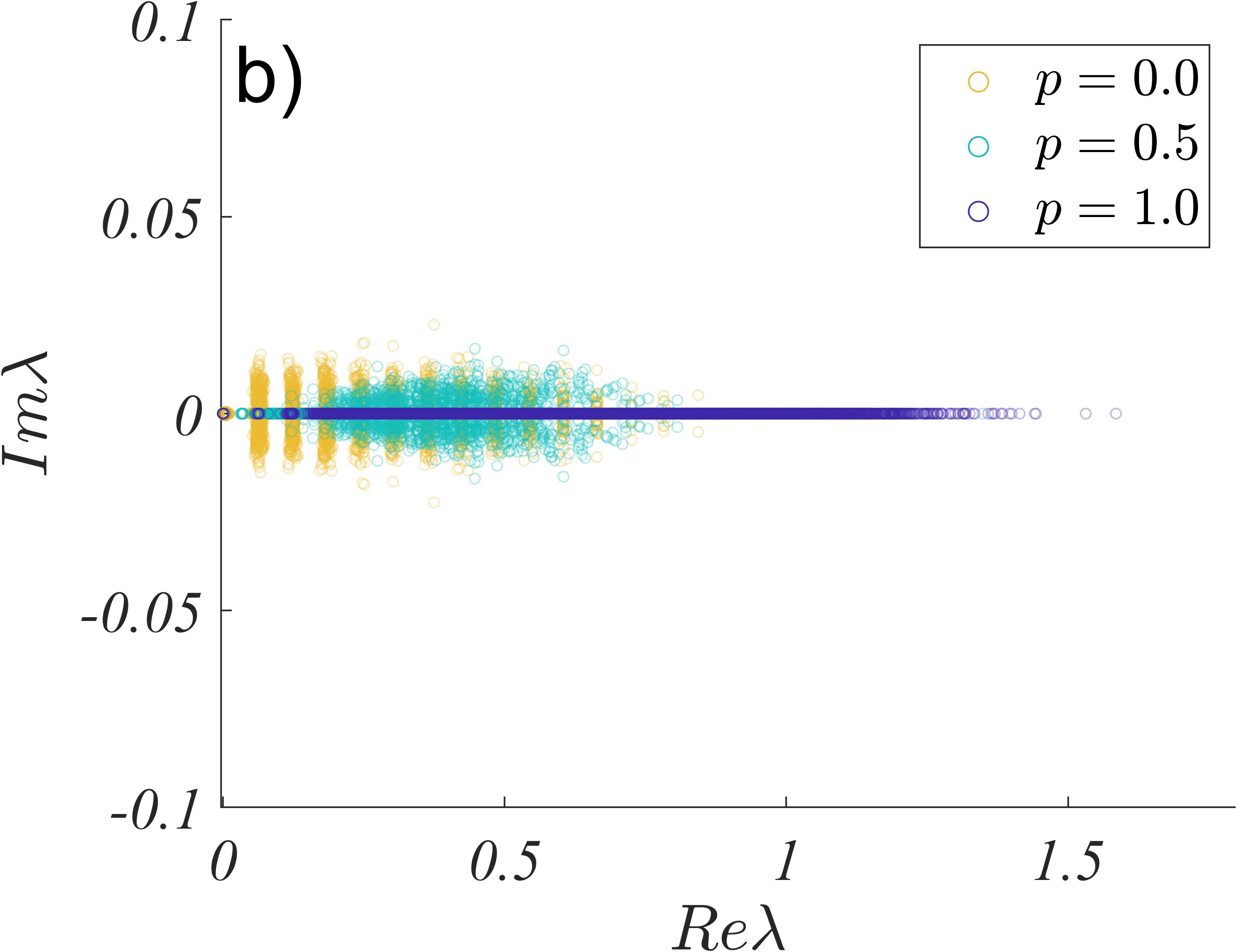}
\includegraphics[width=0.4\linewidth]{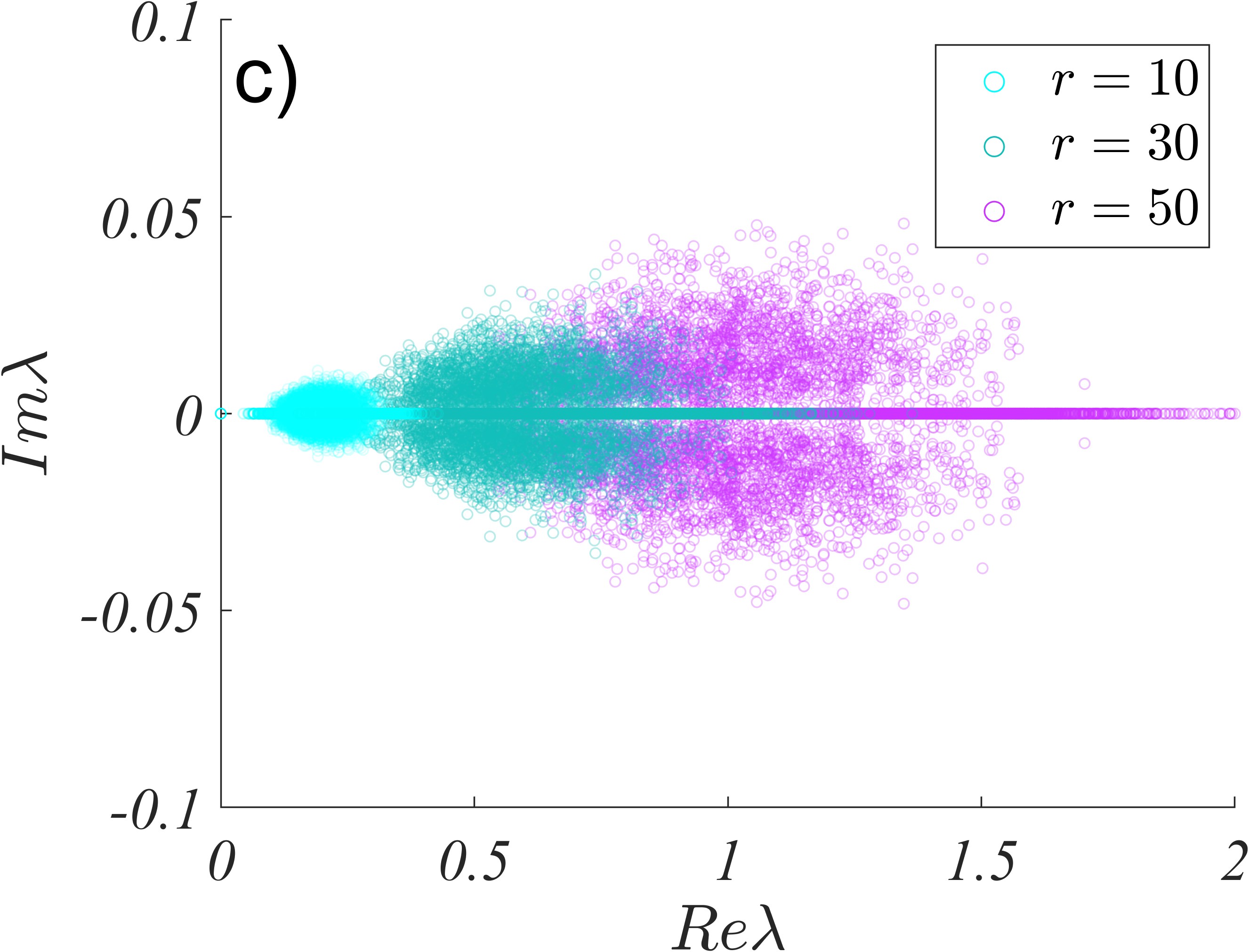}
\includegraphics[width=0.4\linewidth]{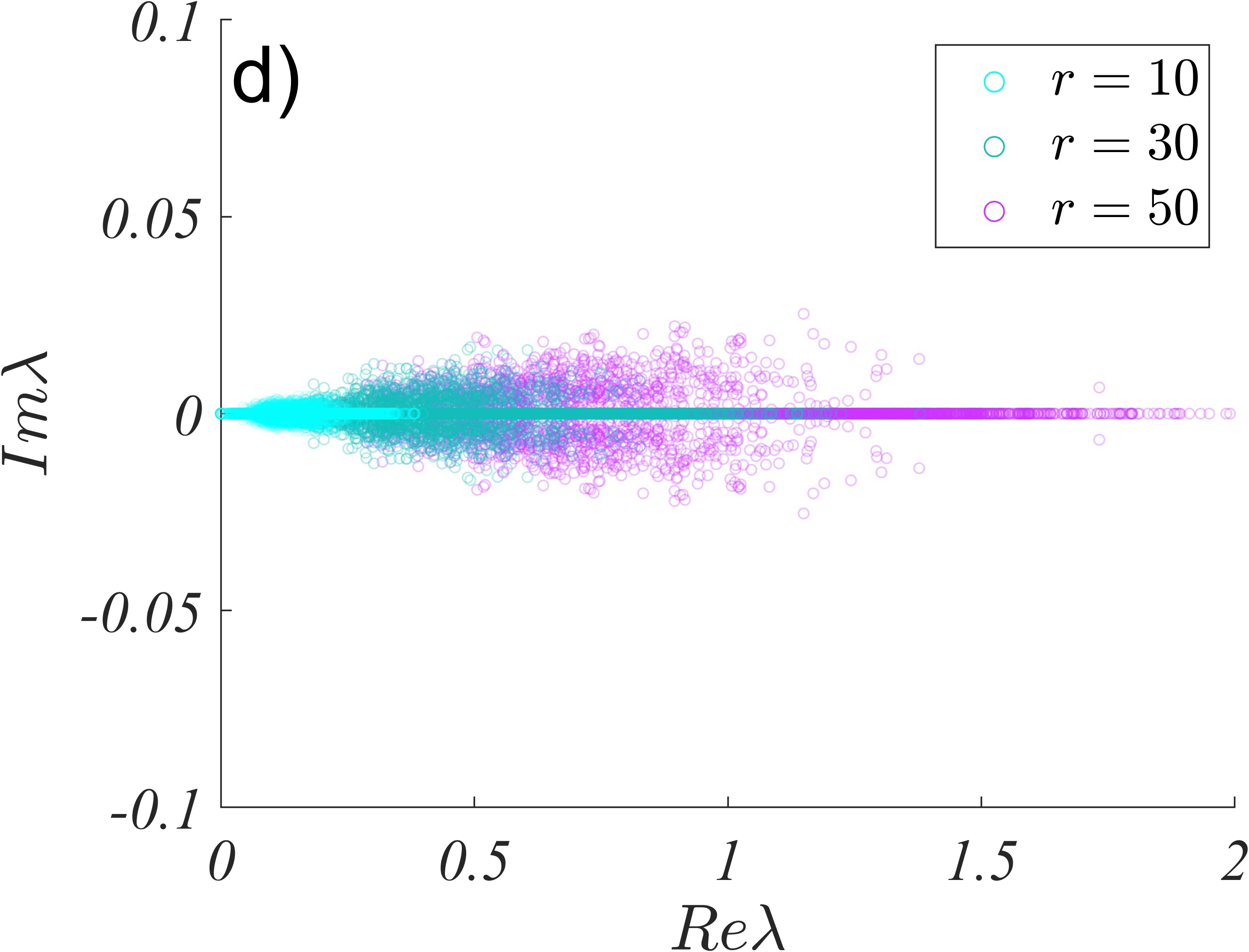}
\caption{\textcolor{black}{\textit{Eigenvalue distribution of random higher-order structures}. a) Variation of the spectrum distributions for the higher-order Newman-Watts model as a function of the symmetry parameter $p$. b) Variation of the distributions for the higher-order Erd\H{o}s-Rényi model as a function of $p$. c) Variation of the distributions for the higher-order Newman-Watts model as a function of the ratio $r$. d) Variation of the distributions for the higher-order Erd\H{o}s-Rényi model as a function of the ratio $r$. In all cases, $\sigma_1 = 0.001$.}}
\label{fig:spectra_random_hypergraphs}
\end{figure}

As the hypergraphs are randomly generated, then the spectrum of the associated matrices $\mathcal{M}$ is also stochastic. Therefore, to understand how synchronization is affected by the hypergraph structure, we need to characterize how the eigenvalues are distributed in the complex plane as a function of the model parameters. In particular, we explore how the spectra of the random hypergraphs vary as a function of the symmetry parameter $p$ and of the ratio $r=\sigma_2/\sigma_1$. In addition, as we also aim at comparing the spectra obtained using the two generative models, we set the model parameters so that the average number of links and the average number of $2$-hyperedges connected to each node are the same for the two algorithms. For each parameter set and for each model, we evaluate the spectrum distribution over $I=1000$ realizations of the hypergraphs.

Fig.~\ref{fig:spectra_random_hypergraphs} displays the eigenvalues of $\mathcal{M}$ as a function of $p$ (panels a) and b) ) and $r$ (panels c) and d) ) for both the NW ( a) and c) ) and the ER ( b) and d) ) higher-order generalization. Typically for $p=0$ there are eigenvalues with nonzero imaginary part such that they are spread into the I and IV quadrants of the complex plane, while these distributions shrink to the real axis for $p=1$ (here we have set $r=30$). We observe that the imaginary part in the NW-like model is generally larger compared to the one of the ER-like model. Similar results are obtained when varying the value of the ratio $r$. For this case, we note that the distributions of eigenvalues remain close to the real axis for small values of $r$, while they spread over the imaginary axis for larger values of $r$ (here we set $p=0.5$). Varying $r$, consistently with what observed above for a fixed value of this parameter, confirms that the eigenvalues in the NW-like model typically have a larger imaginary part compared to their counterparts in the ER-like model.

A comprehensive analysis of how the topological features of a higher-order structure impact on synchronization would require to find the conditions for which the eigenvalues of $\mathcal{M}$ are entirely contained in the stability region. A similar problem appears in the context of pairwise interactions, when directed interactions are considered. Some attempts to elucidate the relationship between eigenvalues of an asymmetric matrix and the emerging synchronous dynamics have been made in \cite{hwang2005synchronization}, but the problem is still open. In the case of higher-order structures, this problem is even more complex as the matrix $\mathcal{M}$ includes contributions from a series of different Laplacian matrices.}

\end{document}